\documentclass[twocolumn]{aastex631}

\usepackage{amsmath,amsfonts,amssymb,amscd,amsthm}
\usepackage{multirow}
\usepackage{xcolor}

\renewcommand{\vec}[1]{{\bf #1}}
\newcommand{\uvec}[1]{\hat{\bf #1}}

\begin{document}

\title{An Evaluation of Different Numerical Methods to Calculate the Pitch-angle\\Diffusion Coefficient from Full-orbit Simulations: disentangling a rope of sand}

\author[0000-0003-1170-1470]{J.P. van den Berg}
\affiliation{Center for Space Research\\
North-West University\\
Potchefstroom, 2522, South Africa}
\correspondingauthor{J.P. van den Berg}
\email{24182869@mynwu.ac.za}

\author[0000-0003-2141-7705]{P.L. Els}
\affiliation{Center for Space Research\\
North-West University\\
Potchefstroom, 2522, South Africa}

\author[0000-0003-3659-7956]{N.E. Engelbrecht}
\altaffiliation{National Institute for Theoretical and Computational Sciences (NITheCS), South Africa}
\affiliation{Center for Space Research\\
North-West University\\
Potchefstroom, 2522, South Africa}

\begin{abstract}

The pitch-angle diffusion coefficient quantifies the effect of pitch-angle scattering on charged particles propagating through turbulent magnetic fields and is a key ingredient in understanding the diffusion of these particles along the background magnetic field. Despite its significance, only a limited number of studies have calculated the pitch-angle diffusion coefficient from test particle simulations in synthetic magnetic turbulence, employing various, often quite different, techniques for this purpose. In this study, we undertake a comparative analysis of nine different methods for calculating the pitch-angle diffusion coefficient from full-orbit simulations. Our objective is to find the strengths and limitations of each method and to determine the most reliable approach. Although all nine methods should theoretically yield comparable results, certain methods may be ill-suited for numerical investigations, while others may not be applicable under conditions of strong turbulence. Through this investigation, we aim to provide recommendations for best practices when employing these methods in future numerical studies of pitch-angle scattering.

\end{abstract}



\section{Introduction}
\label{sec:Intro}

Energetic charged particles diffuse through collisionless plasmas, such as those found in the heliosphere and many astrophysical regions, parallel to some background magnetic field through the process of pitch-angle scattering, where the pitch-angle is the angle between the velocity vector of the particle and the magnetic field vector. Therefore, the pitch-angle diffusion coefficient (PADC) of charged particles in turbulent plasmas must be better understood as it is an essential input for particle transport models used in the study of the transport of such particles  \citep[see, e.g.,][]{Shalchi2009, Shalchi2020, EngelbrechtEA2022} and a key, but often problematic, input for several scattering theories \citep[see, e.g.,][]{FiskEA1974, SakaiKato1984, BieberEA1988, Shalchi2005, Shalchi2009}. Currently, the best way to investigate pitch-angle scattering involves the numerical study of the transport of charged particles in the presence of synthetic turbulence. Although such studies have been extensively done for isotropic  (i.e., pitch-angle independent) diffusion and drift coefficients \citep[see the brief review by][]{EngelbrechtEA2022}, relatively few studies investigating diffusion \textbf{on} a pitch-angle level have been published to date \citep[see, e.g.,][]{QinShalchi2009, QinShalchi2014, TautzEA2013, SunEA2016, PleumpreedapornSnodin2019}.

These studies employ various methods to calculate PADCs, with varying degrees of success. The typical approach to calculate spatial diffusion coefficients from full-orbit simulations is to use the slope of the mean-squared-displacement (MSD) versus time to calculate the diffusion coefficient at late times, where the system should be in a diffusive state (see Sec.~\ref{subsec:MSD} for more details on this approach). At first glance, it would seem relatively simple to extend the same method to the pitch-cosine \citep[see, e.g.,][]{QinShalchi2009, QinShalchi2014, DundovicEA2020}. \citet{RiordanPeer2019}, however, demonstrate that this approach yields a running PADC with initial oscillations that die out and give way to a $1/t$ behaviour at later times \citep[as also illustrated by][]{TautzEA2013}. Additionally, \citet{RiordanPeer2019} show that the limited temporal range over which the PADC might reach a constant value disappears into the initial oscillations and the eventual decay as the turbulence strength is increased, although they then choose to construct their PADC after 20 gyrations. Although \citet{QinShalchi2009} do not explicitly state how they calculate the PADC, they do state that they do not find a constant PADC if the turbulence strength becomes too high. In light of this, it is unclear how \citet{QinShalchi2014} have addressed this issue for the stronger turbulence used in their simulations. Both \citet{IvascenkoEA2016} and \citet{DundovicEA2020} demonstrate how the PADC changes with time when employing the MSD, although \citet{IvascenkoEA2016} do not state at what time they construct the final PADCs they show, while they also show that this approach fails in strong turbulence.

\citet{WeidlEA2015} use the slope of a line from the origin passing through the minimum of the MSD after one gyroperiod to estimate the shape of the PADC. They do this to avoid the asymptotic temporal behaviour of the MSD and to minimize the effect of energy diffusion due to their electric fields. While acknowledging that this approach will not yield the correct scattering amplitude, they expect it to preserve the correct pitch-angle dependence. \citet{PleumpreedapornSnodin2019} find that constructing the PADC from the MSD after a gyroperiod yields larger values than an alternative method \citep[something expected from theoretical considerations by][]{WeidlEA2015}. \citet{TautzEA2013} and \citet{IvascenkoEA2016} point out that this asymptotic temporal behaviour is due to the pitch-angle being bounded, causing the MSD to eventually saturate. This behaviour of the MSD is analytically investigated and further discussed by \citet{Tautz2013}. \citet{TautzEA2013} use quasi-linear theory \citep[QLT; see, e.g.,][]{Jokipii1966, Shalchi2009} to calculate the MSD and then solve the expression for the time when it reaches unity, demonstrating that the time when the PADC could be constructed, assuming that the MSD will start to saturate at later times, is pitch-angle dependent. This estimate, however, cannot be used in situations where QLT might not be valid. \citet{TautzEA2013} therefore choose to construct their PADC after a hundred cyclotron frequencies (i.e., $\sim 16$ gyrations) since the oscillations have then died out.

An alternative method to calculate the PADC is by integrating the diffusion equation for the pitch-angle distribution function (see Sec.~\ref{subsec:TPE} for more details on this approach). \citet{IvascenkoEA2016} present two integration methods, still yielding a running PADC having a constant value over some time interval. They show that the MSD approach yields different results than integrating the pitch-angle diffusion equation due to the effect of resonances of particles with specific wave modes on the MSD. Unfortunately, a smooth distribution function is required since its derivatives are calculated, implying that the distribution function might need to be smoothed or that other schemes might need to be employed to reduce the noise of the derivatives. \citet{PleumpreedapornSnodin2019}, for example, calculate the discrete time derivative over an interval equal to or larger than the gyroperiod and then average this result over some unspecified time. \citet{KaiserEA1978} use a steady-state solution of the diffusion equation to calculate the PADC, where the system is forced to a stationary solution by continuously injecting particles and having absorbing boundaries. \citet{SunEA2016} also employ this method but applies a sliding window and averaging procedure to obtain better results in smaller pitch-cosine bins. This method, however, is not valid close to the source \citep[something that is not addressed by][]{SunEA2016} because the particles must first reach diffusive behaviour.

\textbf{Previous studies on calculating the PADC from full-orbit simulations \citep[see, e.g.,][]{QinShalchi2009, QinShalchi2014, TautzEA2013, SunEA2016, PleumpreedapornSnodin2019} mostly consider low levels of turbulence (i.e., $10^{-4} \le \delta B^2 / B_0^2 \le 1$, although the latter few references do consider stronger turbulence levels, where $\delta B^2$ is the variance of the turbulence and $B_0$ is the background magnetic field strength). It will be seen in the following investigation that all methods considered here present problems to a greater or lesser degree when implemented in high levels of turbulence (i.e., $\delta B^2/B_0^2 \gtrsim 1$). This, however, is not necessarily a limitation in the heliosphere, where both observations \citep[see, e.g.,][]{SmithEA2006, PineEA2020, BurgerEA2022} and turbulence transport modeling \citep[see, e.g.,][]{OughtonEA2011, WiengartenEA2016, UsmanovEA2016, AdhikariEA2021, EngelbrechtEA2022} suggest that $\delta B^2 / B_0^2$ would vary between $\sim 10^{-2}$ and $\sim 0.7$, in the ecliptic plane at least, should the influence of pickup ion-induced waves \cite[see, e.g.,][]{WilliamsZank1994, Isenberg2005} be neglected. This latter omission is a reasonable assumption in the inner heliosphere, and it should also be noted that these fluctuations are not expected to influence the transport of higher-energy charged particles \citep{Engelbrecht2017}. Therefore, numerical test particle simulations can still provide useful insights into the behavior of the PADC for heliospheric particle transport studies if caution is exercised in the choice of the method employed to calculate this quantity and in the interpretation of the obtained results.}

It is clear that several different methods to calculate the PADC from single-particle simulations have been proposed in the literature, each with potential limitations and constraints. Theoretically, it might be expected that these methods should all yield the same result, but in practice, they may not due to the different numerical constraints highlighted above. It is then the aim of this paper to investigate the different methods to calculate the PADC from full-orbit simulations of charged particles in synthetic magnetic turbulence for various turbulence scenarios. The next section introduces the numerical test particle propagation code used in this study. Sec.~\ref{sec:Methods} provides a summary of all the different methods that can be used to calculate the PADC, which are then compared to each other and previous work in Sec.~\ref{sec:Benchmark}. Lastly, the implications of this investigation and recommendations based on it will be discussed in Sec.~\ref{sec:Discuss}.


\section{The Simulation Code}
\label{sec:Code}

The various methods to be investigated for calculating the PADC all use full-orbit simulations in generated synthetic magnetic turbulence, {\bf using the particle pusher code discussed and tested by \citet{ElsEngelbrecht2024}}. All (pseudo) random numbers (PRNs) used in both the generation of synthetic turbulence and for the random aspects of the full-orbit trajectories were tested using the \textsl{dieharder} battery of statistical tests \citep{BrownEA2018}, which compares generated PRNs to statistical noise. An implementation of the Mersenne Twister algorithm by \citet{MatsumotoNishimura1998} is used as the generator for all PRNs in the code. Parallelization is done using the OpenMP API provided by \citet{ChandraEA2001}. All Fourier transforms are done using the FFTW library \citep{FrigoJohnson2005}, which also offers multi-threading capabilities through OpenMP.

For this work, all particle trajectories are calculated using the fully relativistic form of the semi-symplectic method (i.e., the phase-space volume of the particle is preserved even though the method itself is not symplectic) of \citet{Vay2008}, which has been shown to preserve the energy of charged particles and plasma remarkably well over long time intervals \citep[see, e.g.,][]{HigueraCary2017, Vay2020}. The accuracy of the full-orbit simulations has been verified using a set of electromagnetic configurations \citep[similar to those employed by][]{Petri2017, RipperdaEA2018} with either known analytic expressions for the equations of motion of charged particles present in the setup or by comparison with known statistical properties of the particles in such a field. Particle ensembles are constructed by tracking the full-orbit trajectories of each of some number of particles through a particular field realization. Initially, particles are spatially separated and their initial velocity vectors are chosen so to allow the field realization to be fully statistically sampled. If the field realization contains fluctuations, then the particles must remain in the field long enough to achieve a diffusive limit to calculate isotropic diffusion coefficients. However, the time required for this often varies depending on the simulation setup, with a default time of $\sim 2000$ \textit{correlation crossing times} often used. Given that the present study concerns pitch-angle scattering, this aspect will be discussed further below. Simulations are typically repeated for $10$ to $25$ field realizations, and the collection of particle orbits is then used as an ensemble to calculate diffusion coefficients.

\textit{Slab} and \textit{2D} turbulence are generated using Fourier noise to obtain a composite model of synthetic turbulence \citep[see, e.g.,][]{Owens1978, MatthaeusEA1990, BieberEA1996, Qin2002, Minnie2006}. The 1D fluctuating components vary along the background magnetic field direction, while the 2D fluctuating components vary in the plane perpendicular to the background field. The maximum physical dimensions for both are specified along with the numbers defining the resolution of the respective fields (thus determining the nodes per unit length of the field). This, therefore, defines the largest physical fluctuations allowed in the system, $L$, as well as the Nyquist frequency of the fluctuations. The field realizations are treated as periodic beyond this outer boundary (which is smooth on the periodic boundary as a consequence of the Fourier noise algorithm), and the maximal Larmor radius ($R_L$) of a particular particle is kept well above the spacing determined by the grid/Nyquist fluctuations. This ensures that the maximal Larmor radius of the particle remains within the defined spectrum of fluctuations for the turbulence. \textbf{Only magnetostatic turbulence is considered and it is assumed that the background magnetic field is directed along the $z$-axis, such that the transverse fluctuations are directed along the $x$- and $y$-axis. The magnetic field is therefore expressed as $\vec{B} = \delta B_x \, \uvec{e}_x + \delta B_y \, \uvec{e}_y + B_0 \, \uvec{e}_z$, with $\uvec{e}_i$ unit vectors along the Cartesian axes, $\delta B_i$ fluctuating components, and $B_0$ the strength of the background magnetic field. In order to have a well-defined pitch-angle, the pitch-cosine is calculated relative to the background magnetic field, i.e., $\mu = \vec{v} \cdot \vec{B} / v B \approx \vec{v} \cdot \vec{B}_0 / v B_0 = v_z / v$, which should be a valid assumption for low turbulence levels (i.e., $\delta B^2 \ll B_0^2$), where $\vec{v}$ and $v$ is the particle velocity and speed, respectively. Note that the methods discussed in the following section are agnostic to the choice of the magnetic field orientation and that it will be clear where this setup is assumed by the usage of specific Cartesian components.}

The spectral form used to generate slab fluctuations is similar to that employed by, e.g., \citet{RuffoloEA2006}, \citet{MinnieEA2007}, or \citet{TautzEA2013}, in that it consists of a wavenumber-independent energy-containing range and a Kolmogorov inertial range (with spectral index $\nu = 5/3$), and is given by
\begin{equation}
\label{eq:SlabSpectrum}
g_{\rm slab}(k_{\parallel}) \propto \left[ 1 + (l_b^{\rm slab} k_{\parallel})^2 \right]^{-\nu/2} .
\end{equation}
The 2D spectrum used here is again similar to that employed by the aforementioned studies, with the same wavenumber dependence as the slab spectrum, such that
\begin{equation}
\label{eq:2DSpectrum}
g_{\rm 2D}(k_{\perp}) \propto \left[ 1 + (l_b^{\rm 2D} k_{\perp})^2 \right]^{-\nu/2} .
\end{equation}
Here, $l_b$ is the bendover scale at which the inertial range of the power spectrum commences and the fluctuations are normalized such that the area under the spectrum is equal to the specified slab or 2D variance of the fluctuations, $\delta B^2$. The power corresponding to the total (slab+2D) variance is divided among the slab and 2D fluctuations (i.e., $\delta B^2 = \delta B_{\rm slab}^2 + \delta B_{\rm 2D}^2$). The total variance must thus be the area between the grid scale and the largest scale of the simulation box, with the window specified by the fluctuation boundaries cutting out some portion of that power ($l_{\rm max}$ and $l_{\rm min}$ below). In practice, the error that would arise from using $l_{\rm max}$ and $l_{\rm min}$ as the bounds of integration for the power is small since these variables tend to the simulation limits when using high-resolution field realizations. {\bf For more detail, see \citet{ElsEngelbrecht2024}.}

\citet{Qin2002} provides a thorough discussion of the relationships between physical parameters using such a setup. The following must be adhered to when selecting parameters (note that this is not dimension-dependent):
\begin{equation}
L > l_{\rm max} > \lambda_c > l_{\rm min} > h,
\end{equation}
where $h$ is the grid size and $\lambda_c = \sqrt{\pi} \, \Gamma (\nu/2) l_b / \Gamma (\nu/2 - 1/2) \approx 0.747 \, l_b$ is the \emph{correlation length} associated with the turbulence spectrum, with $\Gamma (x)$ the gamma function. Typically, the following values are used for these parameters, unless otherwise specified: $l_{\rm min}^{\rm slab} = 10^{-4}~{\rm au}$ \textbf{(i.e., astronomical units)}, $l_{\rm max}^{\rm slab} = 1~{\rm au}$, $L_z = 10~{\rm au}$, $N_z = 2^{23}$, $l_{\rm min}^{\rm 2D} = 1.5 \times 10^{-4}~{\rm au}$, $l_{\rm max}^{\rm 2D} = 0.15~{\rm au}$, $L_x = L_y = 1~{\rm au}$, $N_x = N_y = 2^{14}$, along with using a fixed time step for the Vay method set to the gyroperiod over $64$ \textbf{($\delta t = P_{\rm gyro} / 64$)}. This implies the grid separation for slab and 2D fluctuations is $1.2 \times 10^{-6}~{\rm au}$ and $6.1 \times 10^{-5}~{\rm au}$, respectively. \textbf{Some of} these parameters are changed in what follows below and will be mentioned and discussed in Sec.~\ref{sec:Benchmark}.


\section{Methods to Calculate the Pitch-angle Diffusion Coefficient}
\label{sec:Methods}

In deriving the focused transport equation (FTE, which is valid for a gyrotropic distribution) from the Kolmogorov equation, the so-called Fokker-Planck \emph{pitch-angle diffusion coefficient} (PADC) is
\begin{subequations}
\begin{align}
\label{eq:DmmFPint}
D_{\mu\mu}(\mu) & = \int_{-2}^2 \! \frac{(\Delta \mu)^2}{2 \Delta t} P(\Delta \mu \, | \, \mu, \Delta t) \, {\rm d}(\Delta \mu) \\
\label{eq:DmmFP}
 & = \left\langle \frac{(\Delta \mu)^2}{2 \Delta t} \right\rangle ,
\end{align}
\end{subequations}
where $P(\Delta \mu \, | \, \mu, \Delta t)$ is the conditional probability distribution function (PDF) of the particles experiencing a change in pitch-cosine $\Delta \mu$ in the time interval $\Delta t$ for a given pitch-cosine $\mu$ and the definition of a weighted average was used \textbf{(the latter expression is referred to as M0 in Sec.~\ref{sec:Benchmark})}. Underlying the Kolmogorov equation is the assumption of a Markov process, meaning that the state of the system after the time interval $\Delta t$ is only dependent on the current state and not the history of the system \citep{Chandrasekhar1943, Jokipii1966}. It is therefore perceivable that there is some condition on the time interval here: if $\Delta t$ is too small, then the pitch-angle scattering will not be resolved and the PDF will be a delta function (i.e., motion appears deterministic); if $\Delta t$ is too large, then there would be no correlation between $\Delta \mu$ and $\mu$, resulting in a uniform distribution for the PDF.

Calculating the PADC from computer simulations, therefore, involves finding a method to determine the average change in pitch-cosine within a given time interval as a function of the pitch-cosine itself, presenting a unique challenge not encountered when calculating spatial diffusion coefficients \citep{TautzEA2013}. In this section, we introduce different methods that can be used for this purpose. We note that methods M1a, \textbf{M3,} M4a, M4b, and M5 have been previously employed to calculate the PADC, while methods M1b and M2a are usually used in theoretical work deriving the PADC. We propose and test method M2b as \textbf{a} new alternative and compare the different methods in Sec.~\ref{sec:Benchmark}.


\subsection{Mean-squared-displacement}
\label{subsec:MSD}

In the case of unbounded and constant spatial diffusion, an initial delta function in time and space evolves into a Normal distribution. Using the resulting probability distribution, it is found that the \emph{mean-squared-displacement} (MSD) increases linearly with time, with the rate of increase dependent on the diffusion coefficient \citep[see, e.g.,][]{Shalchi2009, Tautz2013}. One might be tempted to naively write
\begin{equation}
\label{eq:WrongMSD}
D_{\mu\mu}(\mu(0)) = \lim_{t \longrightarrow \infty} \left\langle \frac{[\mu(t) - \mu(0)]^2}{2 t} \right\rangle ,
\end{equation}
where \textbf{$\mu(t)$ is the temporal evolution of the pitch-angle and} the angle brackets indicate a suitable average \textbf{(mainly an ensemble average of particles having the same initial pitch-cosine $\mu(0)$)}. However, since $\mu \in [-1; 1]$, $|\Delta \mu|$ cannot be larger than 2 (note that this also explains the integration limits in Eq.~\ref{eq:DmmFPint}), implying that $\left\langle [\mu(t) - \mu(0)]^2 / 2 t \right\rangle$ will tend to zero as $t \longrightarrow \infty$ \citep{TautzEA2013, IvascenkoEA2016, PleumpreedapornSnodin2019, RiordanPeer2019}. \citet{Tautz2013} demonstrates this saturation of the MSD analytically by deriving an expression for the MSD from the pitch-angle diffusion equation (see Sec.~\ref{subsec:TPE}) under the assumption of isotropic pitch-angle scattering (i.e., $D_{\mu\mu} \propto 1 - \mu^2$).

It is therefore necessary to find the time interval over which the running PADC,
\begin{equation}
\label{eq:DmmMSDt}
D_{\mu\mu}(\mu(0),t) = \left\langle \frac{[\mu(t) - \mu(0)]^2}{2 t} \right\rangle ,
\end{equation}
is constant, that is, the time where $t_c < t < t_a$, with $t_c$ being the time when the pitch-angle is correlated with its initial value and $t_a$ being the time when the MSD approaches its asymptotic value \citep{Tautz2013, RiordanPeer2019}. Alternatively, \citet{Shalchi2009} defines the running diffusion coefficient as the time derivative of the MSD,
\begin{equation}
\label{eq:DmmMSDdt}
D_{\mu\mu}(\mu(0),t) = \frac{\rm d}{{\rm d}t} \left\langle \frac{[\mu(t) - \mu(0)]^2}{2} \right\rangle ,
\end{equation}
thereby relaxing the assumption that the MSD is linearly proportional to time. These two methods will be referred to as M1a (Eq.~\ref{eq:DmmMSDt}) and M1b (Eq.~\ref{eq:DmmMSDdt}), respectively, and should yield comparable answers, at least for spatial diffusion, if the running diffusion coefficient becomes time-independent \citep{Tautz2013, TautzEA2013}. \citet{DundovicEA2020} claim that the former method takes longer to converge compared to the latter method for spatial diffusion.


\subsection{TGK Approach and Pitch-angle Scattering Time}
\label{subsec:TGK}

\citet{Shalchi2009} demonstrates that for unbounded spatial diffusion, the diffusion coefficient can be related to the speed correlation function by taking the time derivative of the MSD calculated from the \emph{Taylor-Green-Kubo} (TGK) \emph{formalism}. The second method, which will be referred to as M2a, can be written as
\begin{equation}
\label{eq:DmmTGK}
D_{\mu\mu}(\mu(0)) = \int_0^{\infty} \!\!\! \left\langle \dot{\mu}(0) \dot{\mu}(\tau) \right\rangle \, {\rm d}\tau ,
\end{equation}
where $\dot{\mu}$ is the instantaneous rate of change of the pitch-cosine and $\tau$ is the temporal lag. This method should again be comparable to the previous two methods, at least for spatial diffusion, if the running diffusion coefficient becomes time-independent, but this method might be numerically ill-suited due to noise in the correlation function accumulating through the integral \citep{Tautz2013, TautzEA2013, RiordanPeer2019}.

In contrast to spatial diffusion, where the particle velocity is readily available from the simulation, $\dot{\mu}$ is not. However, the definition of the pitch-cosine relative to a static and homogeneous background magnetic field can be used to calculate
\begin{equation}
\dot{\mu}(t) = \frac{\rm d}{{\rm d}t} \left[ \frac{\vec{v} \cdot \vec{B}_0}{v B_0} \right] = \frac{\vec{B}_0}{v B_0} \cdot \frac{{\rm d}\vec{v}}{{\rm d}t} - \frac{\vec{v} \cdot \vec{B}_0}{v^2 B_0} \frac{{\rm d}v}{{\rm d}t} .
\end{equation}
For the magnetic field orientation assumed in the simulations (see Sec.~\ref{sec:Code}), employing ${\rm d}v/{\rm d}t = (\vec{v} / v) \cdot ({\rm d}\vec{v}/{\rm d}t)$ and the Newton-Lorentz equation (${\rm d}\vec{v}/{\rm d}t = q \, \vec{v} \times \vec{B} / m$, where $q$ and $m$ is the particle charge and relativistic mass, respectively), the above expression becomes
\begin{align}
\label{eq:dmudt}
\dot{\mu}(t) = \; & \frac{q}{p} \left\lbrace \left( 1 - \frac{v_z^2}{v^2} \right) (v_x \, \delta B_y - v_y \, \delta B_x) \, - \right. \\
 & \left. \frac{v_z}{p^2} [ v_x (v_y B_0 - v_z \, \delta B_y) + v_y (v_z \, \delta B_x - v_x B_0) ] \right\rbrace \nonumber ,
\end{align}
which can be computed from the simulations.

Somewhat related to the concept of a speed correlation, is the \emph{pitch-angle correlation function},
\begin{equation}
\label{eq:PACF}
C_{\mu\mu}(\mu(0),\tau) = \left\langle \mu(0) \mu(\tau) \right\rangle
\end{equation}
\textbf{for a suitable average over particles with a specific initial pitch-cosine,} from which a \emph{pitch-angle correlation time} can be calculated,
\begin{equation}
\label{eq:PACorrelTime}
T_{\mu}(\mu(0)) = \frac{1}{C_{\mu\mu}(\mu(0),0)} \int_0^{\infty} \!\!\! C_{\mu\mu}(\mu(0),\tau) \, {\rm d}\tau .
\end{equation}
Note that the parallel (pitch-angle independent) \textbf{correlation time} (which will be indicated by $T_{\parallel}$) is usually considered by integrating $3 \left\langle v_{\parallel}(0) v_{\parallel}(\tau) \right\rangle / v^2 \approx 3 \left\langle v_z(0) v_z(\tau) \right\rangle / v^2$ \textbf{(for the simulation setup assuming $\delta B^2 \ll B_0^2$, such that the direction of the total magnetic field can be approximated by the direction of the background magnetic field along the $z$-axis)}, while it might be possible to infer the scattering amplitude from this by deriving its time dependence from the pitch-angle diffusion equation (see Sec.~\ref{subsec:TPE}) if the pitch-angle dependence of the PADC is known \citep[see, e.g.,][for isotropic scattering]{ShalchiEA2012, Tautz2013}. Due to \textbf{statistical} noise in \textbf{constructing} the correlation function at late times, it might be preferable to fit an exponential decay time, but the correlation and $e$-folding times are not the same in the general case if $C_{\mu\mu}$ does not follow an exponential decrease. It is also interesting to note that there is a relation between the MSD and the correlation function,
\begin{equation}
\label{eq:MSD-CF-Relate}
\left\langle [\mu(t) - \mu(0)]^2 \right\rangle = \left\langle \mu^2(0) \right\rangle + \left\langle \mu^2(t) \right\rangle - 2 \left\langle \mu(0) \mu(t) \right\rangle ,
\end{equation}
which is reminiscent of a second-order structure function \cite[see, e.g.,][]{HuangEA2010}.

In the scattering theories of \citet{ShalchiEA2004b} and \citet{leRouxWebb2007}, it is assumed that the pitch-angle scattering rate is related to the PADC \citep[an idea also explored by][for perpendicular diffusion]{Engelbrecht2019}. It may be plausible that this correlation time could be given by
\begin{equation}
\label{eq:DmmCorrelTime}
T_{\mu}(\mu) = \frac{1 - \mu^2}{2 D_{\mu\mu}(\mu)} ,
\end{equation}
where the factor of two was chosen such that the correlation time for isotropic \textbf{pitch-angle} scattering with its scattering rate normalized to the parallel mean-free-path (MFP; $\lambda_{\parallel}$) would be equal to the usual assumption of $T_{\parallel} = \lambda_{\parallel} / v$ for an isotropic distribution \citep[see, e.g.,][]{MatthaeusEA2003}. We note, however, that \citet{RiordanPeer2019} do not assume that the scattering time is inversely proportional to the PADC and that they view \textbf{the scattering and correlation time} as two separate time scales. This method of calculating the PADC (Eq.~\ref{eq:DmmCorrelTime}), which will be referred to as M2b, has the advantage that $C_{\mu\mu}$ could be used (for example) in the pitch-angle dependent perpendicular diffusion coefficient of \citet{RuffoloEA2012}'s random ballistic decorrelation interpretation of the non-linear guiding centre theory \citep[as discussed by][]{Engelbrecht2019}. Note, however, that this is an assumption and not a formal definition, implying that this method would be an empirical method if proved from the simulations to hold.


\subsection{Adapted Giacalone et al. (1999) Method}
\label{subsec:Gea99}

When calculating the off-diagonal elements of the isotropic diffusion tensor, which are anti-symmetric and contain information about drifts in a non-homogeneous magnetic field, the MSD does not yield the correct oddness and \citet{GiacaloneEA1999} therefore proposed a different numerical method to calculate the drift coefficient. Appendix~\ref{apndx:DeriveGea99} demonstrates that a similar derivation to that of \citet{GiacaloneEA1999} \citep[also see the derivation of][for clarifications]{Minnie2006} can be done in momentum space for a gyrotropic distribution to derive \textbf{\citep{TautzEA2014}}
\begin{equation}
\label{eq:DmmGea99}
D_{\mu\mu}(\mu(0),t) = \langle \dot{\mu}(t) [\mu(t) - \mu(0)] \rangle .
\end{equation}
This method\textbf{, which is comparable to that of \citet{TautzEA2014},} is postulated as an analogy to that of \citet{GiacaloneEA1999} and will be referred to as M3.

This expression naturally follows from Eq.~\ref{eq:DmmMSDdt} if the MSD is not viewed as a single function, such that the time derivative and averaging can be interchanged, i.e.,
\begin{align}
\frac{\rm d}{{\rm d}t} \left\langle \frac{[\mu(t) - \mu(0)]^2}{2} \right\rangle & \approx \left\langle [\mu(t) - \mu(0)] \frac{\rm d}{{\rm d}t} [\mu(t) - \mu(0)] \right\rangle \nonumber \\
 & = \left\langle [\mu(t) - \mu(0)] \frac{{\rm d}\mu}{{\rm d}t} \right\rangle .
\end{align}
Similarly, for the TGK approach of Eq.~\ref{eq:DmmTGK},
\begin{align}
\int_0^t \!\! \left\langle \dot{\mu}(t') \dot{\mu}(t'+\tau) \right\rangle \, {\rm d}\tau & \approx \left\langle \dot{\mu}(t') \int_0^t \! \frac{{\rm d}\mu}{{\rm d}\tau} \, {\rm d}\tau \right\rangle \nonumber \\
 & = \left\langle \dot{\mu}(t) [\mu(t) - \mu(0)] \right\rangle ,
\end{align}
where the assumption of temporal homogeneity (valid for homogeneous magnetostatic turbulence) was relaxed to have a more general time dependence \textbf{\citep[see also][where $\dot{\mu}(0)$ is used and not $\dot{\mu}(t)$]{TautzEA2014}}.


\subsection{Time-dependent Solution of the Diffusion Equation}
\label{subsec:TPE}

The \emph{pitch-angle diffusion equation} (alternatively, the FTE without energy changes averaged over space, which is valid for homogeneous magnetostatic turbulence), given by
\begin{subequations}
\label{eq:TPE}
\begin{align}
\label{eq:TPEconserve}
\frac{\partial f}{\partial t} & = \frac{\partial}{\partial \mu} \left[ D_{\mu\mu} \frac{\partial f}{\partial \mu} \right] \\
\label{eq:TPEnonConserve}
 & = \frac{\partial D_{\mu\mu}}{\partial \mu} \frac{\partial f}{\partial \mu} + D_{\mu\mu} \frac{\partial^2 f}{\partial \mu^2} ,
\end{align}
\end{subequations}
describes the effect of pitch-angle scattering on the \emph{pitch-angle distribution function} $f(\mu,t)$. We note that \citet{GiacaloneJokipii1999} fitted analytical solutions of the isotropic diffusion equation to their simulation results to calculate the parallel diffusion coefficient and compare this to the MSD method. Unfortunately, since $D_{\mu\mu}$ is dependent on $\mu$, it is not possible to have an analytical solution for Eq.~\ref{eq:TPE} if the form of $D_{\mu\mu}$ is not known beforehand \citep[][provides an example for isotropic scattering]{Tautz2013}. However, integrating Eq.~\ref{eq:TPEconserve} from $-1$ to $\mu$ allows the PADC to be calculated as \citep{Tautz2013, IvascenkoEA2016}
\begin{equation}
\label{eq:DmmIntDiffEq}
D_{\mu\mu}(\mu,t) = \left[ \int_{-1}^{\mu} \! \frac{\partial f(\eta,t)}{\partial t} \, {\rm d}\eta \right] \left[ \frac{\partial f(\mu,t)}{\partial \mu} \right]^{-1} .
\end{equation}

In practice, the distribution function is approximated by a discrete function $\hat{f}(\mu_m,t_s)$ at several pitch-cosine and temporal bins with midpoints $\mu_m$ and $t_s$, respectively (see Sec.~\ref{subsec:Stationary} for details\textbf{; note that here and in Sec.~\ref{subsec:Stationary}, non-bold-faced quantities with hats indicate discretization of a continuous function by binning; although hats are also used to indicate unit vectors, such quantities are also bold-faced to indicate them being vectors}). \citet{IvascenkoEA2016} wrote out the discrete derivative of Eq.~\ref{eq:TPEnonConserve} for the diffusion coefficient, yielding
\begin{subequations}
\label{eq:DmmMatrixInvertDiffEq}
\begin{align}
\frac{\partial \hat{f}(\mu_0,t_s)}{\partial t} = \; & \frac{\hat{D}_{\mu\mu}(\mu_1,t_s) - \hat{D}_{\mu\mu}(\mu_0,t_s)}{\varpi_{\mu}} \frac{\partial \hat{f}(\mu_0,t_s)}{\partial \mu} \nonumber \\
 & + \hat{D}_{\mu\mu}(\mu_0,t_s) \frac{\partial^2 \hat{f}(\mu_0,t_s)}{\partial \mu^2} \\
\frac{\partial \hat{f}(\mu_m,t_s)}{\partial t} = \; & \frac{\hat{D}_{\mu\mu}(\mu_{m+1},t_s) - \hat{D}_{\mu\mu}(\mu_{m-1},t_s)}{2 \, \varpi_{\mu}} \frac{\partial \hat{f}(\mu_m,t_s)}{\partial \mu} \nonumber \\
 & + \hat{D}_{\mu\mu}(\mu_m,t_s) \frac{\partial^2 \hat{f}(\mu_m,t_s)}{\partial \mu^2} \; \mbox{for} \; m \neq 0, M \\
\frac{\partial \hat{f}(\mu_M,t_s)}{\partial t} = \; & \frac{\hat{D}_{\mu\mu}(\mu_M,t_s) - \hat{D}_{\mu\mu}(\mu_{M-1},t_s)}{\varpi_{\mu}} \frac{\partial \hat{f}(\mu_M,t_s)}{\partial \mu} \nonumber \\
 & + D_{\mu\mu}(\mu_M,t_s) \frac{\partial^2 \hat{f}(\mu_M,t_s)}{\partial \mu^2} ,
\end{align}
\end{subequations}
where $\varpi_{\mu} = \mu_{m+1} - \mu_m$ is the pitch-cosine bin width, $\mu_0$ denotes the mid-point of the first pitch-cosine bin, $M$ is the index of the last pitch-cosine bin (i.e., there are $M+1$ bins), and a central difference method is used on the interior points, while a forward or backward difference method is used at the endpoints. This set of equations forms a matrix equation to be solved for $\hat{D}_{\mu\mu}$, as shown in Appendix~\ref{apndx:MatrixEquation} \citep[][do not seem to take the derivatives at the endpoints into consideration in their matrix equation]{IvascenkoEA2016}.

These methods, which will be referred to as M4a (Eq.~\ref{eq:DmmIntDiffEq}) and M4b (Eq.~\ref{eq:DmmMatrixInvertDiffEq}), respectively, assume the diffusive behaviour of a gyrotropic distribution and will therefore be invalid for calculating the PADC in situations where the distribution is not gyrotropic or where pitch-angle scattering is anomalous. Note that the initial distribution must not be a stationary distribution and that as the distribution isotropizes, both $\partial f / \partial t$ and $\partial f / \partial \mu$ will become zero. This will cause the running PADC to become very noisy as soon as the distribution gets close to isotropy, but the discrete derivatives of the distribution function could potentially also make the answer very noisy, especially the $\partial^2 f / \partial \mu^2$ term. The integral in Eq.~\ref{eq:DmmIntDiffEq}, however, smooths the time derivative to some extent. For these methods to work, the derivatives must, therefore, be non-zero, implying that significant pitch-angle scattering (which changes the distribution function appreciably between evaluation times) might not be resolved with these methods \citep[\textbf{i.e., the temporal resolution of the distribution function must be sufficiently high to capture the diffusion process;}][]{IvascenkoEA2016}.


\begin{table*}[t]
\begin{center}
\caption{\label{tab:Minimap}Summary of the different methods investigated in this study for easy reference.}
\begin{tabular}{|l|cc|cl|}
\hline\hline
{\bf Method} & {\bf Sec.} & {\bf Eq.} & {\bf Figs.} & {\bf Notes for figures} \\
\hline\hline
M0                   & \ref{sec:Methods}                        & \ref{eq:DmmFP}                                  & \ref{fig:BM1padc}                                                            & dashed red line in top left panel \\
\hline\hline
\multirow{4}{*}{M1a} & \multirow{4}{*}{\ref{subsec:MSD}}        & \multirow{4}{*}{\ref{eq:DmmMSDt}}               & \ref{fig:BM1padc}, \ref{fig:BM2padc}                                         & solid cyan line in top left panel \\
                     &                                          &                                                 & \ref{fig:BM1msd}                                                             & right panel in second row \& bottom left panel \\
                     &                                          &                                                 & \ref{fig:BM2msd}                                                             & top right \& bottom left panels \\
                     &                                          &                                                 & \ref{fig:BM3msd}                                                             & top right panel \\
\hline
\multirow{5}{*}{M1b} & \multirow{5}{*}{\ref{subsec:MSD}}        & \multirow{5}{*}{\ref{eq:DmmMSDdt}}              & \ref{fig:BM1padc}, \ref{fig:BM2padc}                                         & solid green line in top left panel \\
                     &                                          &                                                 & \ref{fig:BM1msd}                                                             & right panel in third row \& bottom right panel \\
                     &                                          &                                                 & \ref{fig:BM2msd}, \ref{fig:BM3msd}                                           & bottom right panel \\
                     &                                          &                                                 & \ref{fig:BMCompareMethods}                                                   & red lines \\
\hline\hline
M2a                  & \ref{subsec:TGK}                         & \ref{eq:DmmTGK}                                 & \ref{fig:BM1tgk}                                                             &  \\
\hline
\multirow{4}{*}{M2b} & \multirow{4}{*}{\ref{subsec:TGK}}        & \multirow{4}{*}{\ref{eq:DmmCorrelTime}}         & \ref{fig:BM1padc}, \ref{fig:BM2padc}                                         & solid blue and purple lines in top right panel \\
                     &                                          &                                                 & \ref{fig:BM1Correlation}, \ref{fig:BM2Correlation}, \ref{fig:BM3Correlation} &  \\
                     &                                          &                                                 & \ref{fig:BM3padc}                                                            & solid purple line \\
                     &                                          &                                                 & \ref{fig:BMCompareMethods}                                                   & cyan lines \\
\hline\hline
\multirow{2}{*}{M3}  & \multirow{2}{*}{\ref{subsec:Gea99}}      & \multirow{2}{*}{\ref{eq:DmmGea99}}              & \ref{fig:BM1padc}                                                            & solid grey line in top right panel \\
                     &                                          &                                                 & \ref{fig:BM1Gea99}                                                           &  \\
\hline\hline
\multirow{4}{*}{M4a} & \multirow{4}{*}{\ref{subsec:TPE}}        & \multirow{4}{*}{\ref{eq:DmmIntDiffEq}}          & \ref{fig:BM1padc}, \ref{fig:BM2padc}                                         & solid orange line in bottom left panel \\
                     &                                          &                                                 & \ref{fig:BM1DisFunc}, \ref{fig:BM3DisFunc}                                   & left panel \\
                     &                                          &                                                 & \ref{fig:BM1deDmmT}, \ref{fig:BM2deDmmT}, \ref{fig:BM3deDmmT}                & left panels \\
                     &                                          &                                                 & \ref{fig:BMCompareMethods}                                                   & blue lines \\
\hline
\multirow{4}{*}{M4b} & \multirow{4}{*}{\ref{subsec:TPE}}        & \multirow{4}{*}{\ref{eq:DmmMatrixInvertDiffEq}} & \ref{fig:BM1padc}, \ref{fig:BM2padc}                                         & solid red line in bottom left panel \\
                     &                                          &                                                 & \ref{fig:BM1DisFunc}, \ref{fig:BM3DisFunc}                                   & left panel \\
                     &                                          &                                                 & \ref{fig:BM1deDmmT}, \ref{fig:BM2deDmmT}, \ref{fig:BM3deDmmT}                & right panels \\
                     &                                          &                                                 & \ref{fig:BMCompareMethods}                                                   & purple lines \\
\hline\hline
\multirow{5}{*}{M5}  & \multirow{5}{*}{\ref{subsec:Stationary}} & \multirow{5}{*}{\ref{eq:DmmStationary}}         & \ref{fig:BM1padc}, \ref{fig:BM2padc}                                         & dashed blue line in bottom left panel \\
                     &                                          &                                                 & \ref{fig:BM1DisFunc}, \ref{fig:BM3DisFunc}                                   & right panel \\
                     &                                          &                                                 & \ref{fig:BM3padc}                                                            & dashed blue line \\
                     &                                          &                                                 & \ref{fig:BM3BestPracticeM5}, \ref{fig:BM1BestPracticeM5}                     &  \\
                     &                                          &                                                 & \ref{fig:BMCompareMethods}                                                   & orange lines \\
\hline\hline
\end{tabular}
\end{center}
\end{table*}

\subsection{Stationary Solution of the Diffusion Equation}
\label{subsec:Stationary}

To avoid the temporal evolution of the distribution function, \citet{KaiserEA1978} considered the stationary solution of particles continuously injected at $\mu = \mu_S$ and escaping through two absorbing boundaries located at $\mu = \mu_L < \mu_S$ and $\mu = \mu_R > \mu_S$, with $\mu_L$ and $\mu_R$ being to the left and right of the source, respectively. The system will reach a stationary state as soon as the injection of particles is balanced by the escape of particles through the absorbing boundaries. The distribution function $F(\mu)$ is now only dependent on the pitch-cosine, the governing equation is
\begin{equation}
\label{eq:StationaryTPE}
0 = \frac{\rm d}{{\rm d}\mu} \left[ D_{\mu\mu} \frac{{\rm d}F}{{\rm d}\mu} \right] + \dot{\cal N}_0 \, \delta (\mu - \mu_S) ,
\end{equation}
where $\dot{\cal N}_0$ is the rate at which particles are injected, and the absorbing boundaries imply that $F(\mu_L) = F(\mu_R) = 0$. The solution of this equation\textbf{, which will be referred to as M5,} is
\begin{equation}
\label{eq:DmmStationary}
D_{\mu\mu}(\mu) = \left\lbrace \begin{array}{rl}
\dot{\cal N}_L \left( \frac{{\rm d}F}{{\rm d}\mu} \right)^{-1} & {\rm if} \;\, \mu_L < \mu < \mu_S \\
- \dot{\cal N}_R \left( \frac{{\rm d}F}{{\rm d}\mu} \right)^{-1} & {\rm if} \;\, \mu_S < \mu < \mu_R
\end{array} \right. ,
\end{equation}
where $\dot{\cal N}_L$ and $\dot{\cal N}_R$ is the rate at which particles escape \textbf{from the system} through the left and right boundaries, respectively \citep[][seem to have a typo in the sign between calculating $D_{\mu\mu}$ to the left and right of $\mu_S$]{SunEA2016}. In practice, the escape rates through the boundaries are just the number of particles reaching the boundary divided by the time interval under consideration. This method, however, is not valid close to the source because the system might not be diffusive there and the derivative is not guaranteed to be continuous at the source \citep[something that][do not seem to address or take into consideration]{SunEA2016}. Unfortunately, it is \textbf{unclear} how to calculate from the simulations the criteria given by \citet{KaiserEA1978} for where the diffusion approximation is valid.

If there are no time dependencies in the system, such as temporally evolving fields, then the stationary solution can be approximated by injecting all of the particles at the same time, constructing the distribution function \textbf{in} several discrete \textbf{temporal bins} until all the particles have exited the domain, and adding the distribution over the temporal bins. This is equivalent to constructing the pitch-angle distribution over the entire simulation time without binning in time and can be done because the delta injection in time would be a Green's solution with the true solution found by shifting it in time and taking its convolution with a uniform distribution in time \citep{KaiserEA1978}. \citet{SunEA2016} also employ this method but use a sliding window of $[\mu_L, \mu_R]$ over many small pitch-cosine bins to calculate the PADC and then average the results within larger pitch-cosine bins, but it is not clear exactly how this is done or where they put the source and absorbing boundaries. Both \citet{KaiserEA1978} and \citet{SunEA2016} do not compute ${\rm d}F / {\rm d}\mu$ directly, but rather use a linear regression through several neighbouring bins to find the gradient. This method also assumes diffusive behaviour for a gyrotropic distribution but has the advantage that it is not time-dependent and that the distribution function should be smoothed by the time-independent way of construction, yielding a non-vanishing derivative.

The normalization of this method is important. Let the discrete distribution function for a delta injection of $N_p$ particles be constructed as
\begin{equation}
\hat{f}_{\delta}(\mu_m, t_s) = \frac{N_m^s \, \delta t}{N_p \, \varpi_{\mu} \, \varpi_t} ,
\end{equation}
where $N_m^s$ is the number of particles with pitch-cosines in $[\mu_m - \varpi_{\mu} / 2; \mu_m + \varpi_{\mu} / 2]$ in the temporal bin $[t_s - \varpi_t / 2; t_s + \varpi_t / 2]$ and $\delta t$ is the simulation time step. Here, the distribution function is averaged over the temporal bin $\varpi_t > \delta t$ for better statistics, i.e., $\bar{f}(\mu,t_s) = \int_{t_s - \varpi_t / 2}^{t_s + \varpi_t / 2} f(\mu,t) \, {\rm d}t / \varpi_t$, implying that the particles are binned $\varpi_t / \delta t$ times \textbf{in the temporal bin}, such that the distribution is normalized, i.e., $\sum_{m = 0}^M \hat{f}_{\delta}(\mu_m, t_s) \, \varpi_{\mu} = 1$ (note that the distribution function in Sec.~\ref{subsec:TPE} is constructed in \textbf{a similar} way). The solution for a constant and continuous injection is then
\begin{equation}
\hat{f}_c(\mu_m, t_s) = \sum_{j = 0}^s \hat{f}_{\delta}(\mu_m, t_j) \, \dot{\cal N}_0 \, \varpi_t = \frac{\dot{\cal N}_0 \, \delta t}{N_p \, \varpi_{\mu}} \sum_{j = 0}^s N_m^j ,
\end{equation}
according to the convolution, i.e., $f_c(\mu,t) = \dot{\cal N}_0$ $\int_0^t f_{\delta}(\mu,t-\tau) \, {\rm d}\tau$. Similarly, the escape rate of particles reaching the absorbing boundaries from the delta injection is $\hat{\cal N}_{L/R}^{\delta}(t_s) = N_{L/R}^s / N_p \, \varpi_t$, where $N_{L/R}^s$ is the number of particles escaping through the left or right boundary \textbf{in the temporal bin} $[t_s - \varpi_t / 2; t_s + \varpi_t / 2]$, respectively, such that $\sum_{s = 0}^S [\hat{\cal N}_L^{\delta}(t_s) + \hat{\cal N}_R^{\delta}(t_s)] \, \varpi_t = 1$ if \textbf{it} is assumed that all the particles escape in the $S+1$ time intervals, and the solution for the continuous injection is
\begin{equation}
\hat{\cal N}_{L/R}^c(t_s) = \sum_{j = 0}^s \hat{\cal N}_{L/R}^{\delta}(t_j) \, \dot{\cal N}_0 \, \varpi_t = \frac{\dot{\cal N}_0}{N_p} \sum_{j = 0}^s N_{L/R}^j .
\end{equation}

The stationary solution will be found at time $t_S$, i.e., $F(\mu) = \lim_{t \longrightarrow \infty} f(\mu, t)$ and $\dot{\cal N}_{L/R} = \lim_{t \longrightarrow \infty} \dot{\cal N}_{L/R}(t)$ (note that it might already be reached at earlier times as $\hat{f}_c$ and $\hat{\cal N}_{L/R}^c$ converge to the stationary answer), such that
\begin{equation}
\hat{F}(\mu_m) = \frac{\dot{\cal N}_0 \, \delta t}{N_p \, \varpi_{\mu}} \sum_{j = 0}^S N_m^j = \frac{\dot{\cal N}_0 N_m \, \delta t}{N_p \, \varpi_{\mu}}
\end{equation}
and
\begin{equation}
\hat{\cal N}_{L/R} = \frac{\dot{\cal N}_0}{N_p} \sum_{j = 0}^S N_{L/R}^j = \frac{\dot{\cal N}_0 N_{L/R}}{N_p} ,
\end{equation}
where it is now clear that the temporal bins can be replaced by a single temporal bin over the entire simulation time, such that $N_m$ is the number of particles falling into the $m^{\rm th}$ pitch-cosine bin over the entire simulation (binned \textbf{after} every \textbf{simulation} time step) and $N_{L/R}$ is the number of particles escaping through the left or right absorbing boundary during the simulation time. Applying Eq.~\ref{eq:DmmStationary} to calculate the PADC from $\hat{F}$ and $\hat{\cal N}_{L/R}$ implies that the injection rate $\dot{\cal N}_0$ need not be specified and that the \textbf{exact} number of particles $N_p$ \textbf{is unimportant} (except for yielding better statistics).


\begin{figure*}[t]
\begin{center}
\includegraphics[trim=44mm 11mm 29mm 27mm, clip, width=0.99\textwidth]{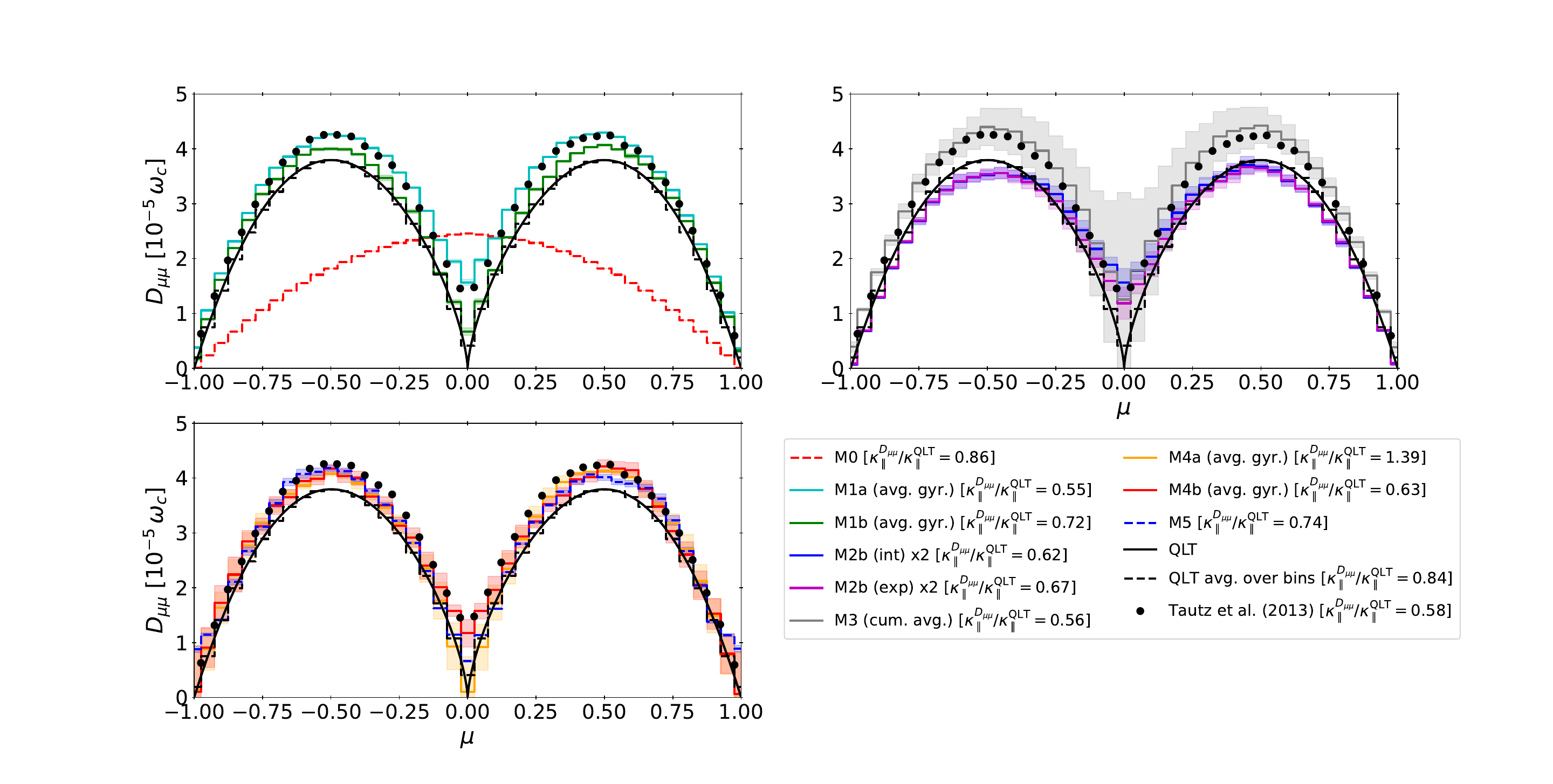}
\caption{\label{fig:BM1padc}PADC as a function of pitch-cosine resulting from Sec.~\ref{subsec:BM1}. The results of M0 (Eq.~\ref{eq:DmmFP}), M1a (\textbf{Eq.~\ref{eq:DmmMSDt};} dots in the \textbf{bottom left panel of Fig.~\ref{fig:BM1msd}}), M1b (\textbf{Eq.~\ref{eq:DmmMSDdt};} horizontal dashed lines in the \textbf{bottom right panel of Fig.~\ref{fig:BM1msd}}), M2b (Eq.~\ref{eq:DmmCorrelTime} multiplied by two for the solid red and blue lines in \textbf{the right panel of Fig.~\ref{fig:BM1Correlation}}), M3 (\textbf{Eq.~\ref{eq:DmmGea99};} dots in Fig.~\ref{fig:BM1Gea99}), M4a and M4b (\textbf{Eq.~\ref{eq:DmmIntDiffEq} and Eq.~\ref{eq:DmmMatrixInvertDiffEq}, respectively;} dashed lines in the bottom panels of Fig.~\ref{fig:BM1deDmmT} after applying a three-point average), and M5 (applying Eq.~\ref{eq:DmmStationary} to the distributions in the \textbf{right} panel of Fig.~\ref{fig:BM1DisFunc} and combining the two results such that the hemisphere containing the source is excluded) are compared to the prediction of QLT (Eq.~\ref{eq:DmmQLT}) and the simulation results of \citet[][black dots]{TautzEA2013}. The dashed black line shows the QLT prediction averaged over the pitch-cosine bins, and the legend gives the ratio of the parallel diffusion coefficient resulting from the constructed PADC (Eq.~\ref{eq:KappaParallel}) to the parallel diffusion coefficient predicted by QLT (Eq.~\ref{eq:MFP_QLT}).  \textbf{Shaded areas indicate uncertainties.}}
\end{center}
\end{figure*}

\section{Comparison of Methods}
\label{sec:Benchmark}

The methods introduced in the previous section are compared to each other and previous studies of protons for three different cases:
\begin{enumerate}
\item Pure slab turbulence with $\delta B_{\rm slab}^2 / B_0^2 = 10^{-3}$, $R_L / l_b^{\rm slab} = 0.1$, and $l_b^{\rm slab} = 0.03~{\rm au}$, in order to compare with \citet{TautzEA2013}.
\item Composite (slab + 2D) turbulence with $\delta B^2 / B_0^2 = 9 \times 10^{-4}$, $\delta B_{\rm slab}^2 = 0.2 \, \delta B^2$, $\delta B_{\rm 2D}^2 = 0.8 \, \delta B^2$, $B_0 = 4~{\rm nT}$, $l_b^{\rm slab} = 0.03~{\rm au}$, $l_b^{\rm 2D} / l_b^{\rm slab} = 0.1$, and $R_L / l_b^{\rm slab} = 0.05$, in order to compare with \citet{QinShalchi2009}.
\item Pure slab turbulence with $\delta B_{\rm slab}^2 / B_0^2 = 1$, $B_0 = 5~{\rm nT}$, $l_b^{\rm slab} = 0.01~{\rm au}$, and kinetic energy $E = 1~{\rm GeV}$, in order to compare with \citet{SunEA2016}.
\end{enumerate}
Since \citet{TautzEA2013} do not state the strength of the background magnetic field they use, $B_0 = 4~{\rm nT}$ will also be assumed for the first case. For the slab spectrum used here, the $l_b^{\rm slab}$ for these simulations yields $\lambda_c^{\rm slab} = 2.24 \times 10^{-2}~{\rm au}$ for the first two simulations and $\lambda_c^{\rm slab} = 7.5 \times 10^{-3}~{\rm au}$ for the last one. For protons, the first two cases translate to kinetic energies of $\sim 144~{\rm MeV}$ and $\sim 38~{\rm MeV}$, respectively, while the last case\textbf{, which is adequately taken into account by using a relativistic integrator (see Sec.~\ref{sec:Code}) and considering the Lorentz factor in calculations (see, e.g., Sec.~\ref{subsec:BM3}),} has $R_L / l_b^{\rm slab} = 0.756$ \citep[note that][do not specify the type of particle used in their simulations, but it is assumed that they used protons]{SunEA2016}. \textbf{Note that the methods, results, and discussions of this work should hold for electrons or any other charged particle as well. We consider protons as they are usually considered in previous work. Additionally, due to the smaller Larmor radii of electrons, the dissipation range is important in their study, while dynamical effects could also be important \citep[see, e.g.,][]{BieberEA1994}. Currently, we are only considering magnetostatic turbulence, as the incorporation of dynamical turbulence is a study in itself. Although the dissipation range could be included in the slab component, it would be more difficult to resolve it numerically for the 2D component due to memory constraints (see the discussion at the beginning of Sec.~\ref{subsec:BM2}).}

For the following results, we do 20 runs with different turbulence realizations and at least $N_p = 8 \times 10^4$ particles per run. The results are averaged over \textbf{temporal bins} (this will be specified for each case separately) and binned according to pitch-angle \textbf{(calculated relative to the background magnetic field along the $z$-axis)} in $41$ pitch-cosine bins \citep[i.e., $\varpi_{\mu} \approx 0.05$; compare this to the $9$ bins used by][yielding $\varpi_{\mu} \approx 0.2$]{QinShalchi2014}. The angle brackets in the expressions of the methods are therefore realized in the results as an average over time, different turbulence realizations, and particles falling into the pitch-cosine bins. An isotropic distribution of particles, which remains isotropic throughout the simulation, is used for all the methods except M4a, M4b, and M5. Since the latter methods require non-stationary initial distributions (see Sec.~\ref{subsec:TPE} and Sec.~\ref{subsec:Stationary} for details), an initially triangular pitch-angle distribution is used for M4a and M4b, while a delta injection is used for M5.

Given the large number of methods to be compared here, Table~\ref{tab:Minimap} provides a quick guide to the different method names and their corresponding sections and equations \textbf{together with the figures showing their results}. In the following results, uncertainties, indicated \textbf{in the figures} by shaded areas, are calculated using standard statistical uncertainty estimates on averages and standard error propagation techniques for subsequent calculations. \textbf{In the figures, the PADC is given in units of the cyclotron frequency $\omega_c$ and time is normalised to the gyroperiod $P_{\rm gyro} = 2 \pi / \omega_c$.} For the low turbulence conditions in magnetostatic turbulence of the first two cases, it is expected from previous work that QLT should be a reasonable approximation, especially for the pure slab turbulence of the first benchmark simulation. Therefore, the simulation results are compared, where possible, to what would be expected from QLT (see Appendix~\ref{apndx:QLT}). In Sec.~\ref{subsec:BM1} and Sec.~\ref{subsec:BM2}, we use the predicted parallel diffusion coefficient from QLT (Eq.~\ref{eq:MFP_QLT}) to calculate the parallel scattering time, not the parallel diffusion coefficient from the simulation itself, as it is unclear if it already converged to a constant value, despite using a long run time. Note that in calculating Eq.~\ref{eq:DmmtQLT} and Eq.~\ref{eq:MSD_QLT}, the integrals were performed from $k_{\rm min}^{\rm slab} = 2 \pi / l_{\rm max}^{\rm slab}$ to $k_{\rm max}^{\rm slab} = 2 \pi / l_{\rm min}^{\rm slab}$ in an attempt to \textbf{mimic better} the simulation setup, where particles experience a spectrum over a limited range of wavenumbers.


\begin{figure*}[t]
\begin{center}
\includegraphics[trim=32mm 122mm 17mm 18mm, clip, width=0.99\textwidth]{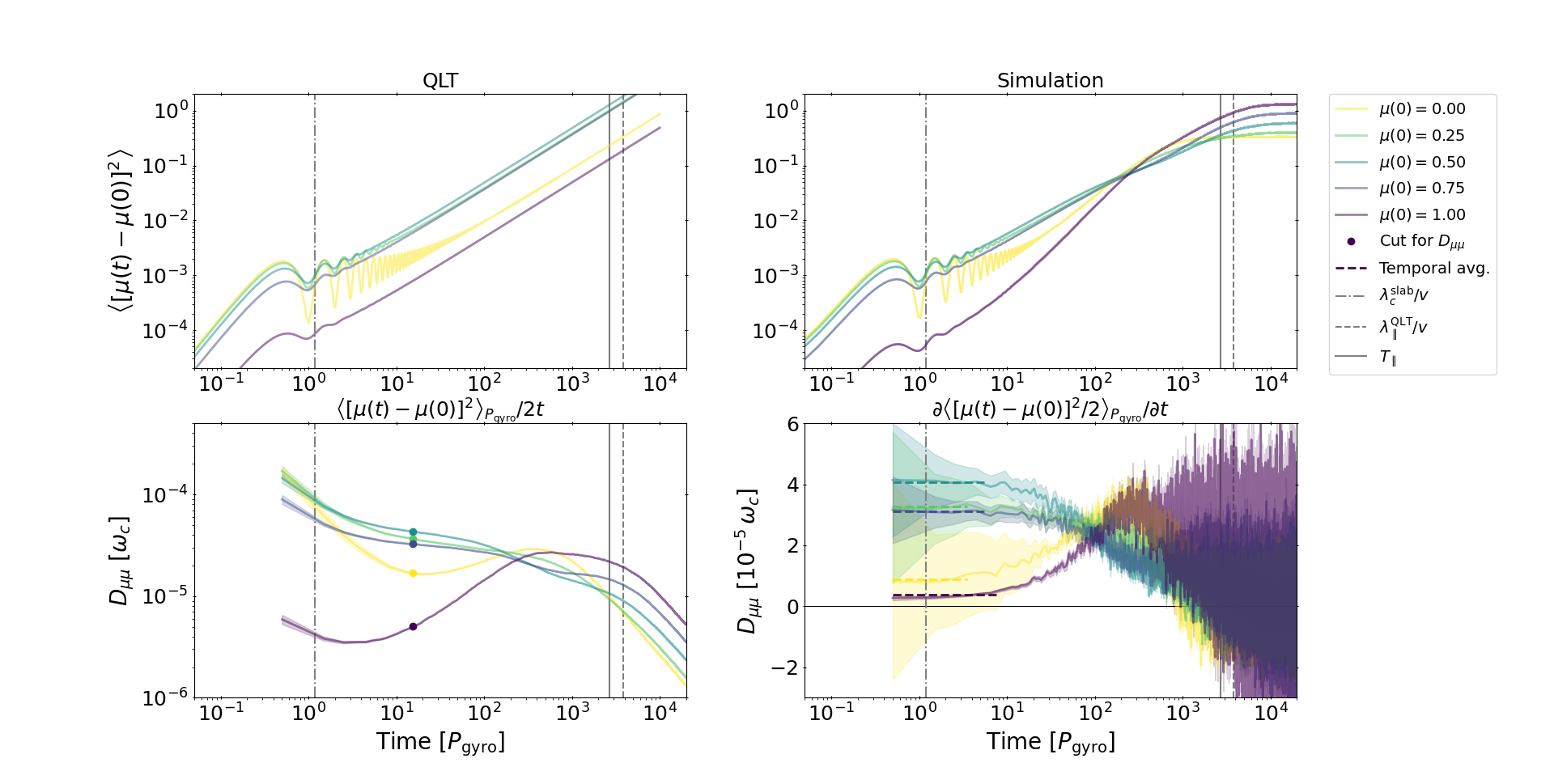}
\includegraphics[trim=32mm 18mm 17mm 22mm, clip, width=0.99\textwidth]{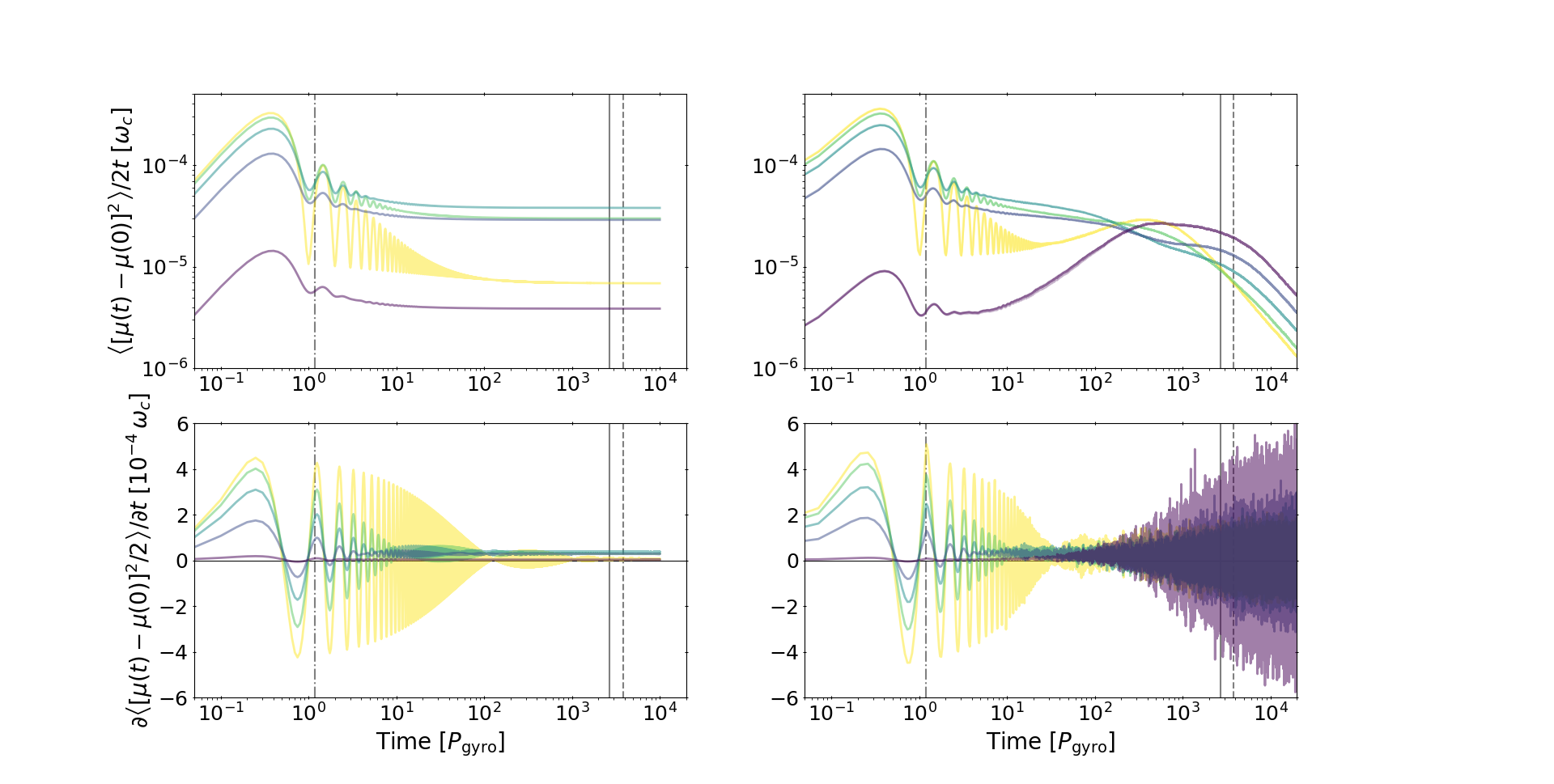}
\includegraphics[trim=32mm 8mm 17mm 124mm, clip, width=0.99\textwidth]{BM1_MSD.png}
\caption{\label{fig:BM1msd}\textbf{Results of M1a (Eq.~\ref{eq:DmmMSDt}) and M1b (Eq.~\ref{eq:DmmMSDdt}) as a function of time for different initial pitch-angles (indicated in the legend by different colors) from Sec.~\ref{subsec:BM1}. \textit{Top row:} Pitch-cosine MSD expected from QLT (Eq.~\ref{eq:MSD_QLT}; \textit{left panel}) and calculated from the simulations (\textit{right panel}). \textit{Second row:} Running PADC for M1a expected from QLT (Eq.~\ref{eq:DmmMSDt} applied to Eq.~\ref{eq:MSD_QLT}; \textit{left panel}) and calculated from the MSD in the top right panel (\textit{right panel}). \textit{Third row:} Running PADC for M1b expected from QLT (Eq.~\ref{eq:DmmtQLT}; \textit{left panel}) and calculated from the MSD in the top right panel (\textit{right panel}). \textit{Bottom row:} Running PADC for M1a (\textit{left panel}) and M1b (\textit{right panel}) after averaging the MSD in the top right panel over a gyration. Dots in the left panel indicate the time (i.e., at 16 gyrations) when the PADC was constructed for M1a, while horizontal dashed lines in the right panel show the time over which the results were averaged to construct the PADC for M1b (i.e., the solid cyan and green lines, respectively, in the top left panel of Fig.~\ref{fig:BM1padc}). Shaded areas visible in the bottom panels indicate uncertainties. The vertical grey lines indicate the turbulent correlation crossing time (dash-dotted line) and the scattering times calculated from either the parallel correlation function (solid line) or the parallel MFP expected from QLT (dashed line). Note that the pitch-cosine has been shifted by a quarter of a bin width when calculating the expectations from QLT for $\mu = 0$ and $\mu = 1$, as QLT predicts zero for both of these pitch-cosine.}}
\end{center}
\end{figure*}

\subsection{Method Comparison in Weak Slab Turbulence}
\label{subsec:BM1}

\begin{figure*}[t]
\begin{center}
\includegraphics[trim=37mm 112mm 25mm 26mm, clip, width=0.99\textwidth]{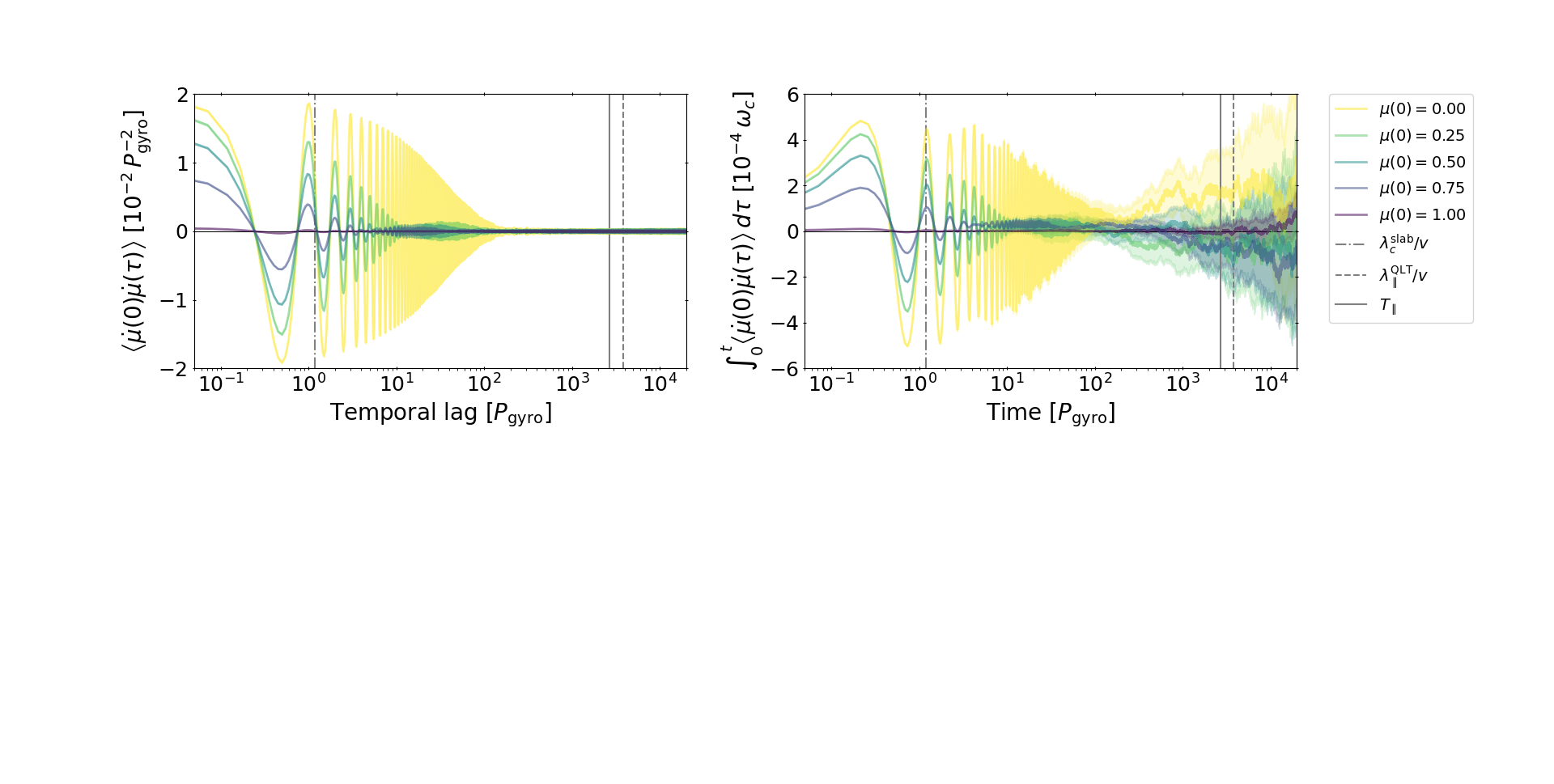}
\caption{\label{fig:BM1tgk}\textbf{Results of M2a (Eq.~\ref{eq:DmmTGK}) as a function of time for different initial pitch-angles (indicated by different colors) from Sec.~\ref{subsec:BM1}. \textit{Left panel:} Pitch-angle rate of change correlation function. \textit{Right panel:} Running PADC calculated from the pitch-angle rate of change correlation function in the left panel. Vertical grey lines are as explained in Fig.~\ref{fig:BM1msd} and shaded areas in the right panel indicate uncertainties.}}
\end{center}
\end{figure*}

\begin{figure*}[t]
\begin{center}
\includegraphics[trim=35mm 103mm 1mm 27mm, clip, width=0.99\textwidth]{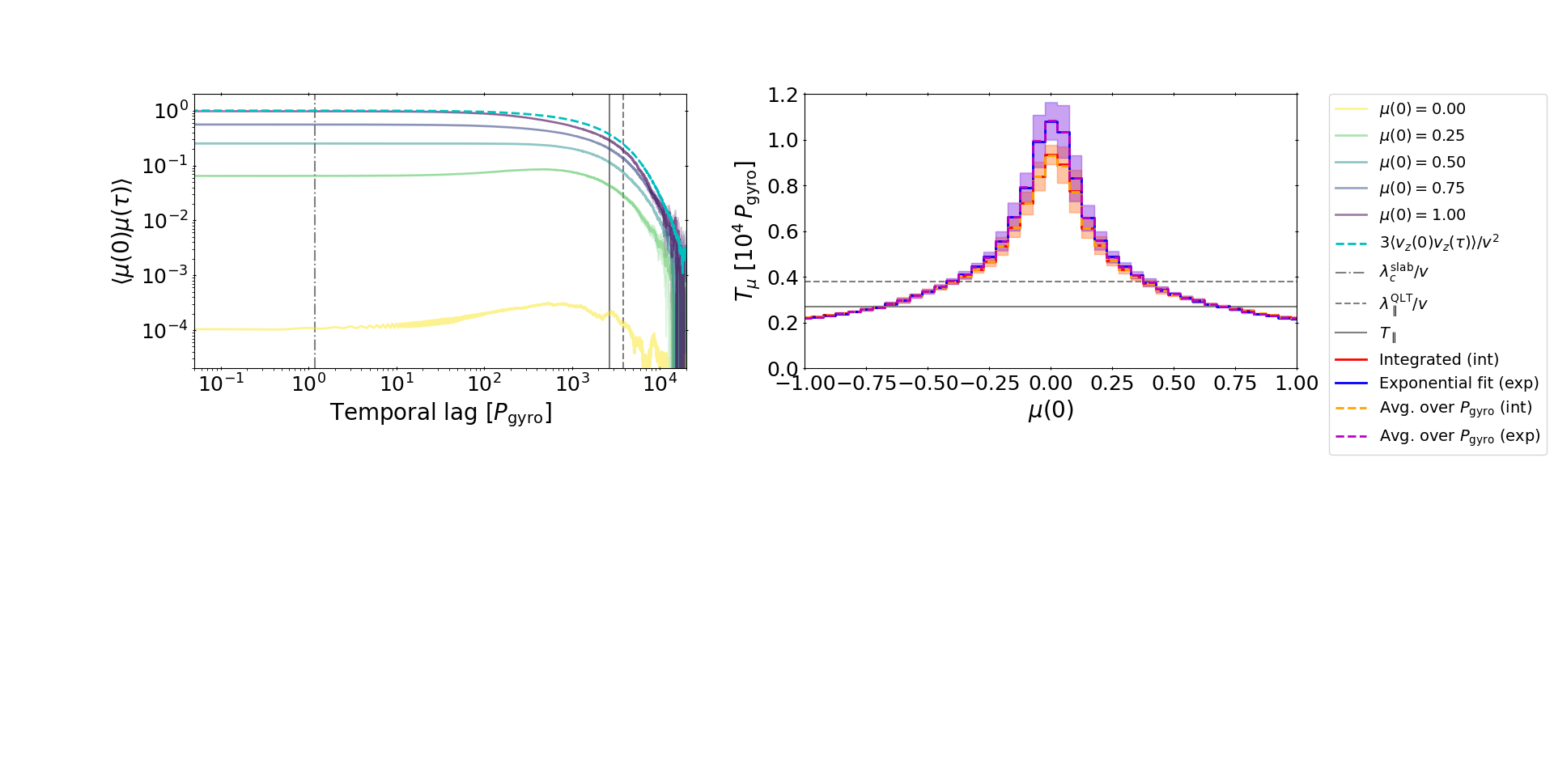}
\caption{\label{fig:BM1Correlation}\textbf{Results of M2b (Eq.~\ref{eq:DmmCorrelTime}) from Sec.~\ref{subsec:BM1}. \textit{Left panel:} Pitch-angle correlation function as a function of time for different initial pitch-angles (different colors). The dashed cyan line represents the parallel velocity correlation function. Vertical grey lines are as explained in Fig.~\ref{fig:BM1msd}. \textit{Right panel:} Pitch-angle correlation time as a function of pitch-cosine for the correlation functions in the left panel. The solid red and blue lines represent the correlation time calculated from the original correlation functions, while the dashed orange and purple lines represent the correlation time calculated after averaging the correlation functions over a gyration. The solid red and dashed orange lines show the correlation time calculated from integrating the correlation function (Eq.~\ref{eq:PACorrelTime}), while the solid blue and dashed purple lines show the correlation time determined by fitting an exponential function to the correlation function. Note that the dashed purple and solid blue lines fall almost on top of one another, while the same is true for the dashed orange and solid red lines. The horizontal solid and dashed grey lines have the same interpretation as their vertical counterparts in the left panel. Shaded areas indicate uncertainties.}}
\end{center}
\end{figure*}

\begin{figure}[t]
\begin{center}
\includegraphics[trim=231mm 112mm 12mm 20mm, clip, width=0.475\textwidth]{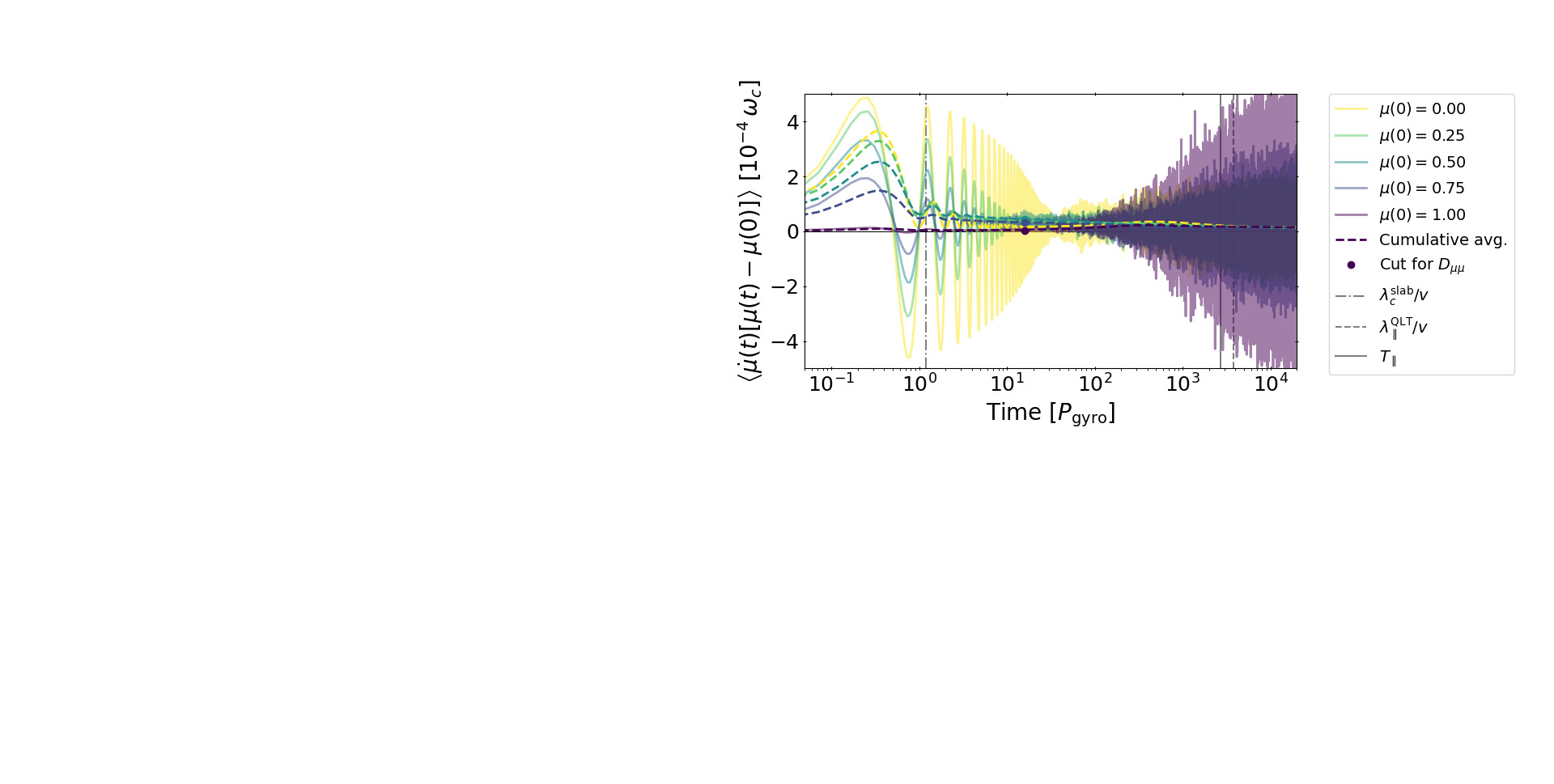}
\caption{\label{fig:BM1Gea99}\textbf{Running PADC of M3 (Eq.~\ref{eq:DmmGea99}) as a function of time for different initial pitch-angles from Sec.~\ref{subsec:BM1}. The dashed lines show the cumulative average up to a given time, while the dots indicate the time when the PADC was constructed from the cumulative average (i.e., the solid grey line in the top right panel of Fig.~\ref{fig:BM1padc}). Vertical grey lines are as explained in Fig.~\ref{fig:BM1msd}.}}
\end{center}
\end{figure}

\begin{figure*}[t]
\begin{center}
\includegraphics[trim=38mm 115mm 17mm 27mm, clip, width=0.99\textwidth]{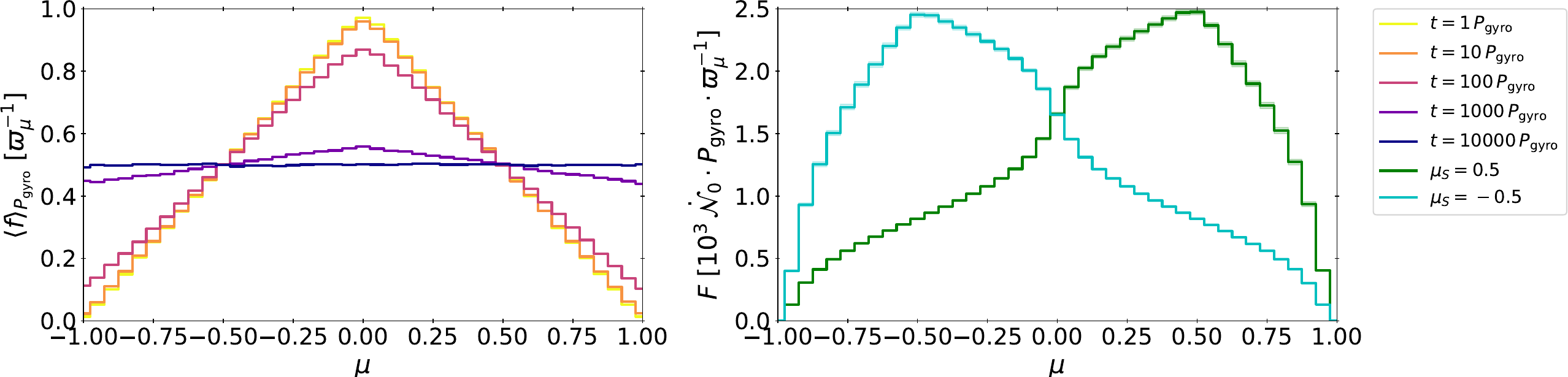}
\caption{\label{fig:BM1DisFunc}Pitch-angle distribution functions as a function of pitch-cosine \textbf{from} Sec.~\ref{subsec:BM1}. \textit{Left panel:} Temporal evolution of an initially triangular distribution averaged over a gyration at different times (indicated in the legend by different colours) for the methods of Sec.~\ref{subsec:TPE}. \textit{Right panel:} Stationary distributions of two different point sources with absorbing boundaries at $\mu = \varpi_{\mu} / 2 - 1$ and $\mu = 1 - \varpi_{\mu} / 2$ for the method of Sec.~\ref{subsec:Stationary}. \textbf{Shaded areas (somewhat visible in the right panel) indicate uncertainties.}}
\end{center}
\end{figure*}

\begin{figure*}[t]
\begin{center}
\includegraphics[trim=35mm 8mm 17mm 22mm, clip, width=0.99\textwidth]{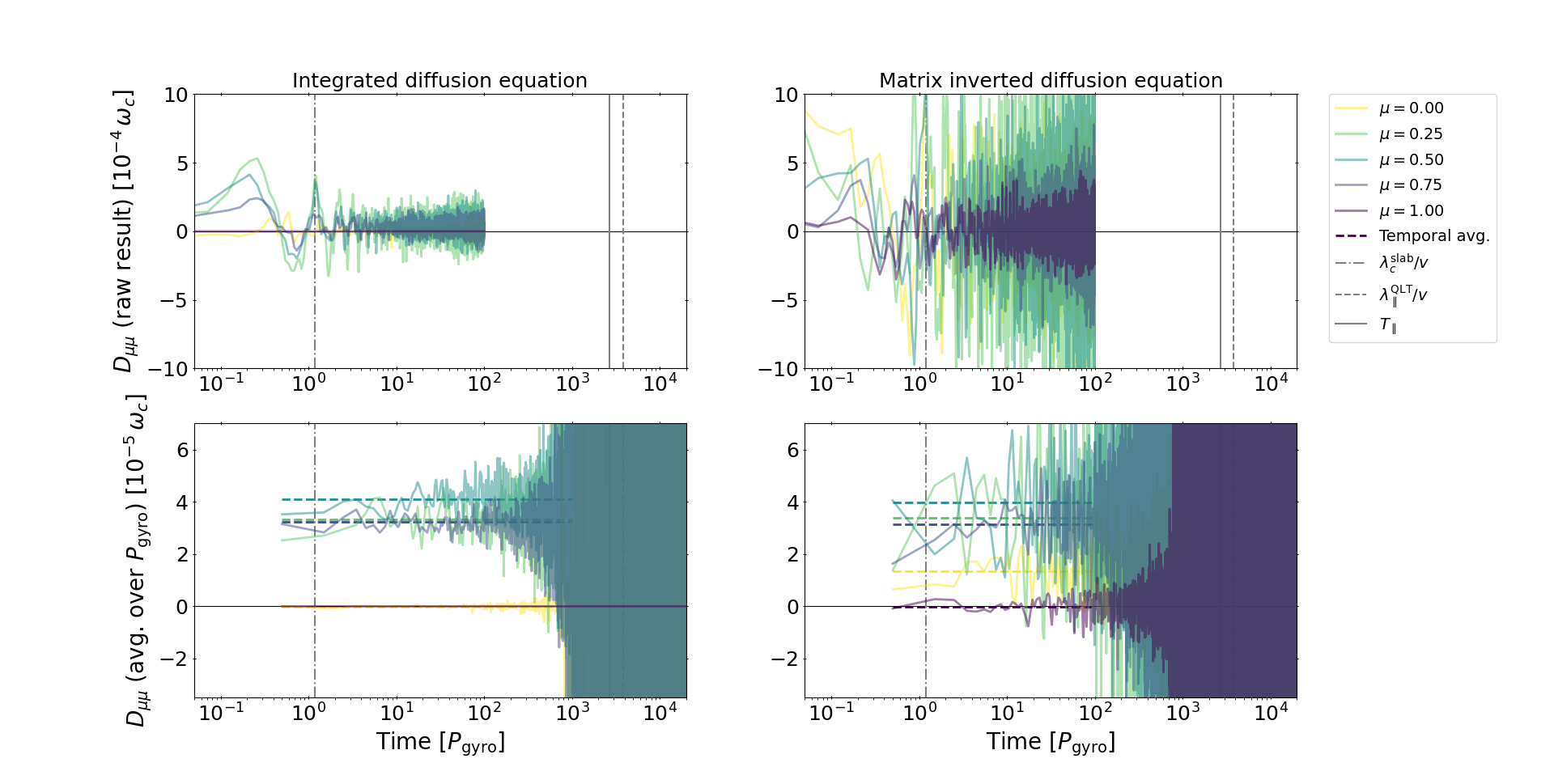}
\caption{\label{fig:BM1deDmmT}Running PADC \textbf{of M4a (Eq.~\ref{eq:DmmIntDiffEq}; \textit{left panels}) and M4b (Eq.~\ref{eq:DmmMatrixInvertDiffEq}; \textit{right panels})} as a function of time for different initial pitch-angles from Sec.~\ref{subsec:BM1}. Both \textbf{methods} are applied to the constructed distribution function (\textit{top panels}) and the distribution function averaged over a gyration (\textbf{left} panel of Fig.~\ref{fig:BM1DisFunc}; \textit{bottom panels}). The results in the top panels are only shown up to a hundred gyrations as the noise becomes extreme thereafter. The horizontal dashed lines in the bottom panels show the \textbf{time over which the results were averaged to construct the PADC (i.e., the solid orange and red lines in the bottom left panel of Fig.~\ref{fig:BM1padc}). Vertical grey lines are as explained in Fig.~\ref{fig:BM1msd}.}}
\end{center}
\end{figure*}

We used a total simulation time of $6 \times 10^4~\lambda_c^{\rm slab} / v$ (in units of the slab correlation crossing time) and averaged the results over a \textbf{temporal bin width} of $\varpi_t = 0.04~\lambda_c^{\rm slab} / v$ (\textbf{$2.56$ times longer than the simulation time step, but} shorter than any characteristic time scale in the system, ensuring no information loss). Fig.~\ref{fig:BM1padc} - Fig.~\ref{fig:BM1deDmmT} show the results of the first benchmark comparison for weak pure slab turbulence. All the PADC results from the first benchmark comparison between the different methods are summarized in Fig.~\ref{fig:BM1padc} (each method will be discussed in the following paragraphs). Eq.~\ref{eq:DmmQLT} of QLT is also shown, together with its average over the pitch-cosine bins and the simulation results from \citet{TautzEA2013}. A more quantitative way to compare the different PADCs is to calculate their parallel diffusion coefficients (see Eq.~\ref{eq:KappaParallel} and indicated in the label of Fig.~\ref{fig:BM1padc} by $\kappa_{\parallel}^{D_{\mu\mu}}$) and compare that to the prediction of QLT (see Eq.~\ref{eq:MFP_QLT} and indicated in the label of Fig.~\ref{fig:BM1padc} by $\kappa_{\parallel}^{\rm QLT}$). The finite pitch-cosine bins introduce some difference between these two quantities, as clearly visible for the theoretical case, mostly because the bin at $\mu = 0$ overestimates the pitch-angle scattering there. To better estimate this bin, we fitted straight lines through the neighbouring two bins on each side and averaged their extrapolations. \textbf{Similarly for the $|\mu| = 1$ bins, we extrapolate a straight line from the neighbouring two bins as some methods yield exactly zero for these bins. These extrapolations} did not change the \textbf{diffusion coefficient ratio} results significantly.

\paragraph{M0}

Simply averaging the square of the pitch-cosine changes during each time step (i.e., Eq.~\ref{eq:DmmFP}) yields a PADC with an isotropic form \textbf{(see the dashed red line in the top left panel of Fig.~\ref{fig:BM1padc})}, which is not expected from theory. This is probably due to the statistical behaviour of $(\Delta \mu)^2$ being incorrect for the chosen \textbf{temporal bin width and/or simulation time step} used to calculate $(\Delta \mu)^2$.

\paragraph{M1a \& M1b}

It is clear from the \textbf{top right} panel of \textbf{Fig.~\ref{fig:BM1msd}} that the MSD has oscillations that die out after $\sim 10$ gyrations or turbulent crossing times (unfortunately, these two times are almost similar in the current setup, making it difficult to ascertain whether features are due to sampling the turbulence enough or \textbf{due} to small pitch-angle changes accumulated over enough gyrations, although these two effects are probably linked), while it saturates at later times. The oscillations die out faster for certain pitch-cosines and can be removed by averaging the results over a gyro-period. The $\mu = 0$ bin saturates at a lower value than the $|\mu| = 1$ bins because a particle starting with $\mu = 0$ can have a maximum $|\Delta \mu| = 1$, while a particle starting with $|\mu| = 1$ can have a maximum $|\Delta \mu| = 2$. Saturation appears to occur after the parallel scattering time. The \textbf{top left} panel of \textbf{Fig.~\ref{fig:BM1msd}} clearly shows that the MSD from the simulation is in good agreement with the QLT prediction (Eq.~\ref{eq:MSD_QLT}) initially. However, the theoretical MSD seems to grow linearly with time and does not saturate. This shows that the idea of \citet{TautzEA2013} to theoretically calculate the saturation time from QLT as the time when $\langle (\Delta \mu)^2 \rangle = 1$ would be invalid, as not all pitch-angles are expected to reach such a large MSD.

The \textbf{right} panel \textbf{in the second row} of \textbf{Fig.~\ref{fig:BM1msd}} shows that the oscillations in the MSD lead to the running PADC from M1a (Eq.~\ref{eq:DmmMSDt}) having initial oscillations that transition to a constant value before decreasing with time due to the saturation of the MSD \citep[in line with the results of][]{TautzEA2013, RiordanPeer2019}. When M1a is applied to the expected MSD from QLT in the top left panel, it seems as if the running PADC reach constant values at late times\textbf{, as shown in the left panel of the second row}, but analytically, Eq.~\ref{eq:MSD_QLT} implies eventual sub-diffusive behaviour since $\left[ 1 - \cos \left( |k_{\parallel} \mu v - n \Omega| t \right) \right] / t$ approaches zero as $t \longrightarrow \infty$ ($\Omega$ is the signed cyclotron frequency and $n = \pm 1$). Even if the MSD is averaged over a gyration to remove the oscillations in the \textbf{bottom left} panel, it remains difficult to discern where the running PADC is constant, especially for certain pitch-angles. While an automated search for this region might be possible, it would fail under stronger turbulence conditions, as this region would disappear \citep[see][]{RiordanPeer2019}. Simply choosing a time to use, as done by \citet{TautzEA2013} and indicated by the dots in the \textbf{bottom left} panel, does not guarantee that the PADC is truly constant there (e.g., it is not clear if the \textbf{$|\mu| = 0.25$, $|\mu| = 0.5$, and $|\mu| = 0.75$} bins show a constant value, while the value for the $|\mu| = 1$ bin is likely overestimated), especially in stronger turbulence conditions. The PADC constructed from M1a (dots in the \textbf{bottom left} panel of \textbf{Fig.~\ref{fig:BM1msd}}) compare well to the results of \citet{TautzEA2013}, as they are constructed at the same time \textbf{(see the solid cyan line in the top left panel of Fig.~\ref{fig:BM1padc})}.

It is clear from \textbf{the third row of Fig.~\ref{fig:BM1msd}} that M1b (Eq.~\ref{eq:DmmMSDdt}; \textbf{right} panel) exhibit oscillations which are qualitatively the same as predicted by QLT (Eq.~\ref{eq:DmmtQLT}; left panel). Quantitatively, there are slight differences, probably due to the finite pitch-cosine bin widths, as the analytical expressions were evaluated at the bin midpoints (except for $\mu = 0$ and $\mu = 1$). From the theoretical prediction, it is apparent that the oscillation amplitudes decrease with time until small oscillations around a constant value remain. Unfortunately, noise in the numerical methods obscures this, making it challenging to infer the constant value from \textbf{this method} to construct the PADC. Simply averaging the results of \textbf{M1b} over a gyration does not generally yield a constant running PADC due to numerical imperfections (e.g., an integer number of temporal bins not perfectly fitting into a gyration or errors introduced by the numerical derivatives or integration schemes) and noise. \textbf{The running PADC predicted by QLT and calculated from M1b exhibit complicated behavior in that it displays both negative values and anomalous diffusion.}

\textbf{Negative values of the (time-independent) PADC for certain pitch-angles have also been predicted by theories incorporating the effect of focusing on the pitch-angle scattering process \citep[see][]{TautzEA2014, Florinski2024}. If the diffusion coefficient is negative, then the pitch-angle diffusion equation (Eq.~\ref{eq:TPEconserve}) can be explicitly written as,
\begin{equation}
- \frac{\partial f}{\partial t} = \frac{\partial}{\partial \mu} \left[ \left| D_{\mu\mu} \right| \frac{\partial f}{\partial \mu} \right] \, \mbox{for} \;\, D_{\mu\mu} < 0 .
\end{equation}
This can be interpreted as a time reversal of the diffusion process and would therefore represent a congregation of particles diffusing in such a way as to be less spread out in pitch-angle. Of course, to satisfy the second law of thermodynamics, the total amount of ‘negative diffusion’ cannot be more than the total amount of ‘positive diffusion’ and we do see this in that the negative area of the oscillating running PADC is less than the positive area \citep{TautzEA2014}. The microphysics of pitch-angle scattering does not seem to be understood well enough to give an explanation at this stage as to why negative diffusion would occur.}

\textbf{In unbounded spatial diffusion, the diffusive behaviour is classified in terms of the time dependence of the MSD and the process is said to be diffusive or ballistic if the time dependence is linear or quadratic, respectively, while being sub- or super-diffusive if the time dependence is between constant and linear or between linear and quadratic, respectively \citep[see, e.g.,][and references therein]{Shalchi2009, ZimbardoEA2015}. It is clear from the initial oscillations and eventual saturation seen in the pitch-angle MSD examples of this and previous work, that the particles experience ballistic, super-diffusive, diffusive, and sub-diffusive motion at different times, while the saturation could be wrongly interpreted as sub-diffusion if the boundedness of the pitch-angle is not considered. The same criteria to identify anomalous diffusion can therefore not be applied to pitch-angle scattering or diffusion in a bounded domain. Perhaps better suited for identifying anomalous pitch-angle scattering, is the correlation function, as \citet{ZimbardoPerri2020} show for isotropic pitch-angle scattering that power-law scattering times could lead to a pitch-angle correlation function decreasing as a power-law and not an exponential. These authors also show that the temporal evolution of the pitch-angle distribution function from a delta injection in pitch-cosine will be different if anomalous diffusion is acting.}

Averaging the MSD over a gyration and then applying M1b yields a fairly constant PADC up to $\sim 10$ gyrations, as shown in the bottom right panel of \textbf{Fig.~\ref{fig:BM1msd}}. Here, an automated procedure was applied to find the average constant value of the PADC for each pitch-cosine bin separately \textbf{(indicated by the horizontal dashed lines)} by calculating the cumulative average until the instantaneous value deviates by more than a factor of 1.5 times the cumulative standard deviation. The factor 1.5 is a free parameter that can be adjusted, but reducing it to 1 yields an average over too short a time (resulting in a larger uncertainty) and increasing it to 2 yields an average over too long a time (changing the average away from the approximate constant value). While one could simply choose a time (e.g., 2 gyrations), the aforementioned method yields smaller uncertainties on the constructed PADC. However, it can be questioned whether a system with such low turbulence levels is already in a diffusive state during the first $10$ gyrations. This average of M1b yield\textbf{s a} PADC between the QLT prediction and the results of \citet{TautzEA2013} \textbf{(see the solid green line in the top left panel of Fig.~\ref{fig:BM1padc})}.

\paragraph{M2a}

The \textbf{left} panel of \textbf{Fig.~\ref{fig:BM1tgk}} shows the pitch-angle rate of change correlation function needed for the TGK approach. The oscillations would not be well resolved if even fewer temporal bins were used, with the oscillations being very important for the TGK approach \textbf{due to the integral over the pitch-angle rate of change correlation function. These correlation functions become noisy at late times due to statistical fluctuations, which add up when integrating them to yield the running PADC in the right panel} (a lot more particles would be needed to reduce the noise at late times). \textbf{M2a (Eq.~\ref{eq:DmmTGK}) is again qualitatively the same as predicted by QLT (compare to the left panel in the third row of Fig.~\ref{fig:BM1msd}) until the noise becomes severe. Since we are unable to construct a constant PADC from this method, no result is shown in Fig.~\ref{fig:BM1padc}.}

\paragraph{M2b}

The pitch-angle correlation function in the \textbf{left} panel of Fig.~\ref{fig:BM1Correlation} is constructed from Eq.~\ref{eq:MSD-CF-Relate} since the MSD is already well-behaved. Unfortunately, the correlation functions become very noisy at late times, making it difficult to verify if they follow an exponential decorrelation. \textbf{The onset of the noise can of course be delayed by using more particles.} It is interesting to note a slight increase in the correlation for the $\mu = 0$ and $|\mu| = 0.25$ bins, but this could also be due to the range of different pitch-angles in the bin compared to the small correlation value at zero lag. The oscillations seen for some pitch-angles (close to $90^{\circ}$) are small, resulting in a correlation function with a similar shape when averaged over a gyration, while still being fairly noisy at late times after the averaging.

The negligible effect of averaging the correlation function over a gyration is also evident in \textbf{the right panel of Fig.~\ref{fig:BM1Correlation}}, which shows the correlation time as a function of pitch-angle (compare the dashed and solid lines). The noise at late times accumulates during the integration of the correlation function (see Eq.~\ref{eq:PACorrelTime}), yielding an incorrect correlation time. Consequently, we also determined the correlation time by fitting the cumulative function (i.e., $\int_0^t C_{\mu\mu}(\tau) \, {\rm d}\tau$) to what would be expected from an exponentially decreasing correlation function (i.e., $\int_0^t e^{- \tau / T_{\mu}} \, {\rm d}\tau = T_{\mu} (1 - e^{- t/T_{\mu}})$), as it is easier to discern where noise is affecting the cumulative function than the correlation function itself. The most significant difference between these two approaches is seen closer to $\mu = 0$, where there is more noise in the correlation functions. The correlation time is longer for pitch-angles close to $90^{\circ}$ as it is expected that particles with these pitch-angles will experience less pitch-angle scattering (see the resulting PADCs in Fig.~\ref{fig:BM1padc}). Note that a three-point average was applied to the correlation time to smooth it and make it slightly more symmetric, which could slightly underestimate the correlation time for the bins around $\mu = 0$. The correlation time found from integrating the parallel velocity correlation function (dashed grey line) is closer to the parallel scattering time estimated from QLT \textbf{(solid grey line)} than the parallel scattering time estimated from the parallel diffusion coefficient calculated during the simulation is to the QLT estimate (hence the aforementioned doubt over whether the simulation has run long enough for the parallel diffusion coefficient to converge).

\textbf{The PADC from M2b (Eq.~\ref{eq:DmmCorrelTime}) is closer to the QLT prediction than the results of \citet{TautzEA2013} (see the solid blue and purple lines in the top right panel of Fig.~\ref{fig:BM1padc}).} Note, however, that Eq.~\ref{eq:DmmCorrelTime} had to be multiplied by 2 for M2b to yield results on the same level as the other methods and QLT. We note that \citet{ShalchiEA2004b} defined the pitch-angle scattering time as in Eq.~\ref{eq:DmmCorrelTime}, while \citet{leRouxWebb2007} defined it without the factor of a half (as is necessary to yield the results in the top right panel of Fig.~\ref{fig:BM1padc}). \textbf{Both assumptions seem equally \textsl{ad hoc}, and if this geometry problem} (i.e., isotropic vs. anisotropic scattering) \textbf{prove to persist in the following comparisons, then the assumption of the scattering time being related to the correlation time must be called into question}.

\paragraph{M3}

It is clear from \textbf{Fig.~\ref{fig:BM1Gea99}} that M3 (Eq.~\ref{eq:DmmGea99}) \textbf{also} exhibit similar oscillations \textbf{to those} predicted by QLT \textbf{(see the left panel in the third row of Fig.~\ref{fig:BM1msd})}. As shown by the dashed lines, it might be possible to find a time interval where the running PADC is constant by calculating the cumulative average. However, automating a procedure to determine where the oscillations have sufficiently died out and where noise starts affecting the average would be difficult, raising questions about the consistency of results acquired in this manner. \textbf{The solid grey line in the top left panel of Fig.~\ref{fig:BM1padc} shows that t}he PADC constructed from this cumulative average of M3 (dots in \textbf{Fig.~\ref{fig:BM1Gea99}}) compare well to the results of \citet{TautzEA2013}, as \textbf{it is also} constructed at the same time (note that the uncertainty yielded by M3 increases close to $\mu = 0$ due to the averaged oscillations having a larger amplitude for these pitch-cosines).

\paragraph{M4a \& M4b}

The distribution function, averaged over a gyration and shown in the \textbf{left} panel of Fig.~\ref{fig:BM1DisFunc}, evolves slowly from a triangular to an isotropic distribution for these low turbulence levels. Therefore, the distribution function averaged over a gyration is very similar to the unaveraged one. A triangular distribution was used to ensure a well-behaved derivative in all pitch-angle bins, except the $\mu = 0$ bin. When calculating the PADC from the distribution function, it would also be possible to do linear fits of the distribution function, but this did not yield less noisy results and will therefore not be considered further. Comparing the evolution of the distribution function between $10^3$ and $10^4$ gyrations to the parallel scattering time and the fact that the pitch-angle decorrelation occurs around the same time that the MSD saturates imply that the system has had enough time to relax to an isotropic distribution (if it were to start in a non-stationary state) and would probably be in a truly diffusive state.

The top left panel of Fig.~\ref{fig:BM1deDmmT} shows that applying M4a (Eq.~\ref{eq:DmmIntDiffEq}) yields oscillations similar to what is expected from QLT \textbf{(see again the left panel in the third row of Fig.~\ref{fig:BM1msd})} but with a lot of noise. Note that we applied M4a from $\mu = -1$ to $\mu = 0$ and from $\mu = 1$ down to $\mu = 0$ to avoid the pitch-angle derivative being close to zero at $\mu = 0$ and to prevent noise in the negative hemisphere of the distribution function accumulating and affecting the positive hemisphere when calculating the integral of the distribution function. However, this approach results in the $\mu = 0$ bin having a near-zero value for the symmetric distribution and PADC. If the method is not applied in this way, then the result for $\mu = 0$ is too noisy due to the derivative being close to zero. Note that M4a yields zero for the $|\mu| = 1$ bins due to the cumulative distribution being zero for the first bin. M4b (Eq.~\ref{eq:DmmMatrixInvertDiffEq}), shown in the top right panel \textbf{of Fig.~\ref{fig:BM1deDmmT}}, is even noisier, as might be expected from its use of a second-order derivative. An initially constant running PADC can be found from both M4a and M4b if the distribution function is first averaged over a gyration \textbf{(this reduces the temporal resolution of the distribution function, but only removes the oscillations, while still capturing the diffusion process)}, as shown in the bottom panels. This improvement in identifying the PADC when averaging the distribution function is somewhat expected, given that the diffusion equation is averaged over gyrophase for a gyrotropic distribution. Since the results fluctuate around a seemingly constant value, the temporal average of M4a and M4b was taken up to $10^3$ and $10^2$ gyrations, respectively (the noise becomes too much at later times; these constant values are indicated by the horizontally dashed lines). \textbf{The solid orange and red lines in the bottom left panel of Fig.~\ref{fig:BM1padc} result after applying a} three-point average to reduce the noise. \textbf{Both M4a and M4b} yield PADCs between the QLT prediction and the results of \citet{TautzEA2013}, but M4a yields less pitch-angle scattering at $90^{\circ}$ pitch-angles than the QLT prediction.

\begin{figure*}[t]
\begin{center}
\includegraphics[trim=37mm 11mm 24mm 27mm, clip, width=0.99\textwidth]{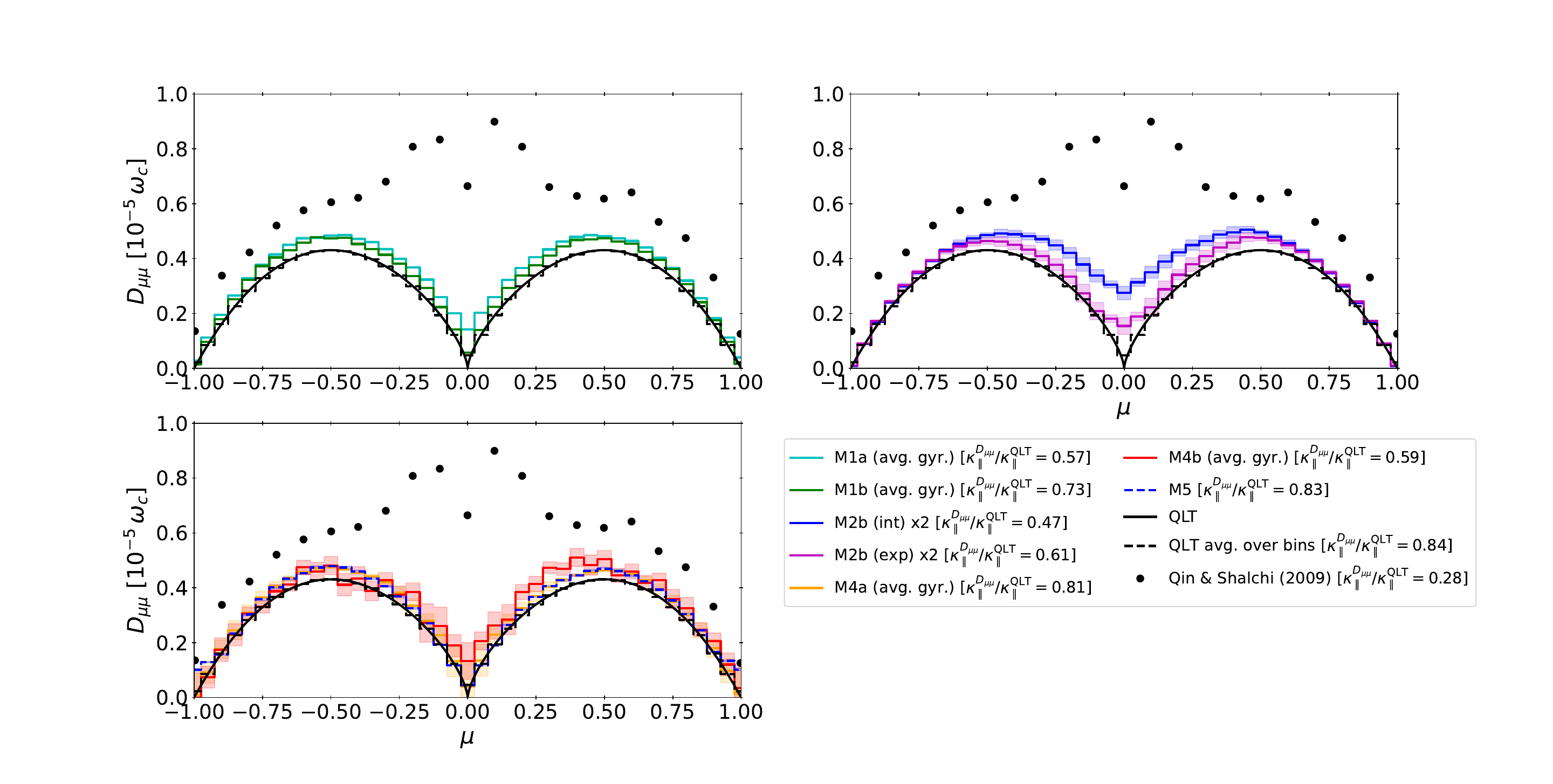}
\caption{\label{fig:BM2padc}Similar to Fig.~\ref{fig:BM1padc}, but for Sec.~\ref{subsec:BM2} and without the results of M0 (Eq.~\ref{eq:DmmFP}) and M3 (Eq.~\ref{eq:DmmGea99}). Here, the results are compared to the simulation results of \citet[][black dots]{QinShalchi2009}. M2b (Eq.~\ref{eq:DmmCorrelTime}) had to be multiplied by two again. \textbf{Shaded areas indicate uncertainties.}}
\end{center}
\end{figure*}

\paragraph{M5}

For M5 \textbf{(Eq.~\ref{eq:DmmStationary})}, the simulation is only run until all the particles escape through the absorbing boundaries. This makes the method computationally much faster, especially under conditions where the particles experience more scattering, but if pitch-angle scattering is inefficient, then this method could be computationally slow. Since this method is not valid close to the source of particles, we ran the method twice, injecting particles at $\mu_S = 0.5$ and $\mu_S = -0.5$, so that the solution of the former can be used for $\mu < 0$ and the latter for $\mu > 0$. This is done in the hope that $|\mu_S|$ would be far enough from the other \textbf{pitch-cosine} hemisphere for the method to be valid. The absorbing boundaries were placed at $\mu = \varpi_{\mu} / 2 - 1$ and $\mu = 1 - \varpi_{\mu} / 2$, implying that the pitch-cosine bins $\mu = \pm 1$ will not have any particles. The two stationary distributions resulting from \textbf{this} are shown in the \textbf{right} panel of Fig.~\ref{fig:BM1DisFunc}. Due to the way in which the stationary distribution function and escape rates are calculated in this method (see Sec.~\ref{subsec:Stationary} for details), even fewer particles (up to a factor of ten less) could be used to still yield an acceptable result. Since this method uses a stationary solution, there is no running PADC, and the final result is shown in \textbf{the bottom left panel of} Fig.~\ref{fig:BM1padc} \textbf{as the dashed blue line}. Having no data in the first and last pitch-cosine bins, we estimated \textbf{a value} for these bins by extrapolating a straight line from the adjacent bins. \textbf{The resulting PADC is also} between the QLT prediction and the results of \citet{TautzEA2013}.


\subsection{Method Comparison in Weak Composite Turbulence}
\label{subsec:BM2}

\begin{figure*}[t]
\begin{center}
\includegraphics[trim=30mm 8mm 1mm 20mm, clip, width=0.99\textwidth]{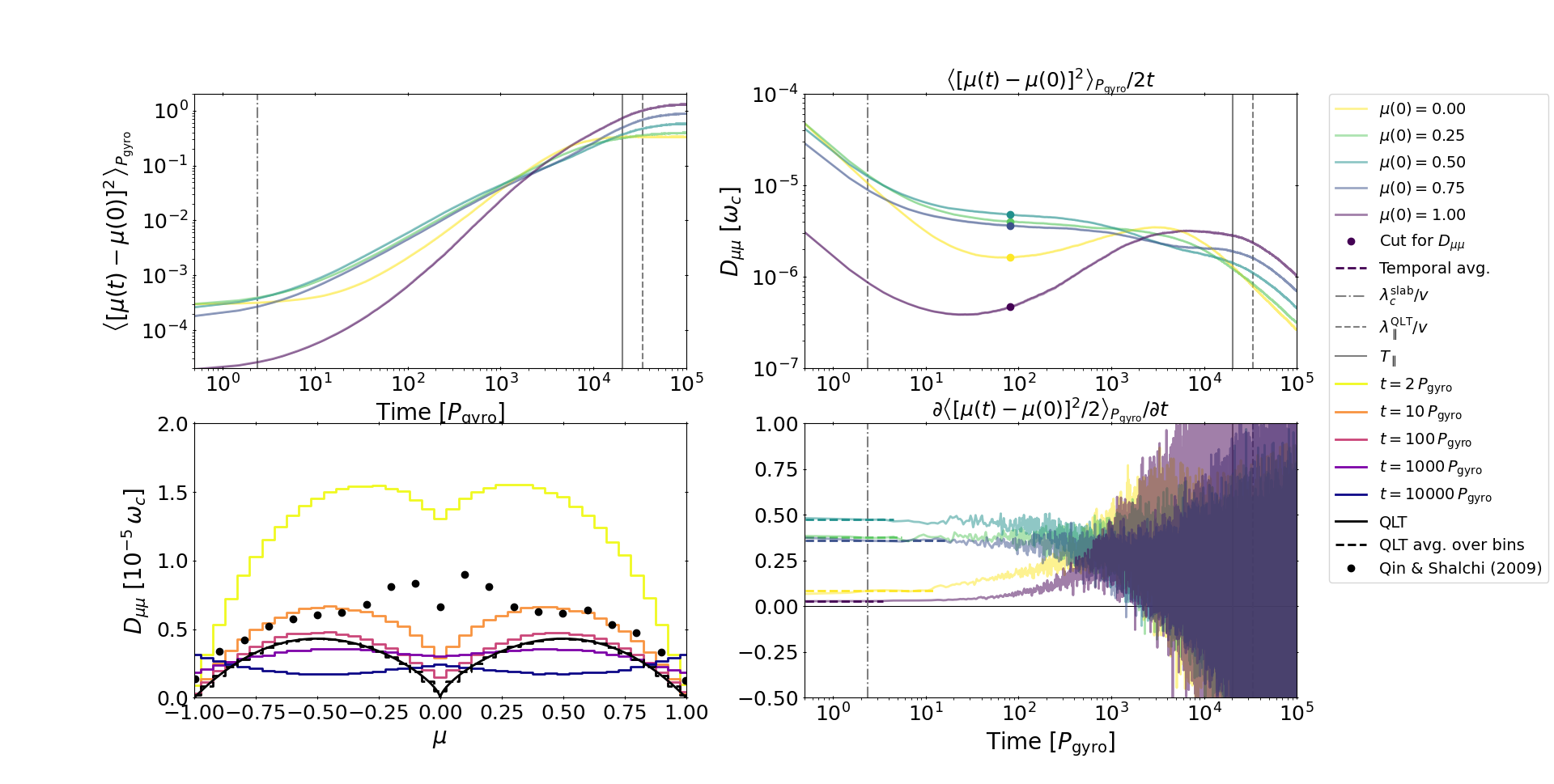}
\caption{\label{fig:BM2msd}\textbf{Results of M1a (Eq.~\ref{eq:DmmMSDt}) and M1b (Eq.~\ref{eq:DmmMSDdt}) as a function of time for different initial pitch-angles (indicated in the legend by different colors) from Sec.~\ref{subsec:BM2}. \textit{Top left panel:} Pitch-cosine MSD calculated from the simulations and averaged over a gyration. \textit{Right column:} Running PADC for M1a (\textit{top panel}) and M1b (\textit{bottom panel}) calculated from the MSD in the top left panel. Dots in the top right panel indicate the time (i.e., at 80 gyrations) when the PADC was constructed for M1a, while horizontal dashed lines in the bottom right panel show the time over which the results were averaged to construct the PADC for M1b (i.e., the solid cyan and green lines, respectively, in the top left panel of Fig.~\ref{fig:BM2padc}). The vertical grey lines are as explained in Fig.~\ref{fig:BM1msd}. \textit{Bottom left panel:} Similar to the top left panel of Fig.~\ref{fig:BM2padc}, but for the PADC constructed from M1a at different times (indicated by the different colors in the legend).}}
\end{center}
\end{figure*}

\begin{figure}[t]
\begin{center}
\includegraphics[trim=224mm 111mm 12mm 24mm, clip, width=0.475\textwidth]{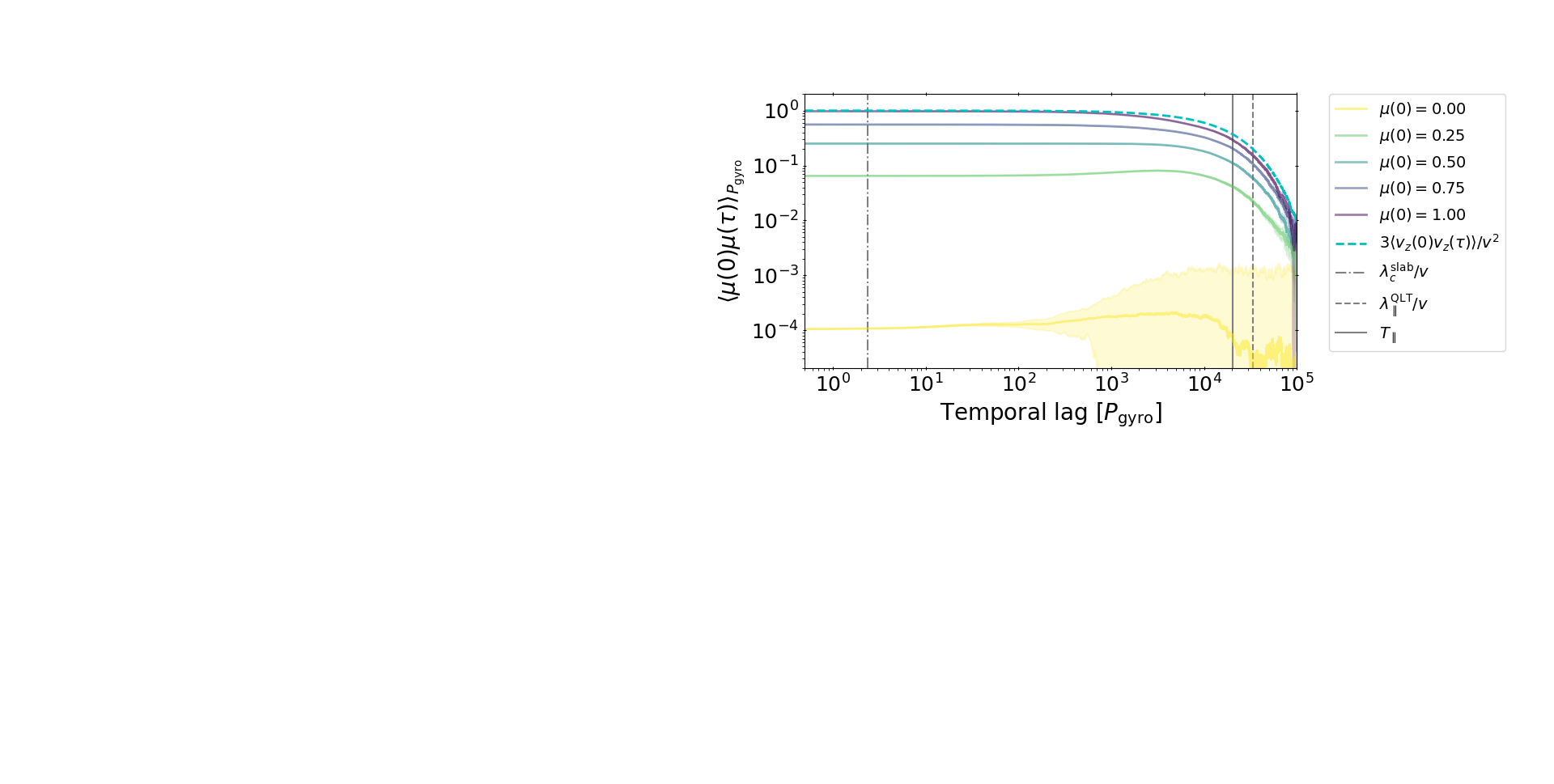}
\caption{\label{fig:BM2Correlation}\textbf{Similar to Fig.~\ref{fig:BM1Correlation}, but for Sec.~\ref{subsec:BM2} and without the pitch-angle correlation time as a function of pitch-angle. These correlation functions have been averaged over a gyration. Vertical grey lines are as explained in Fig.~\ref{fig:BM1msd} and shaded areas indicate uncertainties.}}
\end{center}
\end{figure}

\begin{figure*}[t]
\begin{center}
\includegraphics[trim=30mm 112mm 17mm 22mm, clip, width=0.99\textwidth]{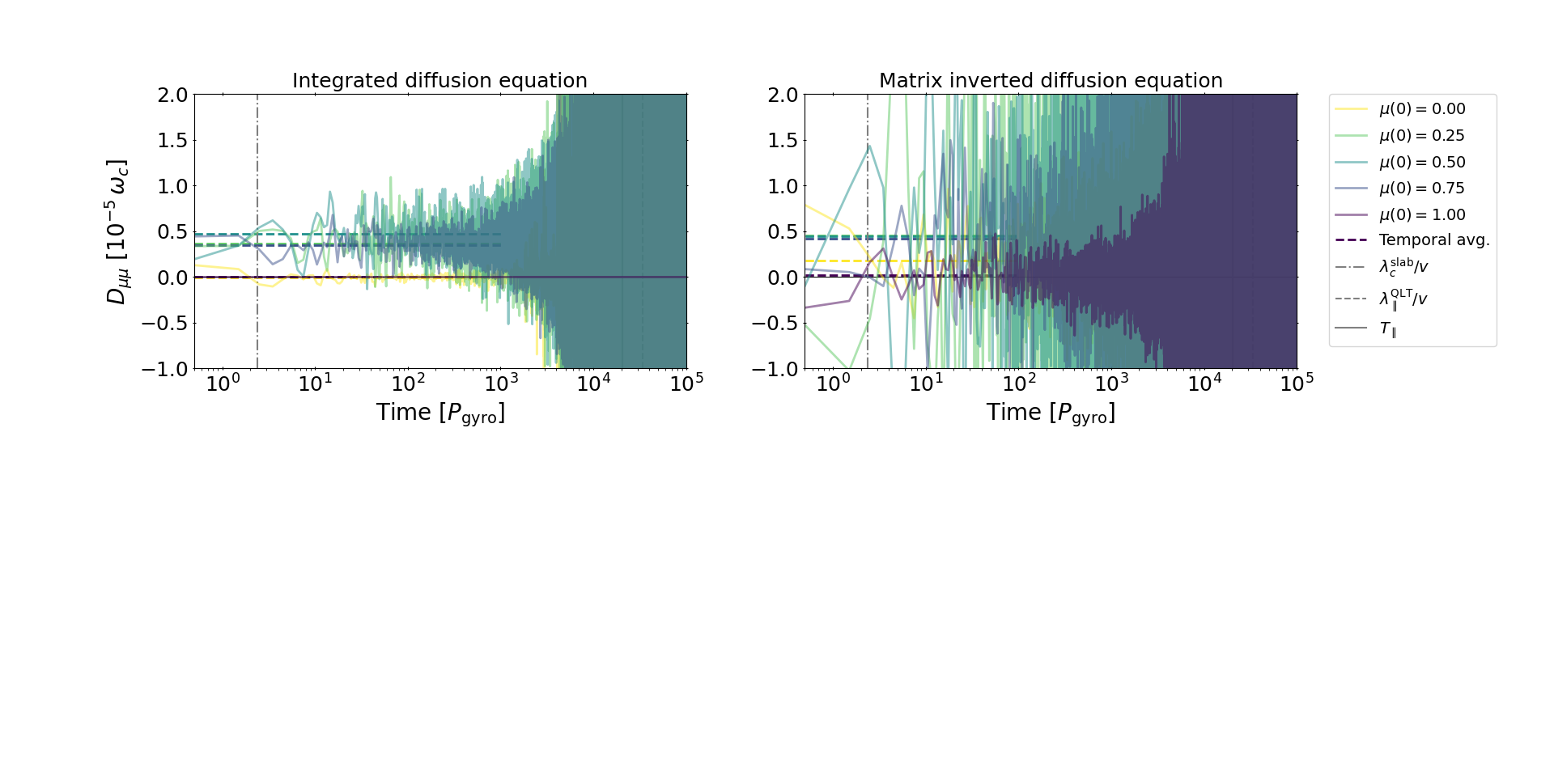}
\caption{\label{fig:BM2deDmmT}\textbf{Similar to the bottom row of Fig.~\ref{fig:BM1deDmmT}, but for Sec.~\ref{subsec:BM2}. The horizontal dashed lines show the time over which the results were averaged to construct the PADC (i.e., the solid orange and red lines
in the bottom left panel of Fig.~\ref{fig:BM2padc}).}}
\end{center}
\end{figure*}

In light of the slightly lower turbulence level for the second case, the total simulation time is increased here to $10^5~\lambda_c^{\rm slab} / v$. Fig.~\ref{fig:BM2padc} - Fig.~\ref{fig:BM2deDmmT} show the results of the second benchmark comparison for weak composite turbulence. Given the discussion of the results from the previous section, the results of this section are only shown averaged over a gyration \textbf{(i.e., the temporal bin width $\varpi_t$ is now 64 times larger than the simulation time step $\delta t$)}. This was also necessary due to memory constraints since the \textbf{number of} grid \textbf{points} used for 2D turbulence\textbf{, $N_x \times N_y = 2^{14} \times 2^{14} = 2^{28}$,} is much larger than the \textbf{number of} grid \textbf{points} used for slab turbulence\textbf{, $N_z = 2^{23}$}. Methods yielding an inherently oscillatory running diffusion coefficient (i.e., M2a and M3) are therefore not considered \textbf{as the oscillations cannot be resolved}. All the PADC results from the second benchmark comparison between the different methods are summarized in Fig.~\ref{fig:BM2padc} together with the expectation from QLT (Eq.~\ref{eq:DmmQLT}) and the simulation results of \citet{QinShalchi2009}. The results from M0 (Eq.~\ref{eq:DmmFP}) and M3 (Eq.~\ref{eq:DmmGea99}) are not shown, but both these methods yielded PADCs that were too large (i.e., both yielded $\kappa_{\parallel}^{D_{\mu\mu}} / \kappa_{\parallel}^{\rm QLT} = 0.11$). This is probably because \textbf{the statistics of} $|\Delta \mu|$ \textbf{is different} for the larger temporal bins used in the current setup. M1a, M1b, M4a, M4b, and M5 are again in very good agreement with one another and QLT. The good agreement between the different methods and QLT seems to indicate that 2D turbulence does not have much of an influence on pitch-angle scattering, but this should be tested further for other ratios of $R_L / l_b^{\rm slab}$ and stronger turbulence levels.

It is clear \textbf{from Fig.~\ref{fig:BM2padc}} that none of the methods yielded PADCs that are comparable to the simulation results of \citet{QinShalchi2009}. It is unclear how these authors constructed their PADCs, and the MSD does not seem to behave in such a way that their specific results can be recovered even if the PADC is constructed for each pitch-angle at a different time \textbf{(also see the next paragraph)}. It might be mentioned at this point that \citet{QinShalchi2009} find nearly diffusive behaviour in pure slab turbulence, but not in pure 2D turbulence, which is in direct contrast with their result of finding more pitch-angle scattering through $90^{\circ}$ pitch-angles in composite turbulence (as shown in Fig.~\ref{fig:BM2padc}). A careful consideration of their examples, however, reveals that they show the sub-diffusive behaviour for $90^{\circ}$ pitch-angles in pure 2D turbulence on a log-log scale up to a time of $100-1000~\lambda_c^{\rm slab} / v$ while showing diffusive behaviour for different pitch-angles in pure slab turbulence on a linear scale up to a time of only $25~\lambda_c^{\rm slab} / v$. The question must then be asked if the sub-diffusive behaviour seen in pure 2D turbulence is not just an artefact of the method used in constructing the PADC (presumably by using M1a) which will also manifest in pure slab turbulence if the running PADC is investigated over a longer time interval on a log-log scale.

\paragraph{M1a \& M1b}

The pitch-angle MSD shown in the top left panel of \textbf{Fig.~\ref{fig:BM2msd} also saturates as in} the previous section \textbf{(the oscillations seen in the top row of Fig.~\ref{fig:BM1msd} is not visible here due to the results being averaged over a gyration).} The running PADCs resulting from M1a (Eq.~\ref{eq:DmmMSDt}) and M1b (Eq.~\ref{eq:DmmMSDdt}) are shown in the \textbf{top right and bottom right} panels of \textbf{Fig.~\ref{fig:BM2msd}}, respectively. The PADC \textbf{shown in the top left panel of Fig.~\ref{fig:BM2padc}} was constructed from M1a after $16 \times 5 = 80$ gyrations since the running PADC is not yet constant for most pitch-angles after $16$ gyrations \textbf{(note that if the fraction of energy in the slab component is $5$ times less than in the previous section, it could be expected that the pitch-angle scattering process would be $\sim 5$ times slower). To illustrate how the PADC from M1a depends on the time chosen to construct it, the bottom left panel of Fig.~\ref{fig:BM2msd} shows the PADC at five different times. It can be seen from the yellow and orange lines that if the PADC is constructed too early, then the amount of pitch-angle scattering will be overestimated. The purple and blue lines show that if the PADC is constructed too late, then the shape of the PADC will be wrong and the scattering amplitude will be underestimated.} To construct the PADC from M1b, the instantaneous value of each pitch-angle was again compared to the cumulative average until it deviated by more than 1.5 times the cumulative standard deviation. \textbf{Similar to the previous section, this method, for even lower levels of slab turbulence, gives a constant running diffusion coefficient within the first 10 gyrations.}

\paragraph{M2b}

The results of the pitch-angle correlation function shown in \textbf{Fig.~\ref{fig:BM2Correlation}} are all very similar to the results of the previous section: the correlation function decays exponentially over the time shown and before noise starts masking features, pitch-angles close to $90^{\circ}$ are correlated longer, and calculating the correlation time by fitting an exponential function yields better results than simply integrating the correlation function. The results for the correlation time are qualitatively similar to the results of the previous section and therefore not shown here. Quantitatively, the correlation time is longer due to the lower fraction of slab turbulence. Eq.~\ref{eq:DmmCorrelTime} had to be multiplied by 2 again \textbf{for the PADC shown in the top right panel of Fig.~\ref{fig:BM2padc} to be comparable to what would be expected from QLT.}

\paragraph{M4a \& M4b}

\textbf{The temporal evolution of the pitch-angle distribution function is not shown here (neither is the stationary distribution function) because it is} qualitatively similar to the results of the previous section\textbf{, while q}uantitatively, the distribution function isotropizes slower due to the lower fraction of slab turbulence. The running PADCs resulting from M4a (Eq.~\ref{eq:DmmIntDiffEq}) and M4b (Eq.~\ref{eq:DmmMatrixInvertDiffEq}) are shown in the \textbf{left and right} panels of Fig.~\ref{fig:BM2deDmmT}, respectively. The results from M4a and M4b were again averaged over $10^3$ and $10^2$ gyrations, respectively, to construct the PADC \textbf{shown in the bottom left panel of Fig.~\ref{fig:BM2padc}, where a three-point pitch-angle average was also applied to reduce the noise.}

\begin{figure*}[t]
\begin{center}
\includegraphics[trim=33mm 115mm 31mm 26mm, clip, width=0.99\textwidth]{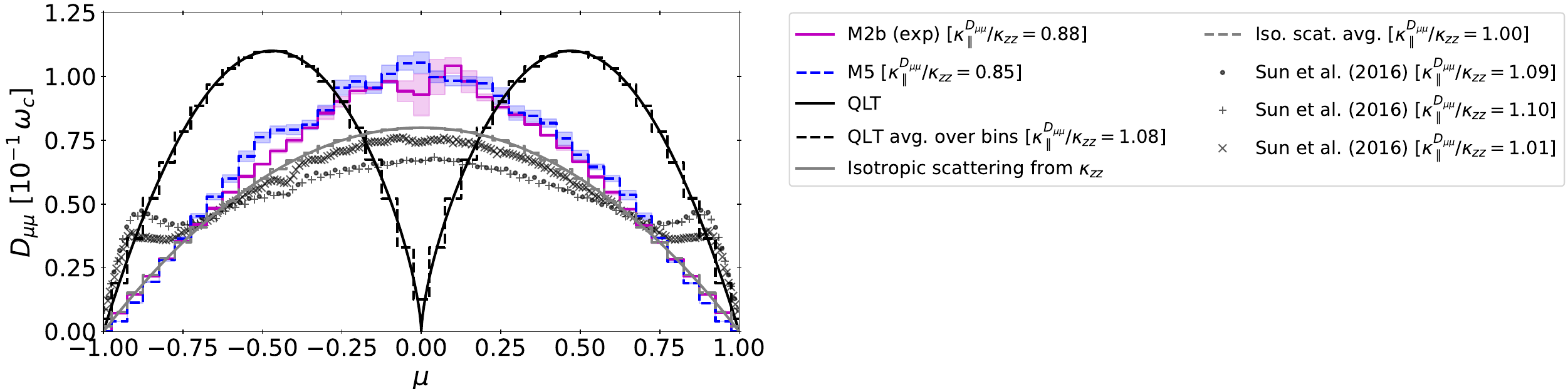}
\caption{\label{fig:BM3padc}PADC as a function of pitch-cosine resulting from Sec.~\ref{subsec:BM3}. The results of M2b (Eq.~\ref{eq:DmmCorrelTime} applied to the blue line in the right panel of Fig.~\ref{fig:BM3Correlation}) and M5 (Eq.~\ref{eq:DmmStationary} applied to the distributions in the right panel of Fig.~\ref{fig:BM3DisFunc}) are compared to the prediction of QLT (Eq.~\ref{eq:DmmQLT}), isotropic scattering with the scattering amplitude calculated from the diffusion coefficient along the background magnetic field, and the simulation results of \citet[][symbols]{SunEA2016}. The dashed lines show the theoretical predictions averaged over the pitch-cosine bins, and the legend gives the ratio of the parallel diffusion coefficient resulting from the constructed PADC (Eq.~\ref{eq:KappaParallel}) to the diffusion coefficient along the background magnetic field calculated from the simulations (denoted by $\kappa_{zz}$). \textbf{Shaded areas indicate uncertainties.}}
\end{center}
\end{figure*}

\begin{figure*}[t]
\begin{center}
\includegraphics[trim=33mm 9mm 25mm 21mm, clip, width=0.99\textwidth]{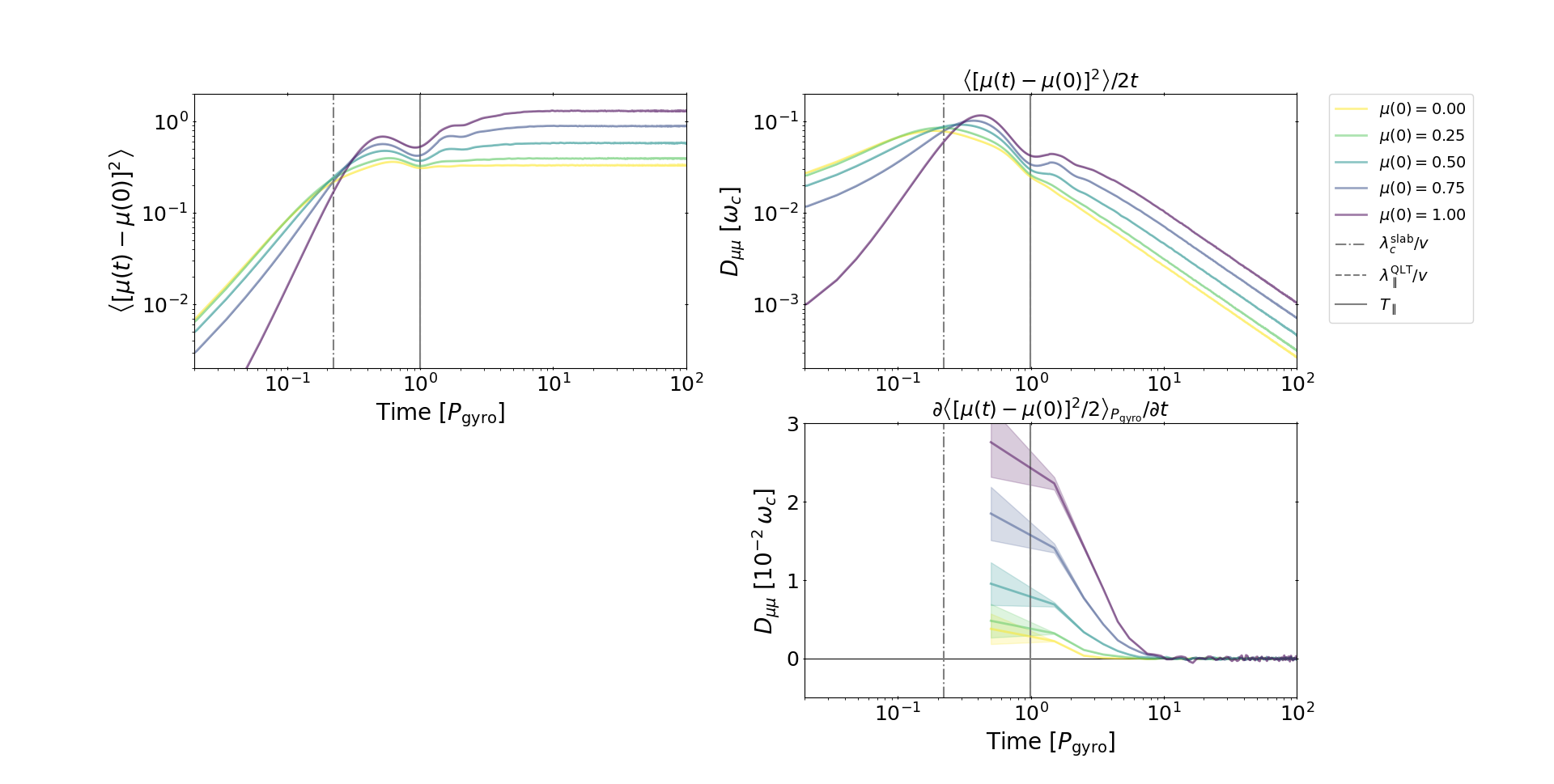}
\caption{\label{fig:BM3msd}\textbf{Similar to Fig.~\ref{fig:BM2msd}, but for Sec.~\ref{subsec:BM3} and without the PADC constructed at different times because the MSD saturates to quickly to calculate a diffusion coefficient from it. Vertical grey lines are as explained in Fig.~\ref{fig:BM1msd}, but the solid and dashed lines coincide. Shaded areas visible in the bottom right panel indicate uncertainties.}}
\end{center}
\end{figure*}


\subsection{Method Comparison in Strong Slab Turbulence}
\label{subsec:BM3}

For the strong turbulence conditions of this section, the total simulation time was reduced to $10^3~\lambda_c^{\rm slab} / v$, and the results were binned over $\varpi_t = 6.25 \times 10^{-2}~\lambda_c^{\rm slab} / v$ \textbf{(4 times longer than the simulation time step)}. The much shorter simulation time also allowed more particles to be used, i.e., $N_p = 1.6 \times 10^5$. The parallel diffusion coefficient calculated directly from the simulation, and not QLT, is used to calculate the parallel scattering time as it is clear that the result has converged by the end of the simulation. For this section, the cyclotron frequency and gyroperiod are defined with respect to the total magnetic field, i.e., $\omega_c = |q| B_0 \sqrt{1 + \delta B^2 / B_0^2} / \gamma m_0$ for the cyclotron frequency, with $\gamma$ and $m_0$ the Lorentz factor and rest mass of the particle, respectively. \textbf{Note that the definition of the pitch-cosine relative to the background magnetic field is not ideal for strong turbulence levels, as the turbulence will greatly perturb the background magnetic field, such that the direction of the total and background magnetic field could be quite different. This, however, is a standard assumption in transport modeling in order to have a well-defined pitch-angle.}

\begin{figure*}[t]
\begin{center}
\includegraphics[trim=35mm 112mm 14mm 27mm, clip, width=0.99\textwidth]{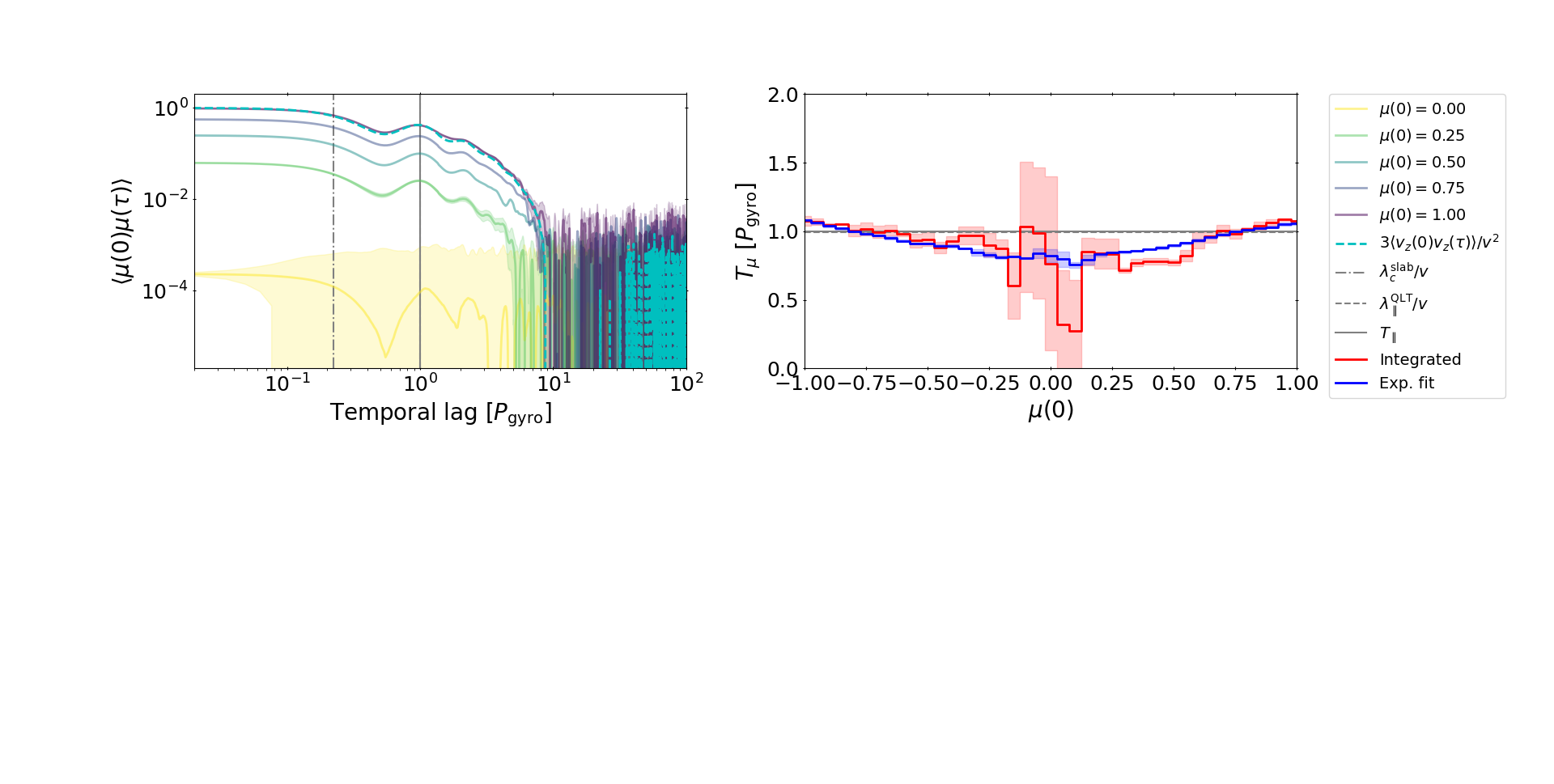}
\caption{\label{fig:BM3Correlation}\textbf{Similar to Fig.~\ref{fig:BM1Correlation}, but for Sec.~\ref{subsec:BM3}. Note that the pitch-angle correlation time is not calculated from the correlation functions after averaging them over a gyration because the correlation functions change too much during a gyration to average them. The grey lines are as explained in Fig.~\ref{fig:BM1Correlation} (the right panel show how the dashed and solid grey lines are almost coinciding) and shaded areas indicate uncertainties.}}
\end{center}
\end{figure*}

\begin{figure*}[t]
\begin{center}
\includegraphics[trim=38mm 115mm 21mm 27mm, clip, width=0.99\textwidth]{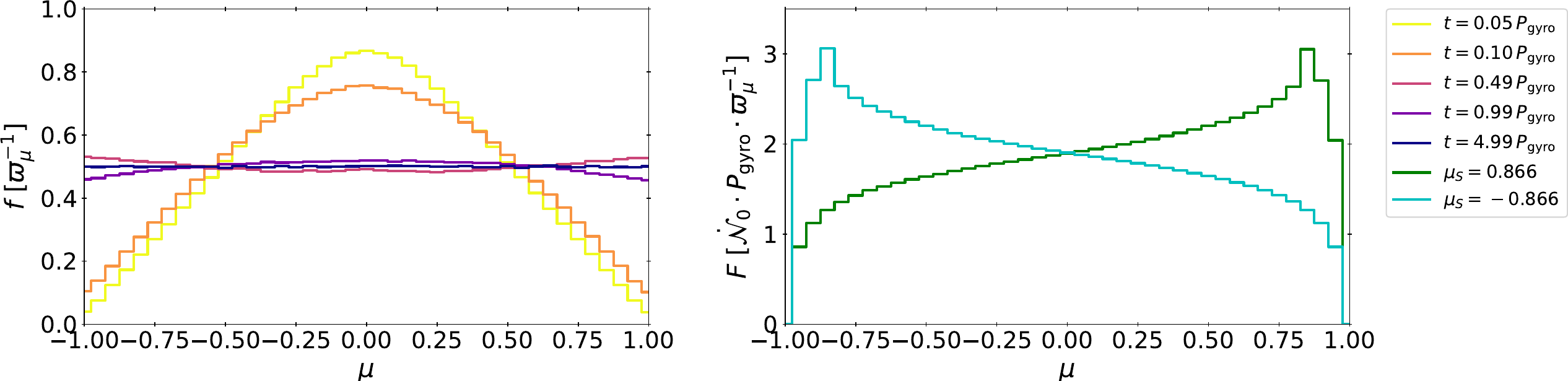}
\caption{\label{fig:BM3DisFunc}\textbf{Similar to Fig.~\ref{fig:BM1DisFunc}, but for Sec.~\ref{subsec:BM3}.}}
\end{center}
\end{figure*}

\begin{figure*}[t]
\begin{center}
\includegraphics[trim=36mm 112mm 25mm 22mm, clip, width=0.99\textwidth]{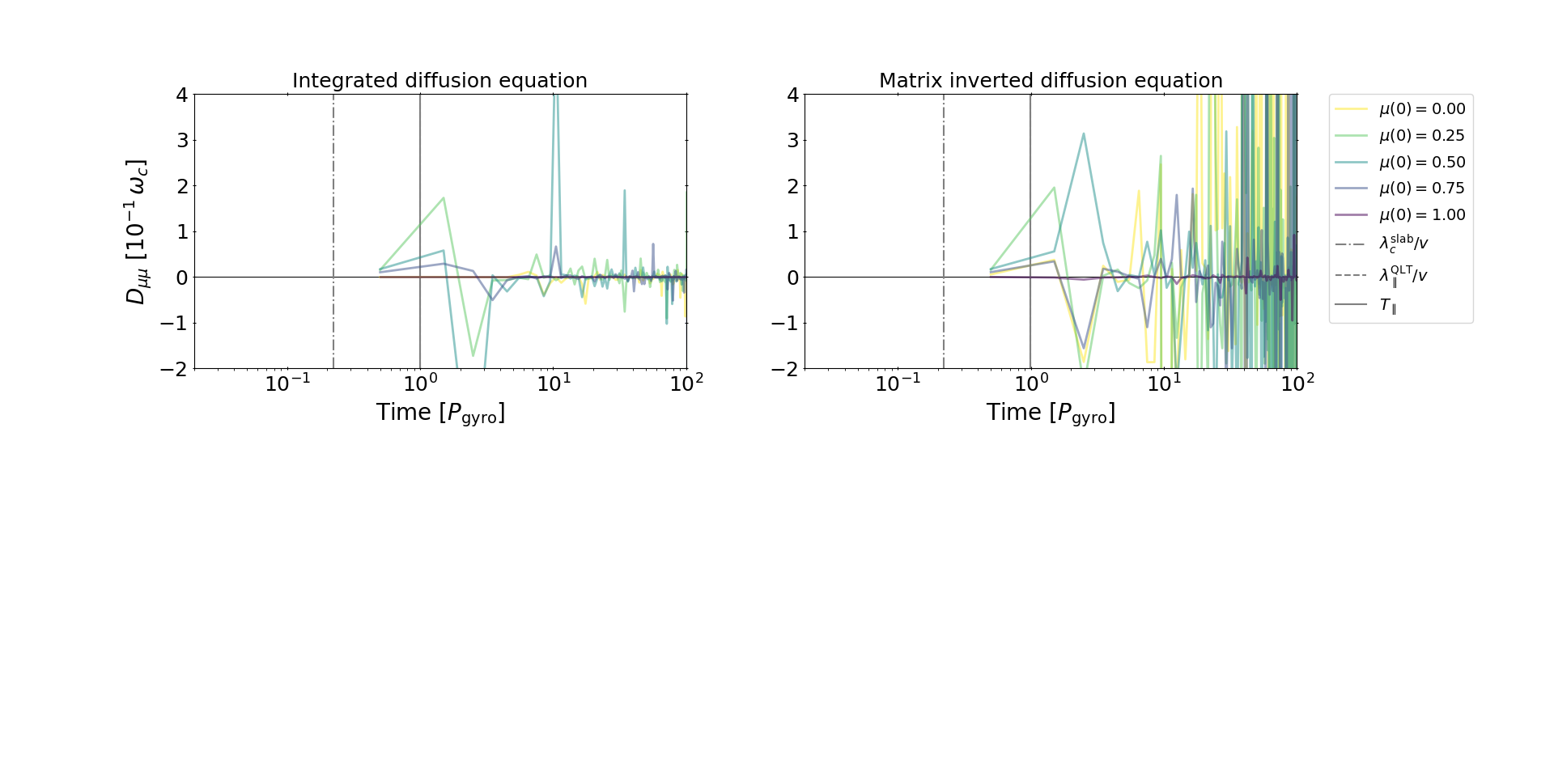}
\caption{\label{fig:BM3deDmmT}\textbf{Similar to Fig.~\ref{fig:BM2deDmmT}, but for Sec.~\ref{subsec:BM3}.}}
\end{center}
\end{figure*}

The results of this third benchmark comparison for strong pure slab turbulence are shown in Fig.~\ref{fig:BM3padc} - Fig.~\ref{fig:BM3deDmmT}. Note, for these results, that the parallel scattering time is approximately equal to the gyro-period, implying an extreme case of pitch-angle scattering. All the PADC results from the third benchmark comparison between the different methods are summarized in Fig.~\ref{fig:BM3padc} together with theoretical expectations and the simulation results of \citet[][the three different sets of simulation results are for the three different field normalizations used in their simulations]{SunEA2016}. We note that \citet{SunEA2016} does not use the same spectrum as this work, i.e., compare $g_{\rm slab} \propto (1 + l_b^{\rm slab} k_{\parallel})^{-\nu}$ with Eq.~\ref{eq:SlabSpectrum}. Therefore, the results presented here cannot be compared quantitatively to their results, but the overall magnitude and shape of the results might still be comparable. While the expectation from QLT (Eq.~\ref{eq:DmmQLT}) is included for reference, it is not expected to be valid in this \textbf{strong turbulence} case. Additionally, based on the results of \citet{SunEA2016} the expectation from isotropic pitch-angle scattering is also shown, where the scattering amplitude is normalized to the diffusion coefficient along the $z$-axis calculated in the simulation ($\kappa_{zz}$) using Eq.~\ref{eq:KappaParallel}.

\paragraph{M1a \& M1b}

The MSD is shown in the top \textbf{left} panel of \textbf{Fig.~\ref{fig:BM3msd}} and saturates again after the parallel scattering time. The running PADCs from M1a (Eq.~\ref{eq:DmmMSDt}) are shown in the top \textbf{right} panel. It is evident that the MSD saturates so quickly that there is hardly any time interval where the diffusion coefficients remain constant. It can be seen in the bottom \textbf{right} panel that averaging the MSD over a gyration and applying M1b (Eq.~\ref{eq:DmmMSDdt}) does not yield an initially constant running PADC. The general shape of the PADC during the first few gyrations is also not theoretically expected and is overall too small compared to the results of other methods. Averaging the MSD over a turbulent correlation crossing time also does not yield a useful result (not shown). \textbf{No PADC could be constructed from the MSD since M1a and M1b} break down in strong turbulence.

\paragraph{M2b}

The \textbf{left} panel of \textbf{Fig.~\ref{fig:BM3Correlation}} shows the pitch-angle correlation function, now exhibiting damped oscillations for all pitch-cosine bins. The oscillations and late-time noise \textbf{(note the amount of noise still present even though the number of particles has been doubled)} make it difficult to determine whether the decorrelation trend is exponential. Comparing the cumulative correlation function to that of an exponential decorrelation (not shown here) seems to indicate a possible difference, at least due to the oscillations. The pitch-angle correlation time is displayed in \textbf{the right panel}. The cumulative correlation function was used only up to 10 gyrations to calculate the correlation time from fitting an exponential function to avoid excessive influence from noise on the results. The correlation time exhibits only a slight pitch-angle dependence.

\textbf{The PADC resulting} from M2b \textbf{follows} the expectation from isotropic scattering for $|\mu| > 0.7$, while yielding more scattering closer to $\mu = 0$. Note that Eq.~\ref{eq:DmmCorrelTime} did not need to be multiplied by a factor of two, as in the previous two sections. \textbf{We note that the theoretical derivation by \citet{ShalchiEA2012} for the parallel velocity correlation function due to isotropic pitch-angle scattering from the pitch-angle diffusion equation (Eq.~\ref{eq:TPE}) can be generalized to derive the expected pitch-angle correlation function and time \citep[see also][]{Tautz2013}. In this case it can be verified that the pitch-angle correlation time is indeed related to the PADC by Eq.~\ref{eq:DmmCorrelTime}, but we are not aware of similar work existing for anisotropic pitch-angle scattering.} Regardless of the problems of this method, it is still shown here for comparison to other methods.

\paragraph{M4a \& M4b}

The temporal evolution of an initially triangular distribution is shown in the \textbf{left} panel of \textbf{Fig.~\ref{fig:BM3DisFunc}}. The distribution function changes very quickly in these strong turbulence conditions and is already close to \textbf{isotropy} after only half a gyration. \textbf{Fig.~\ref{fig:BM3deDmmT}} shows the running PADCs from M4a \textbf{(Eq.~\ref{eq:DmmIntDiffEq})} and M4b \textbf{(Eq.~\ref{eq:DmmMatrixInvertDiffEq})}. The results are too noisy to average \textbf{them} over more than two gyrations, which is somewhat expected given the rapid changes in the distribution function\textbf{, and do not yield a result in agreement with either M2b or M5}. Once again, averaging the distribution function only over a turbulent correlation crossing time does not yield a more useful result \textbf{(i.e., it is difficult to pick the correct temporal resolution for the distribution function if it is changing too rapidly)}.

\paragraph{M5}

Since M5 \textbf{(Eq.~\ref{eq:DmmStationary})} is sensitive to the pitch-angle reaching an absorbing boundary, $2^{15}$ steps were taken per gyration for this method to correctly identify when the particle reaches such a boundary. The injection was also moved to $|\mu_S| = 0.866$ to ensure the source is far enough away from the hemisphere where the PADC should be constructed since the diffusion rate is much higher, giving particles access to the other hemisphere more easily. The stationary distribution functions for a delta injection and absorbing boundaries are shown in the \textbf{right} panel of Fig.~\ref{fig:BM3DisFunc}. The distribution displays large intervals around $\mu = 0$ with a fairly constant slope, unlike the changing distributions of the previous two sections (compare with the \textbf{right} panel of Fig.~\ref{fig:BM1DisFunc}), indicating that pitch-angle scattering does not have a very strong pitch-angle dependence around $90^{\circ}$ pitch-angles. \textbf{The PADC in Fig.~\ref{fig:BM3padc} resulting from} M5 \textbf{is} in good agreement \textbf{with the result of M2b} and follow the expectation from isotropic scattering for $|\mu| > 0.7$, while yielding more scattering closer to $\mu = 0$.


\subsection{Best Practices for M5}
\label{subsec:BestPractice}

\textbf{Unfortunately, it is not clear how far from the source M5 is valid (since the system might not be diffusive near the source and the derivative is not guaranteed to be continuous at the source), so we investigate the effect of the source location when using M5. Firstly, we discuss a possible way to estimate where in the pitch-cosine domain the particles are diffusive, together with a possible improvement of the method to include both diffusive and streaming flux. However, both are argued to be inadequate and we end this section with a possible `best practice' approach for M5.}

\textbf{Firstly, one might think that the average propagation time of particles from the source to a pitch-cosine bin could be used to judge if the particles have had enough time to reach a diffusive limit, but this does not seem to be a reliable measure: in the case of strong turbulence (Sec.~\ref{subsec:BM3}), particles were able to propagate from $\mu_S = \pm 0.866$ to $\mu = \mp 0.175$ in less than a gyroperiod on average, while the bins further than $\mu = \mp 0.375$ had average propagation times longer than the parallel scattering time; in the case of weak turbulence (Sec.~\ref{subsec:BM1}), particles took on average more than $10$ gyroperiods to propagate to the bins next to the bin containing the source, but only the bin next to the furthest absorbing boundary had an average propagation time longer than the parallel scattering time (these propagation times are not shown and are only reported here). In the {\it latter} case, one might naively think that the particles had enough time to reach diffusive behaviour, but such weak turbulence levels do not guarantee enough pitch-angle scattering for diffusive behaviour. In the {\it former} case, one might naively think that the particles have not had enough time to reach diffusive behaviour, but such strong turbulence levels should quickly lead to diffusive behaviour.}

\textbf{Secondly, we note that Eq.~\ref{eq:DmmStationary} is reminiscent of Fick's law, i.e., the diffusive flux is proportional to the diffusion coefficient times the gradient. However, in the general case, the total flux is due to both a `streaming' (in pitch-cosine, not configuration space) and diffusive flux, i.e., $\dot{\cal N} = \langle \dot \mu \rangle F - D_{\mu\mu} ({\rm d}F / {\rm d}\mu)$ (this should also hold for the time-dependent distribution function $f$). Such an equation should be valid closer to the source to calculate the PADC, or the comparison between the streaming and diffusive flux could be used to infer where M5 might be valid, but there are several problems with using such an approach. If this is applied to the simulation setup in M5, it should be taken into account that particles continuously diffuse from the source to the absorbing boundaries, and this motion might appear as a bulk movement of particles in a preferred direction. In applying M5 in both Sec.~\ref{subsec:BM1} and Sec.~\ref{subsec:BM3}, we indeed found the streaming flux to be approximately equal to the flux through the absorbing boundaries if the simulation time step is small enough (not shown). Under different circumstances, the streaming flux is expected to be zero, because starting with an isotropic distribution that remains isotropic in the absence of sources or sinks of particles, it holds that $\dot{\cal N} = 0$ and ${\rm d}F / {\rm d}\mu = 0$, such that the streaming flux should also be zero.}

\begin{figure*}[t]
\begin{center}
\includegraphics[trim=35mm 11mm 44mm 27mm, clip, width=0.99\textwidth]{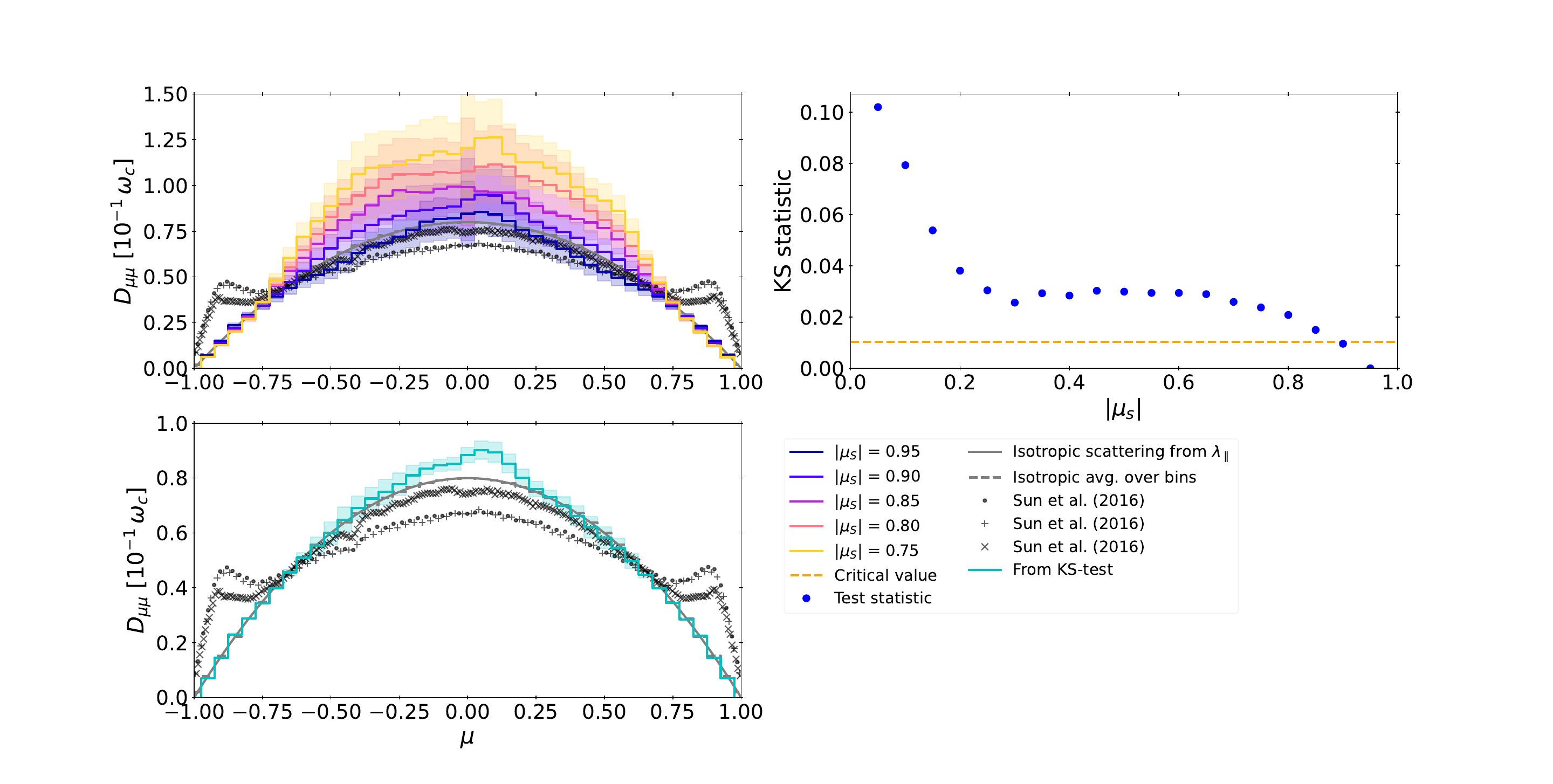}
\caption{\label{fig:BM3BestPracticeM5}\textbf{\textit{Top left panel:} PADC as a function of pitch-cosine resulting from M5 (Eq.~\ref{eq:DmmStationary}) in Sec.~\ref{subsec:BM3} for different source locations (different colors indicated in the legend). \textit{Top right panel:} The KS test statistic of the PADCs for the different sources compared to the result of $|\mu_S| = 0.95$. The horizontal dashed orange line indicates the critical value. \textit{Bottom left panel:} The PADC yielded by averaging the results of the different sources that are statistically similar according to the KS test. The results in the left column are compared to the prediction of isotropic scattering with the scattering amplitude calculated from the diffusion coefficient along the background magnetic field (solid grey line) and the simulation results of \citet[][symbols]{SunEA2016}. The dashed grey line shows the theoretical predictions averaged over the pitch-cosine bins and shaded areas indicate uncertainties.}}
\end{center}
\end{figure*}

\textbf{Conversely, one could also imagine that for a non-constant diffusion coefficient, particles will diffuse away from regions where the diffusion coefficient is larger and will spend more time where the diffusion coefficient is smaller, giving the appearance of some preferred direction of motion for the particles. Hence, if the streaming flux is non-zero in a system where only diffusion operates, then the streaming flux must be somehow related to the varying diffusion coefficient. Indeed, \citet{Jokipii1966} already pointed out that the Fokker-Planck coefficient $\langle \Delta \mu / \Delta t \rangle$ must be related to $\langle (\Delta \mu)^2 / 2 \Delta t \rangle$ for pitch-angle scattering, as explained by \citet{vandenBerg2023}: ``given a long enough time, scattering would cause the distribution function to become isotropic; if scattering did not lead to isotropy, then an initially isotropic distribution could relax to an anisotropic distribution; this, however, would violate Liouville’s theorem, that is, for an isotropic distribution to remain isotropic, there must be a balance between the first and second order processes.'' If this is applied to the Fokker-Plank equation, one finds $\langle \Delta \mu / \Delta t \rangle = \partial [\langle (\Delta \mu)^2 / 2 \Delta t \rangle] / \partial \mu$. Whether or not this can be directly translated to the streaming flux is unclear because it follows from the Fokker-Planck equation \citep[see, e.g.,][]{vandenBerg2023} that
\begin{align}
\frac{\partial f}{\partial t} & = - \frac{\partial}{\partial \mu} \left[ \left\langle \frac{\Delta \mu}{\Delta t} \right\rangle f \right] + \frac{\partial^2}{\partial^2 \mu} \left[ \left\langle \frac{(\Delta \mu)^2}{2\Delta t} \right\rangle f \right] \nonumber \\
 & = \frac{\partial}{\partial \mu} \left[ \left\langle \frac{(\Delta \mu)^2}{2 \Delta t} \right\rangle \frac{\partial f}{\partial \mu} \right] \nonumber \\
 & = - \frac{\partial \dot{\cal N}}{\partial \mu},
\end{align}
implying a total flux of $\dot{\cal N} = - [\langle (\Delta \mu)^2 / 2 \Delta t \rangle] (\partial f / \partial \mu)$, which is again just Fick's law. Hence, if it is unknown exactly how such a `diffusion-induced streaming' flux is related to the diffusion coefficient, then $\dot{\cal N} = \dot{\cal N}_{\rm stream} - D_{\mu\mu} ({\rm d}F / {\rm d}\mu)$ cannot be solved for the diffusion coefficient. Additionally, the streaming flux can be calculated independently (through $\langle \dot \mu \rangle F$), but the diffusive flux cannot, and so one has to assume that $\dot{\cal N}_{\rm diff} = \dot{\cal N} - \dot{\cal N}_{\rm stream}$.}

\begin{figure*}[t]
\begin{center}
\includegraphics[trim=44mm 11mm 44mm 27mm, clip, width=0.99\textwidth]{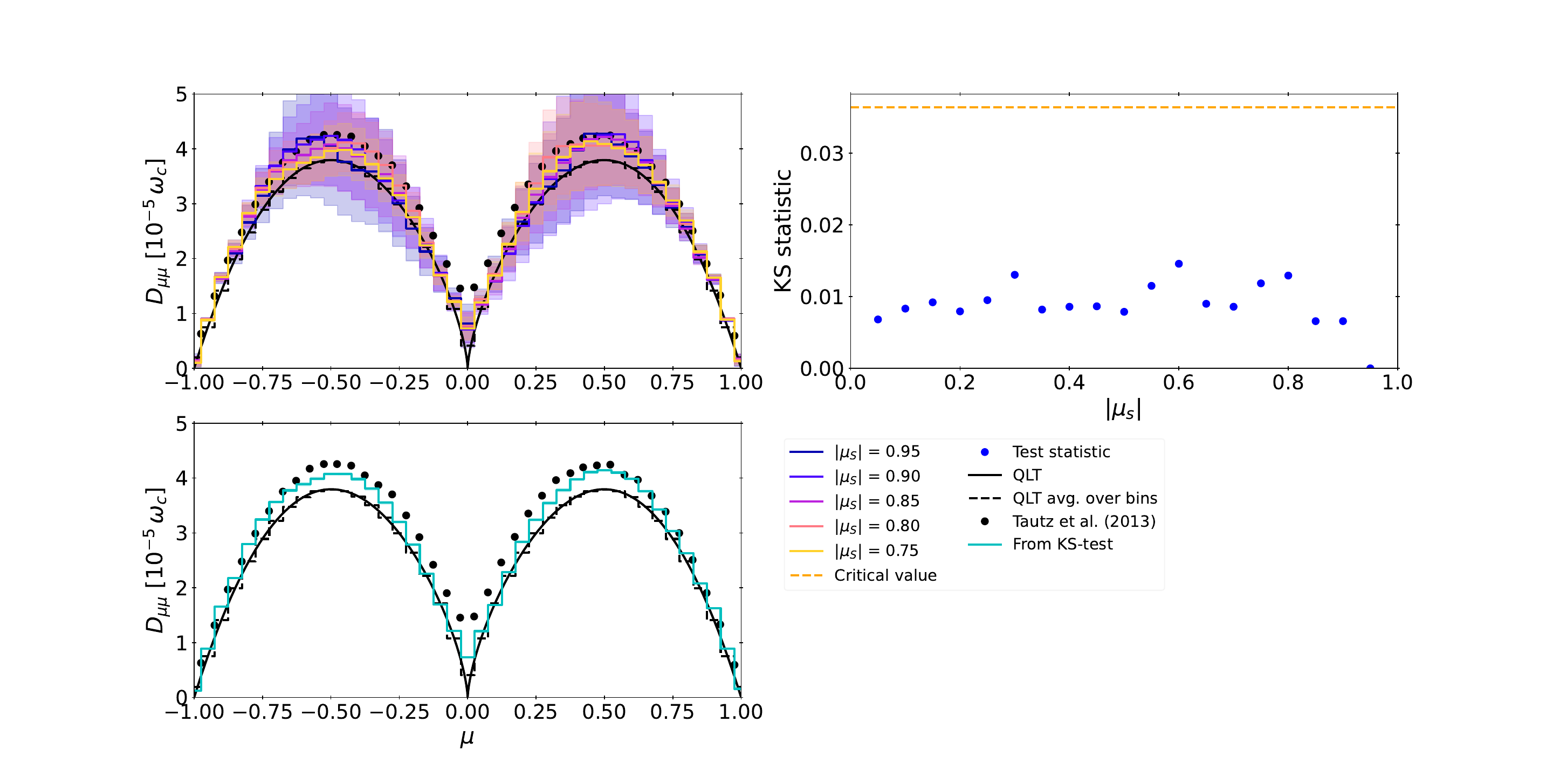}
\caption{\label{fig:BM1BestPracticeM5}\textbf{Similar to Fig.~\ref{fig:BM3BestPracticeM5}, but for Sec.~\ref{subsec:BM1} and comparing the results in the left column to the prediction from QLT (Eq.~\ref{eq:DmmQLT}) and the simulation results of \citet[][symbols]{TautzEA2013}.}}
\end{center}
\end{figure*}

\textbf{In the previous section, the source location was moved for M5 to ensure that the source is far enough from the hemisphere where the PADC is constructed. The top left panel of Fig.~\ref{fig:BM3BestPracticeM5} clearly shows how much the results change if the source is moved for the strong turbulence case of Sec.~\ref{subsec:BM3}, while the top left panel of Fig.~\ref{fig:BM1BestPracticeM5} shows that the results are not sensitive to the source location in the weak turbulence case of Sec.~\ref{subsec:BM1}. In order to quantify whether the results of two different sources are statistically similar, we propose to apply a two-sample Kolmogorov–Smirnov (KS) test: assume that the PADC resulting from the source being the closest to the absorbing boundary ($|\mu_S| = 0.95$ in these examples) is the closest one can come to the true solution (i.e., the null hypothesis) since the source is as far from the hemisphere where the PADC is constructed as possible; compare the PADCs resulting from different source locations to the null hypothesis with the KS test; all PADCs resulting from sources with test statistics smaller than the critical value are statistically similar and can be averaged together. The KS test statistic for Sec.~\ref{subsec:BM3} is shown in the top right panel of Fig.~\ref{fig:BM3BestPracticeM5} and it is clear that the results change significantly with a change in the source position. The PADC from only the first two source locations are statistically similar according to the KS test and their resulting average is shown in the bottom left panel of Fig.~\ref{fig:BM3BestPracticeM5}. Repeating the KS test for Sec.~\ref{subsec:BM1} yields test statistics for the different sources that are all below the critical value, as shown in the top right panel of Fig.~\ref{fig:BM1BestPracticeM5}. It is also clear from visual inspection of the top left panel that the PADC resulting from the different injections are very similar. Note that, compared to the general values of the test statistic, the critical value is not sensitive to the significance level, but is set by the number of particles used in the simulations.}

\section{Discussion and Conclusion}
\label{sec:Discuss}

This paper investigated nine different methods to calculate the pitch-angle diffusion coefficient (PADC) from full-orbit simulations of charged particles in synthetic magnetic turbulence. The single-particle simulation code is summarized in Sec.~\ref{sec:Code} and the methods are described in Sec.~\ref{sec:Methods}. These methods were compared in Sec.~\ref{sec:Benchmark} to one another, previous simulation results, and theoretical expectations where applicable, for the cases of both weak pure slab (Sec.~\ref{subsec:BM1}) and composite (slab+2D; Sec.~\ref{subsec:BM2}) turbulence and strong pure slab turbulence (Sec.~\ref{subsec:BM3}). Although all of the methods should theoretically yield comparable results, it was found that some of the methods are not well-suited for numerical work. The reasons for this vary from being too noisy to presenting difficulty in finding the time interval over which the running PADC is constant. Other methods were found to break down when strong turbulence conditions were considered. Details of this will be summarized and our recommendation for best practices will be discussed here.

It should be noted that the effect of the pitch-cosine bin size was not investigated here. Small bins are desirable but often yield poor statistics, resulting in noisy outcomes and increased uncertainty. Conversely, large bins offer better statistics, leading to less noise, but information is lost about the shape of the PADC when averaging over broader pitch-cosine \textbf{bins}, thereby also increasing the uncertainty. The most important region for differentiating between scattering theories is likely around $90^{\circ}$ pitch-angles, which also proves to be numerically challenging for constructing the PADC. Given the theoretical expectation \textbf{of} anisotropic pitch-angle scattering through $\mu = 0$, where the PADC changes rapidly ($\mu$ is the cosine of the particle pitch-angle), one might expect the highest uncertainty near $\mu = 0$. However, the uncertainty might also be significant where the PADC have maxima, as the pitch-angles change the most rapidly for those pitch-angles. Nonetheless, when comparing simulation results to theory, we recommend averaging the theoretical PADC over the pitch-cosine bins before doing a quantitative comparison.

\begin{figure}[t]
\begin{center}
\includegraphics[trim=40mm 120mm 243mm 22mm, clip, width=0.475\textwidth]{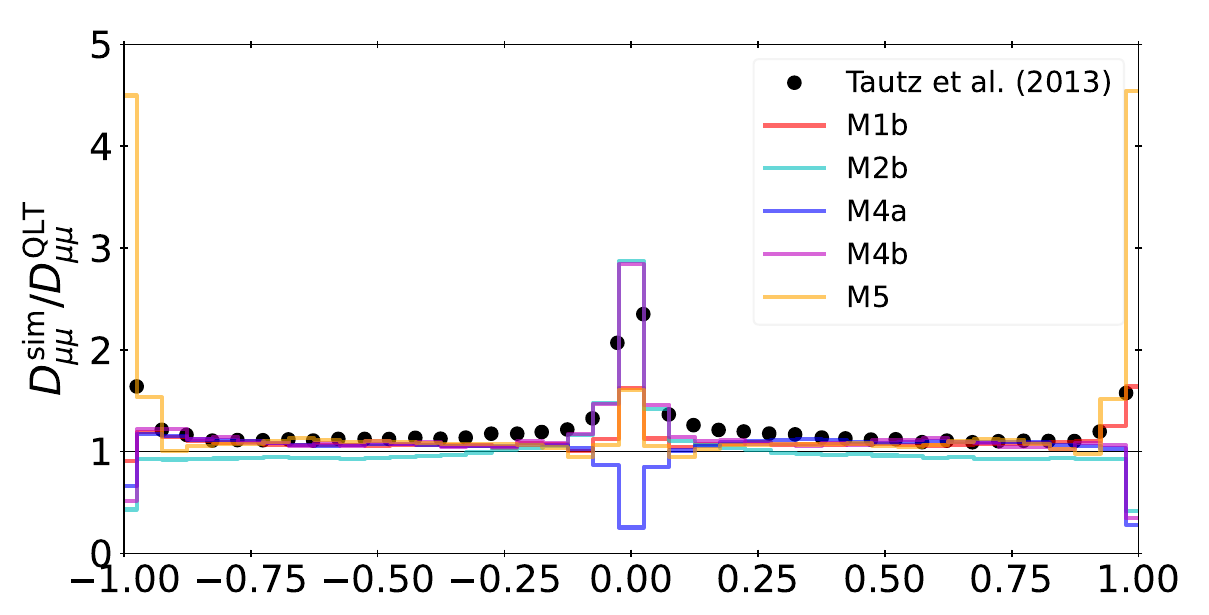}
\includegraphics[trim=40mm 120mm 243mm 22mm, clip, width=0.475\textwidth]{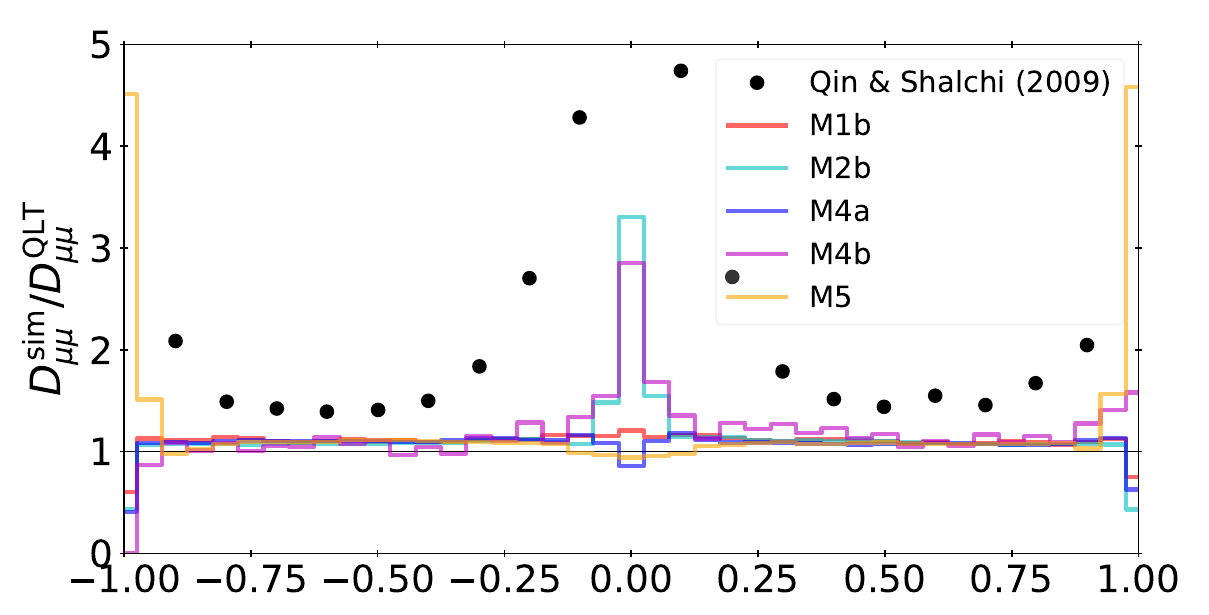}
\includegraphics[trim=40mm 115mm 243mm 22mm, clip, width=0.475\textwidth]{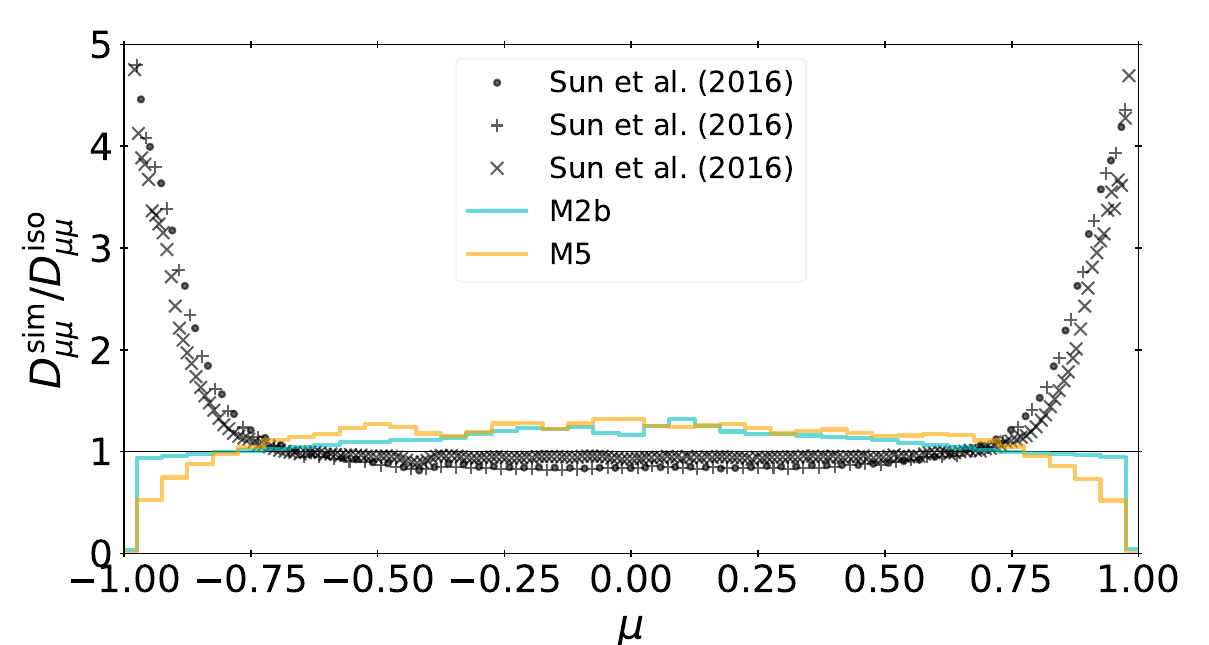}
\caption{\label{fig:BMCompareMethods}Comparison of the PADCs resulting from M1b \textbf{(Eq.~\ref{eq:DmmMSDdt})}, M2b \textbf{(Eq.~\ref{eq:DmmCorrelTime})}, M4a \textbf{(Eq.~\ref{eq:DmmIntDiffEq})}, M4b \textbf{(Eq.~\ref{eq:DmmMatrixInvertDiffEq})}, and M5 \textbf{(Eq.~\ref{eq:DmmStationary})} for Sec.~\ref{subsec:BM1} (\textit{top panel}), Sec.~\ref{subsec:BM2} (\textit{middle panel}), and Sec.~\ref{subsec:BM3} (\textit{bottom panel}). To remove the general shape of the PADC and focus on the differences between methods, the PADCs from the different methods in the top and middle panels are divided by the QLT prediction (Eq.~\ref{eq:DmmQLT}) averaged over the pitch-cosine bins (i.e., the dashed black lines in Fig.~\ref{fig:BM1padc} and Fig.~\ref{fig:BM2padc}), while the PADCs in the bottom panel are divided by the isotropic scattering result averaged over the pitch-cosine bins (i.e., dashed grey lines in Fig.~\ref{fig:BM3padc}). M1b\textbf{, M4a, and M4b are} not shown in the bottom panel since \textbf{they did not yield an usable} result.}
\end{center}
\end{figure}

Based on the results of the method comparisons in Sec.~\ref{sec:Benchmark}, we do not deem M0 \textbf{(Eq.~\ref{eq:DmmFP})}, M1a \textbf{(Eq.~\ref{eq:DmmMSDt})}, M2a \textbf{(Eq.~\ref{eq:DmmTGK})}, and M3 \textbf{(Eq.~\ref{eq:DmmGea99})} viable methods to construct the PADC from simulations. Specifically: M0 and M3 are dependent on the time interval used to construct $\Delta \mu$, thereby biasing the statistics of $\Delta \mu$ \textbf{(see the dashed red line in the top left panel of Fig.~\ref{fig:BM1padc} for M0)}; it is challenging to identify the time interval where M1a yields a constant value, particularly due to the initial oscillations and the faster saturation of the MSD with increased turbulence strength \textbf{(see the right panel in the second row and the bottom left panel of Fig.~\ref{fig:BM1msd} and compare this to the top right panel of Fig.~\ref{fig:BM3msd})}; M2a \textbf{and M3} become too noisy by the time the answer starts oscillating around a constant value \textbf{(see the right panel of Fig.~\ref{fig:BM1tgk} for M2a and Fig.~\ref{fig:BM1Gea99} for M3)}. M2b \textbf{(Eq.~\ref{eq:DmmCorrelTime})} is based on theoretical assumptions and seems to work empirically, but it suffers from a geometry problem: it appears to be correct for isotropic pitch-angle scattering \textbf{(see the solid purple line in Fig.~\ref{fig:BM3padc})}, where the PADC only has a $1 - \mu^2$ dependence, but \textbf{its assumption fails in other situations (as it was a factor of two too small} for anisotropic scattering\textbf{; see the discussion surrounding the solid blue and purple lines in the top right panels of Fig.~\ref{fig:BM1padc} and Fig.~\ref{fig:BM2padc})}. \textbf{Hence, \citet{RiordanPeer2019} must be followed in that the assumption of the scattering and correlation time being related by Eq.~\ref{eq:DmmCorrelTime} is generally not valid and seemingly only valid by coincidence for isotropic scattering.} Therefore, we do not recommend using this method, especially not for comparison to theory, although it could provide an idea of the general shape and magnitude of the PADC if it is ensured that noise in the correlation function does not significantly influence the correlation time. Given the possible theoretical interest of the pitch-angle correlation function and time in scattering theories (see the discussion around Eq.~\ref{eq:DmmCorrelTime}), we defer a theoretical and numerical investigation of this to future work.

M1b \textbf{(Eq.~\ref{eq:DmmMSDdt})}, M4a \textbf{(Eq.~\ref{eq:DmmIntDiffEq})}, and M4b \textbf{(Eq.~\ref{eq:DmmMatrixInvertDiffEq})} could be used in low levels of turbulence if the MSD (for M1b\textbf{; compare the right panel in the third row to the bottom right panel of Fig.~\ref{fig:BM1msd}}) or distribution function (for M4a and M4b\textbf{; compare the top panels to the bottom panels of Fig.~\ref{fig:BM1deDmmT}}) is first averaged over a gyration. Specifically, these methods can be averaged over a relatively easily identifiable time interval where the running PADC is fairly time-independent, and they are the closest to the prediction of QLT and the simulation results of \citet{TautzEA2013}. M1b still changes with time \textbf{(see the bottom right panels of Fig.~\ref{fig:BM1msd} and Fig.~\ref{fig:BM2msd})}, requiring a scheme to find the initially constant values for each pitch-angle, which might compromise the results as they may not always be reproducible. \textbf{Additionally, this method yields a constant running PADC within the first 10 gyrations for the parameters considered here, and it can be questioned whether the system is already in a diffusive state. However, since the results of M1b is consistent with the results of M4a, M4b, and M5 \textbf{(Eq.~\ref{eq:DmmStationary})}, it seems that this is a reliable result.}

\begin{table*}[t]
\begin{center}
\caption{\label{tab:MethodSummary}\textbf{Strengths} and \textbf{weaknesses} of the different methods investigated in this study.}
\begin{tabular}{|l|l|l|}
\hline\hline
{\bf Method} & {\bf Pros} & {\bf Cons} \\
\hline\hline
\multirow{3}{*}{M0}  & Valid for anomalous diffusion. Applicable to an                    & Not guaranteed to have the correct statistical \\
                     & isotropic distribution. Time-independent.                          & behaviour for any given time \textbf{interval} to fulfil the \\
                     &                                                                    & assumed Markovian behaviour. \\
\hline\hline
\multirow{4}{*}{M1a} & Valid for anomalous diffusion, even though diffusive               & Constant only over a limited time interval. Breaks \\
                     & behaviour is assumed due to the division by time.                  & down in strong turbulence and at late times. Might \\
                     & Applicable to an isotropic distribution.                           & show unexpected behaviour due to resonance \\
                     &                                                                    & interactions, requiring further interpretation. \\
\hline
\multirow{4}{*}{M1b} & More fundamental than simply dividing by time.                     & Constant only over a limited time interval. Breaks \\
                     & Valid for anomalous diffusion. Applicable to an                    & down in strong turbulence. Might show unexpected \\
                     & isotropic distribution. Yields time-independent                    & behaviour due to resonance interactions. \\
                     & result initially if MSD is first averaged over a                   & Numerically volatile at late times. \\
                     & gyration.                                                          &  \\
\hline\hline
\multirow{3}{*}{M2a} & Limit of $t \longrightarrow \infty$ can be taken. Applicable to an & Unclear if valid for anomalous diffusion. Requires \\
                     & isotropic distribution.                                            & temporal homogeneity. Numerically volatile at \\
                     &                                                                    & late times. $\dot{\mu}(t)$ must be calculated. \\
\hline
\multirow{5}{*}{M2b} & Correlation function is of interest for $D_{\perp}(\mu)$ and       & Correlation function becomes numerically volatile \\
                     & gives information about when the pitch-angles are                  & at late times (when information becomes available \\
                     & decorrelated, possibly indicating normal or                        & about normal/anomalous diffusion). Empirical \\
                     & anomalous diffusion. Applicable to an isotropic                    & relation with unknown proportionality constant. \\
                     & distribution and strong turbulence.                                &  \\
\hline\hline
\multirow{4}{*}{M3}  & \multirow{4}{*}{Applicable to an isotropic distribution.}          & Not guaranteed to have the correct statistical \\
                     &                                                                    & behaviour for any given time \textbf{interval}. Unclear if \\
                     &                                                                    & valid for anomalous diffusion. Numerically volatile \\
                     &                                                                    & at late times. $\dot{\mu}(t)$ must be calculated. \\
\hline\hline
\multirow{3}{*}{M4a} & Integral smooths the time derivative. Only first-                  & Assumes a diffusive system. Requires a non- \\
                     & order derivatives are needed. Yields time-                         & stationary distribution function. Quite noisy, \\
                     & independent result initially if distribution function              & especially at late times. Unreliable in strong \\
                     & is first averaged over a gyration.                                 & turbulence. \\
\hline
\multirow{4}{*}{M4b} & Yields time-independent result initially if                        & Assumes a diffusive system. Requires a non- \\
                     & distribution function is first averaged over a                     & stationary distribution function. Very noisy, \\
                     & gyration.                                                          & especially at late times. Cumbersome matrix \\
                     &                                                                    & inversion. Unreliable in strong turbulence. \\
\hline\hline
\multirow{4}{*}{M5}  &                                                                 & Assumes a diffusive system. Requires a delta- \\
                     & Time-independent. Computationally efficient.                    & injection and absorbing boundaries (i.e., a \\
                     & Works for strong turbulence.                                    & specialized setup rendering code unsuitable \\
                     &                                                                 & to calculate isotropic diffusion coefficients). \\
\hline\hline
\end{tabular}
\end{center}
\end{table*}

M4a and M4b can be averaged over much longer times than M1b \textbf{(compare the dashed horizontal lines in the bottom right panel of Fig.~\ref{fig:BM1msd} to the bottom panels of Fig.~\ref{fig:BM1deDmmT} or the bottom right panel of Fig.~\ref{fig:BM2msd} to both panels of Fig.~\ref{fig:BM2deDmmT})}, providing more confidence in the applicability of a diffusion equation, but these two methods are noisier than M1b (a three-point average in pitch-cosine was needed after the long temporal average to further reduce the noise to an acceptable level). Furthermore, M4a is easier to implement than M4b, as it does not require a matrix inversion, and better results are found for \textbf{the initially triangular distribution if M4a is} applied separately from $\mu = -1$ to $\mu = 0$ and from $\mu = 1$ to $\mu = 0$, although this could make the result at $\mu = 0$ unreliable. Unfortunately, these three methods all fail if the turbulence strength is increased \textbf{(see the bottom right panel of Fig.~\ref{fig:BM3msd} and both panels of Fig.~\ref{fig:BM3deDmmT})}. Only M2b and M5 yielded comparable results in the case of strong pure slab turbulence \textbf{(see Fig.~\ref{fig:BM3padc}). Since we found that M5 is sensitive to the source location in strong turbulence (see the top left panel of Fig.~\ref{fig:BM3BestPracticeM5}), we showed how a Kolmogorov-Smirnov test can be applied to this method to provide more confidence in the source location.}

\begin{table*}[t]
\begin{center}
\caption{\label{tab:CriteriaSummary}Criteria in selecting which of the different methods are suitable for a numerical investigation of the PADC.}
\begin{tabular}{|l|ccccccccc|}
\hline\hline
{\bf Criteria} & {\bf M0} & {\bf M1a} & {\bf M1b} & {\bf M2a} & {\bf M2b} & {\bf M3} & {\bf M4a} & {\bf M4b} & {\bf M5} \\
\hline\hline
Reproducible and viable for a large range of particle & \multirow{2}{*}{No} & \multirow{2}{*}{No} & \multirow{2}{*}{No}  & \multirow{2}{*}{No} & \multirow{2}{*}{No}      & \multirow{2}{*}{No} & \multirow{2}{*}{No}  & \multirow{2}{*}{No}  & \multirow{2}{*}{Yes} \\
energies and turbulence conditions?                   &                     &                     &                      &                     &                          &                     &                            &                      &  \\
\hline
Are results in consensus with other methods and       & \multirow{2}{*}{No} & \multirow{2}{*}{No} & \multirow{2}{*}{Yes} & \multirow{2}{*}{No} & \multirow{2}{*}{\bf Yes} & \multirow{2}{*}{No} & \multirow{2}{*}{Yes} & \multirow{2}{*}{Yes} & \multirow{2}{*}{\bf Yes} \\
previous simulation results?${}^{*}$                  &                     &                     &                      &                     &                          &                     &                            &                      &  \\
\hline
Relatively easy to implement without compromising     & \multirow{2}{*}{No} & \multirow{2}{*}{No} & \multirow{2}{*}{No}  & \multirow{2}{*}{No} & \multirow{2}{*}{No}      & \multirow{2}{*}{No} & \multirow{2}{*}{Yes} & \multirow{2}{*}{No}  & \multirow{2}{*}{Yes} \\
on the quality of results?                            &                     &                     &                      &                     &                          &                     &                            &                      &  \\
\hline
In fair agreement with QLT for low turbulence levels  & \multirow{2}{*}{No} & \multirow{2}{*}{No} & \multirow{2}{*}{Yes}  & \multirow{2}{*}{No} & \multirow{2}{*}{No}     & \multirow{2}{*}{No} & \multirow{2}{*}{Yes} & \multirow{2}{*}{Yes} & \multirow{2}{*}{Yes} \\
away from $\mu = 0$?                                  &                     &                     &                      &                     &                          &                     &                            &                      &  \\
\hline\hline
Recommended for heliospheric turbulence conditions? & No & No & No & No & No & No & Yes & No & Yes \\
\hline\hline
\end{tabular}
\end{center}
${}^{*}$ Only the bold-faced methods agree with each other for high turbulence levels.
\end{table*}

It is challenging to compare M1b, M2b, M4a, M4b, and M5 with one another using Fig.~\ref{fig:BM1padc}, Fig.~\ref{fig:BM2padc}, or Fig.~\ref{fig:BM3padc}. Hence, Fig.~\ref{fig:BMCompareMethods} shows the results of these methods from all three sections divided by the prediction from theory to remove the general trend and focus on the differences between the methods. All these methods seem to deviate close to $|\mu| = 1$, but the PADC is expected to follow a $1 - \mu^2$ dependence close to $|\mu| = 1$ and the important region is around $\mu = 0$. Fortunately, it is possible to find slightly better estimates for $|\mu| = 1$ and $\mu = 0$ by extrapolating straight lines from the neighbouring bins. Comparing M4b to the other four methods, it appears to have the most noise and \textbf{large} values at $\mu = 0$ for anisotropic scattering (top and middle panels). Along with the complexity of performing a matrix inversion, M4b seems computationally less attractive than the other three methods (not counting M2b). \textbf{M1b,} M4a, and M4b break down for strong turbulence levels (bottom panel\textbf{; also see the bottom right panel of Fig.~\ref{fig:BM3msd} and both panels of Fig.~\ref{fig:BM3deDmmT}}). Additionally, compared to the computational speed of M5, the other three methods (M1b, M2b, and M4a) also seem computationally less attractive. Unfortunately, the simulation setup needed for these methods are all different (i.e., an isotropic initial distribution for M1b and M2b, a non-stationary initial distribution for M4a and M4b, and a continuous point source with absorbing boundaries for M5), otherwise, it would have been easy to apply all these methods simultaneously to a simulation. M4a and M4b could use an isotropic distribution if each particle is given a weight according to its initial pitch-cosine and the desired non-stationary initial distribution while binning the particle weights. However, this would make the distribution function noisy as the statistics in a bin will be governed by the particles with the largest weights.

Users of M4a, M4b, and M5 are cautioned to check either the temporal evolution of the pitch-angle distribution function from a delta injection or the pitch-angle correlation function for possible signatures of anomalous diffusion when utilizing methods based on diffusion equations \citep[see][for the possible signatures in these quantities]{ZimbardoPerri2020}. The pros and cons of all the methods are summarised in Table~\ref{tab:MethodSummary}. To decide which method to ultimately use, we require that the method should: 1) be consistent in terms of being reproducible and viable for a large range of particle energies and turbulence conditions; 2) yield results in consensus with other methods and previous simulation results, unless the simulation results of a particular study are not reproducible by any method; 3) be relatively easy to implement without yielding results that are significantly different from other methods; and finally 4) be in fair agreement with QLT, at least for low levels of pure slab turbulence and away from $\mu = 0$. Only M5 seems to satisfy the first criterion\textbf{, with strong turbulence being the most limiting factor here}; the second criterion is generally satisfied by M1b, M2b, M4a, M4b, and M5, except in the strong turbulence case where only M2b and M5 yield comparable results; M5 certainly fulfils the third criterion, but so does M4a if one is willing to spend some time to get rid of the noise; \textbf{M1b,} M4a, \textbf{M4b,} and M5 satisfy the last criterion. \textbf{A summary of this is given in Table~\ref{tab:CriteriaSummary}.} M5 would be the most broadly applicable method if it can be verified that the results are due to truly diffusive behaviour \textbf{(although we recommend a statistical procedure, as illustrated in Sec.~\ref{subsec:BestPractice}, if diffusive behaviour cannot be verified)}, with the added benefit of enhanced \textbf{computational} speed for higher levels of turbulence. M4a also represents a robust method, but its use should be limited to cases where lower levels of turbulence are assumed. Ideally, however, both methods can be implemented \textbf{when doing an investigation of the PADC under heliospheric conditions}, with \textbf{M1b or} M2b providing \textbf{an alternative comparison. To summarize: the calculation of PADCs in pre-specified turbulence conditions using numerical test particle simulations is a delicate procedure, requiring careful modeling, analysis, and interpretation of results and is, as such, akin to the disentangling of the idiomatic rope of sand. Nevertheless, in this case it is indeed possible to glean some information as to this fundamental quantity from simulations, should sufficient care be exercised, and the caveats listed in this study be taken into account. This will be the subject of future work.}


\begin{acknowledgments}

This work is based on the research supported in part by the National Research Foundation of South Africa (NRF grant number 137793). Opinions expressed and conclusions arrived at are those of the authors and are not necessarily to be attributed to the NRF. JPvdB wishes to thank H. Fichtner, F. Effenberger, R.D. Strauss, and J. Light for discussions on these methods. PLE acknowledges the National Institute for Theoretical and Computational Sciences (NITheCS) South Africa for funding support during this research. \textbf{This research was supported by the International Space Science Institute (ISSI) in Bern, through ISSI International Team project \#24-608. We thank the anonymous reviewer for valuable comments that improved the overall readability and layout of this article.}

\end{acknowledgments}

\software{FFTW \citep{FrigoJohnson2005},
          Matplotlib \citep{Hunter2007},
          NumPy \citep{HarrisEA2020},
          OpenMP \citep{ChandraEA2001},
          SciPy \citep{VirtanenEA2020}}


\appendix

\section{Adaption of the Giacalone et al. (1999) Method to Momentum Space}
\label{apndx:DeriveGea99}

The following derivation \textbf{of Eq.~\ref{eq:DmmGea99} in Sec.~\ref{subsec:Gea99}} is based on that of \citet{GiacaloneEA1999} and \citet{Minnie2006}. The distribution function at a time $t + \Delta t$, $f(\vec{p}', t + \Delta t)$, after the momentum has changed from $\vec{p}$ to $\vec{p}'$ by an amount $\Delta \vec{p} = \vec{p}' - \vec{p}$, is related to the distribution function at time $t$, $f(\vec{p}, t)$, by
\begin{subequations}
\begin{align}
f(\vec{p}', t + \Delta t) & \approx f(\vec{p}, t) + \Delta \vec{p} \cdot \vec{\nabla}_p f(\vec{p}, t) \\
 & = f(p, \alpha, \varphi, t) + \Delta p \frac{\partial f}{\partial p} + \frac{\Delta \alpha}{p} \frac{\partial f}{\partial \alpha} + \frac{\Delta \varphi}{p \sin \alpha} \frac{\partial f}{\partial \varphi} \\
\label{eq:TaylorExpandDistFunc}
 & = f(p, \mu, \varphi, t) + \Delta p \frac{\partial f}{\partial p} + \frac{\Delta \mu}{p} \frac{\partial f}{\partial \mu} + \frac{\Delta \varphi}{p \sqrt{1 - \mu^2}} \frac{\partial f}{\partial \varphi} ,
\end{align}
\end{subequations}
assuming that $\Delta t$ and $\Delta \vec{p}$ are small. A coordinate transformation was performed in momentum space from Cartesian $(p_x, p_y, p_z)$ to spherical $(p, \alpha, \varphi)$ coordinates, where $p$ is the magnitude of the momentum vector, $\alpha$ is the pitch-angle, and $\varphi$ is the gyro-phase, and the pitch-cosine, $\mu = \cos \alpha$, was used to calculate ${\rm d}\mu = - \sin \alpha \, {\rm d}\alpha \Longrightarrow {\rm d}\alpha = - {\rm d}\mu / \sqrt{1 - \mu^2}$. The gyro-phase-averaged streaming flux in momentum space can be written as
\begin{equation}
\vec{\Gamma}_p(\vec{p},t) = \frac{1}{2 \pi} \int_0^{2 \pi} \! \frac{{\rm d}\vec{p}}{{\rm d}t} f(\vec{p},t) \, {\rm d}\varphi
\end{equation}
with
\begin{subequations}
\begin{align}
\frac{{\rm d}\vec{p}}{{\rm d}t} & = \frac{\rm d}{{\rm d}t} \left[ p \left( \sqrt{1 - \mu^2} \cos \varphi \, \uvec{e}_{\perp 1} + \sqrt{1 - \mu^2} \sin \varphi \, \uvec{e}_{\perp 2} + \mu \, \uvec{b} \right) \right] \\
 & = \frac{\vec{p}}{p} \dot{p} + p \sqrt{1 - \mu^2} \left( \cos \varphi \, \uvec{e}_{\perp 2} - \sin \varphi \, \uvec{e}_{\perp 1} \right) \dot{\varphi} + p \left( \uvec{b} - \mu \frac{\cos \varphi \, \uvec{e}_{\perp 1} + \sin \varphi \, \uvec{e}_{\perp 2}}{\sqrt{1 - \mu^2}} \right) \dot{\mu}
\end{align}
\end{subequations}
for a stationary magnetic field, where $\uvec{e}_{\perp 1}$ and $\uvec{e}_{\perp 2}$ are two orthogonal unit vectors perpendicular to the unit vector along the magnetic field $\uvec{b}$ and the dot indicates a temporal derivative. Assuming magnetostatic turbulence, such that $\Delta p = \dot{p} = 0$ (making this valid for any process causing small-angle elastic pitch-angle scattering), and a gyrotropic distribution, such that $f$ is independent of $\varphi$, the flux is then
\begin{subequations}
\begin{align}
\vec{\Gamma}_p & \approx \frac{1}{2 \pi} \int_0^{2 \pi} \left[ \vec{0} + p \sqrt{1 - \mu^2} \left( \cos \varphi \, \uvec{e}_{\perp 2} - \sin \varphi \, \uvec{e}_{\perp 1} \right) \dot{\varphi} + p \left( \uvec{b} - \mu \frac{\cos \varphi \, \uvec{e}_{\perp 1} + \sin \varphi \, \uvec{e}_{\perp 2}}{\sqrt{1 - \mu^2}} \right) \dot{\mu} \right] f(p,\mu,t) \, {\rm d}\varphi \\
 & \approx \vec{0} + \frac{p}{2 \pi} \int_0^{2 \pi} \! \left( \uvec{b} - \mu \frac{\cos \varphi \, \uvec{e}_{\perp 1} + \sin \varphi \, \uvec{e}_{\perp 2}}{\sqrt{1 - \mu^2}} \right) \dot{\mu} \left[ f(p', \mu', t + \Delta t) - \frac{\partial f(p, \mu, t)}{\partial \mu} \frac{\Delta \mu}{p} - 0 - 0 \right] {\rm d}\varphi \\
 & \approx \vec{0} - \frac{1}{2 \pi} \int_0^{2 \pi} \!\! \dot{\mu} \, \Delta \mu \, {\rm d}\varphi \, (\uvec{b} - \vec{0}) \frac{\partial f(p, \mu, t)}{\partial \mu} ,
\end{align}
\end{subequations}
where it is assumed that $\dot{\varphi}$ is independent of $\varphi$\textbf{, Eq.~\ref{eq:TaylorExpandDistFunc} was substituted,} and that ${\rm d}\vec{p}/{\rm d}t$ and $f(\vec{p}', t + \Delta t)$ are independent of each other, such that their gyro-average is zero. Assuming that the pitch-angle changes will be diffusive for long enough times, Fick's law,
\begin{equation}
\vec{\Gamma}_p(\vec{p},t) = - D_{ij} \frac{\partial f(\vec{p}, t)}{\partial p_j} ,
\end{equation}
will apply and implies that the PADC can be identified as
\begin{equation}
D_{\mu\mu}(\mu,t) \approx \left\langle \dot{\mu}(t) [\mu(t + \Delta t) - \mu(t)] \right\rangle .
\end{equation}


\section{Matrix Equation of the Discrete Derivative Form of the Diffusion Equation}
\label{apndx:MatrixEquation}

Eq.~\ref{eq:DmmMatrixInvertDiffEq} in Sec.~\ref{subsec:TPE} forms a matrix equation given by
\begin{equation}
\begin{bmatrix}
\partial_{\mu\mu} \hat{f}_0^s \! - \! \frac{\partial_{\mu} \hat{f}_0^s}{\varpi_{\mu}} & \frac{\partial_{\mu} \hat{f}_0^s}{\varpi_{\mu}} & 0 & 0 & \cdots & 0 & 0 & 0 & 0\\
- \frac{\partial_{\mu} \hat{f}_1^s}{2 \, \varpi_{\mu}} & \partial_{\mu\mu} \hat{f}_1^s & \frac{\partial_{\mu} \hat{f}_1^s}{2 \, \varpi_{\mu}} & 0 & \cdots & 0 & 0 & 0 & 0 \\
0 & \ddots & \ddots & \ddots & \vdots & \vdots & \vdots & \vdots & \vdots \\
0 & \cdots & 0 & - \frac{\partial_{\mu} \hat{f}_m^s}{2 \, \varpi_{\mu}} & \partial_{\mu\mu} \hat{f}_m^s & \frac{\partial_{\mu} \hat{f}_m^s}{2 \, \varpi_{\mu}} & 0 & \cdots & 0 \\
\vdots & \vdots & \vdots & \vdots & \vdots & \ddots & \ddots & \ddots & 0 \\
0 & 0 & 0 & 0 & \cdots & 0 & - \frac{\partial_{\mu} \hat{f}_{M-1}^s}{2 \, \varpi_{\mu}} & \partial_{\mu\mu} \hat{f}_{M-1}^s & \frac{\partial_{\mu} \hat{f}_{M-1}^s}{2 \, \varpi_{\mu}} \\
0 & 0 & 0 & 0 & \cdots & 0 & 0 & - \frac{\partial_{\mu} \hat{f}_M^s}{\varpi_{\mu}} & \partial_{\mu\mu} \hat{f}_M^s \! + \! \frac{\partial_{\mu} \hat{f}_M^s}{\varpi_{\mu}}
\end{bmatrix} \cdot \begin{bmatrix}
\hat{D}_{\mu\mu}(\mu_0, t_s) \\
\hat{D}_{\mu\mu}(\mu_1, t_s) \\
\vdots \\
\hat{D}_{\mu\mu}(\mu_m, t_s) \\
\vdots \\
\hat{D}_{\mu\mu}(\mu_{M-1}, t_s) \\
\hat{D}_{\mu\mu}(\mu_M, t_s)
\end{bmatrix} \!\! = \! \begin{bmatrix}
\partial_t \hat{f}_0^s \\
\partial_t \hat{f}_1^s \\
\vdots \\
\partial_t \hat{f}_m^s \\
\vdots \\
\partial_t \hat{f}_{M-1}^s \\
\partial_t \hat{f}_M^s
\end{bmatrix} ,
\end{equation}
where the shorthand notation $\partial_q \hat{f}_m^s$ is used to denote a discrete partial derivative of $\hat{f}$ with respect to the variable $q$ at the point $(\mu_m, t_s)$.


\section{Quasi-linear Theory}
\label{apndx:QLT}

The most basic and widely used theory for pitch-angle scattering is that of quasi-linear theory \citep[QLT;][]{Jokipii1966, Shalchi2009}. As \textbf{its} derivation can be found in numerous references \citep[see, e.g.,][]{Schlickeiser1989, TeufelSchlickeiser2002, Shalchi2009, QinShalchi2009}, we present here the final expression for a quick comparison to the results \textbf{of Sec.~\ref{sec:Benchmark}}. QLT predicts no pitch-angle scattering due to 2D turbulence in the magnetostatic case and the PADC from slab turbulence as
\begin{equation}
\label{eq:DmmQLT}
D_{\mu\mu}(\mu) = \frac{2 \pi^2 \Omega^2 (1 - \mu^2)}{B_0^2 |\mu| v} g_{\rm slab} \left( \left| \frac{\Omega}{\mu v} \right| \right) = \frac{\sqrt{\pi} \, \Gamma (\nu/2) \, v}{2 \, \Gamma (\nu/2 - 1/2) \, l_b^{\rm slab}} \frac{\delta B_{\rm slab}^2}{B_0^2} \, R^{\nu-2} (1 - \mu^2) |\mu|^{\nu-1} (1 + \mu^2 R^2)^{-\nu/2}
\end{equation}
for the slab spectrum of Eq.~\ref{eq:SlabSpectrum}, where $\Omega = q B_0 / \gamma m_0$ is the (signed) cyclotron frequency and $R = R_L / l_b^{\rm slab}$, with $R_L = v / |\Omega|$ the maximal Larmor radius. Since the results are inherently time-dependent in many of the methods discussed here, it would be useful to compare them to the expected running PADC calculated from the TGK approach (Eq.~\ref{eq:DmmTGK}),
\begin{subequations}
\label{eq:DmmtQLT}
\begin{align}
D_{\mu\mu}(\mu, t) & = \frac{2 \pi \Omega^2 (1 - \mu^2)}{B_0^2} \! \int_0^{\infty} \!\! g_{\rm slab}(k_{\parallel}) \! \sum_{n = \pm 1} \! \frac{\sin \left( |k_{\parallel} \mu v - n \Omega| t \right)}{|k_{\parallel} \mu v - n \Omega|} \, {\rm d}k_{\parallel} \\
 & = \frac{\Gamma(\nu/2) \, v}{2 \sqrt{\pi} \, \Gamma(\nu/2 - 1/2) \, l_b^{\rm slab}} \frac{\delta B_{\rm slab}^2}{B_0^2} \frac{1 - \mu^2}{|\mu| R^2} \int_0^{\infty} \! \left[ 1 + \left( l_b^{\rm slab} k_{\parallel} \right)^2 \right]^{-\nu/2} \!\! \sum_{n = \pm 1} \! \frac{\sin \left( |\mu| v t |k_{\parallel} - n \Omega / \mu v| \right)}{|k_{\parallel} - n \Omega / \mu v|} \, {\rm d}k_{\parallel} .
\end{align}
\end{subequations}
Unfortunately, this expression does not reduce analytically to the former when the limit of $t \longrightarrow \infty$ is taken, but it does yield small oscillations around the values from the time-independent result (see, e.g., the \textbf{right panel in the third row of Fig.~\ref{fig:BM1msd}}). Additionally, the pitch-angle MSD is \citep{Shalchi2005, TautzEA2013}
\begin{subequations}
\label{eq:MSD_QLT}
\begin{align}
\left\langle (\Delta \mu)^2 \right\rangle & = \frac{4 \pi \Omega^2 (1 - \mu^2)}{B_0^2} \! \int_0^{\infty} \!\!\! g_{\rm slab}(k_{\parallel}) \sum_{n = \pm 1} \frac{1 - \cos [(k_{\parallel} \mu v - n \Omega) t]}{(k_{\parallel} \mu v - n \Omega)^2} \, {\rm d}k_{\parallel} \\
 & = \frac{\Gamma(\nu/2)}{\sqrt{\pi} \, \Gamma(\nu/2 - 1/2) \, l_b^{\rm slab}} \frac{\delta B_{\rm slab}^2}{B_0^2} \frac{1 - \mu^2}{\mu^2 R^2} \int_0^{\infty} \! \left[ 1 + \left( l_b^{\rm slab} k_{\parallel} \right)^2 \right]^{-\nu/2} \!\! \sum_{n = \pm 1} \! \frac{1 - \cos \left( |\mu| v t |k_{\parallel} - n \Omega / \mu v| \right)}{(k_{\parallel} - n \Omega / \mu v)^2} \, {\rm d}k_{\parallel} .
\end{align}
\end{subequations}
It is interesting to note that applying Eq.~\ref{eq:DmmMSDdt} to this yields the same running PADC as in Eq.~\ref{eq:DmmtQLT} from the TGK approach, but that Eq.~\ref{eq:DmmMSDt} yields sub-diffusive behaviour analytically if the limit $t \longrightarrow \infty$ is taken. The resulting parallel diffusion coefficient is given by
\begin{subequations}
\begin{align}
\label{eq:KappaParallel}
\kappa_{\parallel} & = \frac{v^2}{8} \int_{-1}^1 \! \frac{(1 - \mu^2)^2}{D_{\mu\mu}(\mu)} \, {\rm d}\mu \\
\label{eq:MFP_QLT}
 & = \frac{\Gamma (\nu/2 - 1/2) \, l_b^{\rm slab} v}{2 \sqrt{\pi} \, \Gamma (\nu/2)} \frac{B_0^2}{\delta B_{\rm slab}^2} \, R^{2-\nu} \left[ \frac{1}{\nu-4} \, {}_2F_1 \left( 2 - \frac{\nu}{2}, - \frac{\nu}{2}; 3 - \frac{\nu}{2}; - R^2 \right) - \frac{1}{\nu-2} \, {}_2F_1 \left( 1 - \frac{\nu}{2}, - \frac{\nu}{2}; 2 - \frac{\nu}{2}; - R^2 \right) \right] ,
\end{align}
\end{subequations}
where ${}_2F_1(\alpha, \beta; \gamma; z)$ is the hypergeometric function \citep{ShalchiEA2004a, Shalchi2009}.



\bibliography{sample631}{}

\begin{thebibliography}{}
\expandafter\ifx\csname natexlab\endcsname\relax\def\natexlab#1{#1}\fi
\providecommand{\url}[1]{\href{#1}{#1}}
\providecommand{\dodoi}[1]{doi:~\href{http://doi.org/#1}{\nolinkurl{#1}}}
\providecommand{\doeprint}[1]{\href{http://ascl.net/#1}{\nolinkurl{http://ascl.net/#1}}}
\providecommand{\doarXiv}[1]{\href{https://arxiv.org/abs/#1}{\nolinkurl{https://arxiv.org/abs/#1}}}

\bibitem[{Adhikari {et~al.}(2021)Adhikari, Zank, \& Zhao}]{AdhikariEA2021}
Adhikari, L., Zank, G.~P., \& Zhao, L. 2021, Fluids, 6, 368,
  \dodoi{10.3390/fluids6100368}

\bibitem[{Bieber {et~al.}(1994)Bieber, Matthaeus, Smith, Wanner, Kallenrode, \&
  Wibberenz}]{BieberEA1994}
Bieber, J.~W., Matthaeus, W.~H., Smith, C.~W., {et~al.} 1994, \apj, 420, 294,
  \dodoi{10.1086/173559}

\bibitem[{Bieber {et~al.}(1988)Bieber, Smith, \& Matthaeus}]{BieberEA1988}
Bieber, J.~W., Smith, C.~W., \& Matthaeus, W.~H. 1988, \apj, 334, 470,
  \dodoi{10.1086/166851}

\bibitem[{Bieber {et~al.}(1996)Bieber, Wanner, \& Matthaeus}]{BieberEA1996}
Bieber, J.~W., Wanner, W., \& Matthaeus, W.~H. 1996, \jgr, 101, 2511,
  \dodoi{10.1029/95JA02588}

\bibitem[{Brown {et~al.}(2018)Brown, Eddelbuettel, \& Bauer}]{BrownEA2018}
Brown, R.~G., Eddelbuettel, D., \& Bauer, D. 2018, Dieharder, Tech. rep., Duke
  University Physics Department Durham, NC

\bibitem[{Burger {et~al.}(2022)Burger, Nel, \& Engelbrecht}]{BurgerEA2022}
Burger, R.~A., Nel, A.~E., \& Engelbrecht, N.~E. 2022, \apj, 926, 128,
  \dodoi{10.3847/1538-4357/ac4741}

\bibitem[{Chandra {et~al.}(2001)Chandra, Dagum, Kohr, Menon, Maydan, \&
  {McDonald}}]{ChandraEA2001}
Chandra, R., Dagum, L., Kohr, D., {et~al.} 2001, {Parallel programming in
  OpenMP} (Morgan Kaufmann)

\bibitem[{Chandrasekhar(1943)}]{Chandrasekhar1943}
Chandrasekhar, S. 1943, Reviews of Modern Physics, 15, 1,
  \dodoi{10.1103/RevModPhys.15.1}

\bibitem[{Dundovic {et~al.}(2020)Dundovic, Pezzi, Blasi, Evoli, \&
  Matthaeus}]{DundovicEA2020}
Dundovic, A., Pezzi, O., Blasi, P., Evoli, C., \& Matthaeus, W.~H. 2020, \prd,
  102, 103016, \dodoi{10.1103/PhysRevD.102.103016}

\bibitem[{Els \& Engelbrecht(2024)}]{ElsEngelbrecht2024}
Els, P.~L., \& Engelbrecht, N.~E. 2024, \apj, 969, 51,
  \dodoi{10.3847/1538-4357/ad479c}

\bibitem[{Engelbrecht(2017)}]{Engelbrecht2017}
Engelbrecht, N.~E. 2017, \apjl, 849, L15, \dodoi{10.3847/2041-8213/aa9372}

\bibitem[{Engelbrecht(2019)}]{Engelbrecht2019}
---. 2019, \apj, 880, 60, \dodoi{10.3847/1538-4357/ab2871}

\bibitem[{Engelbrecht {et~al.}(2022)Engelbrecht, Effenberger, Florinski,
  Potgieter, Ruffolo, Chhiber, Usmanov, Rankin, \& Els}]{EngelbrechtEA2022}
Engelbrecht, N.~E., Effenberger, F., Florinski, V., {et~al.} 2022, \ssr, 218,
  33, \dodoi{10.1007/s11214-022-00896-1}

\bibitem[{Fisk {et~al.}(1974)Fisk, Goldstein, Klimas, \& Sandri}]{FiskEA1974}
Fisk, L.~A., Goldstein, M.~L., Klimas, A.~J., \& Sandri, G. 1974, \apj, 190,
  417, \dodoi{10.1086/152893}

\bibitem[{Florinski(2024)}]{Florinski2024}
Florinski, V. 2024, Journal of Geophysical Research (Space Physics), 129,
  e2024JA032579, \dodoi{10.1029/2024JA032579}

\bibitem[{Frigo \& Johnson(2005)}]{FrigoJohnson2005}
Frigo, M., \& Johnson, S.~G. 2005, in Proceedings of the IEEE, Vol.~93, Special
  issue on ``Program Generation, Optimization, and Platform Adaptation'',
  216--231

\bibitem[{Giacalone \& Jokipii(1999)}]{GiacaloneJokipii1999}
Giacalone, J., \& Jokipii, J.~R. 1999, \apj, 520, 204, \dodoi{10.1086/307452}

\bibitem[{Giacalone {et~al.}(1999)Giacalone, Jokipii, \&
  K\'ota}]{GiacaloneEA1999}
Giacalone, J., Jokipii, J.~R., \& K\'ota, J. 1999, in International Cosmic Ray
  Conference, Vol.~7, 26th International Cosmic Ray Conference (ICRC26), Volume
  7, 37

\bibitem[{Harris {et~al.}(2020)Harris, Millman, {van der Walt}, Gommers,
  Virtanen, Cournapeau, Wieser, Taylor, Berg, Smith, Kern, Picus, Hoyer, {van
  Kerkwijk}, Brett, Haldane, {del R{\'\i}o}, Wiebe, Peterson,
  {G{\'e}rard-Marchant}, Sheppard, Reddy, Weckesser, Abbasi, Gohlke, \&
  Oliphant}]{HarrisEA2020}
Harris, C.~R., Millman, K.~J., {van der Walt}, S.~J., {et~al.} 2020, \nat, 585,
  357, \dodoi{10.1038/s41586-020-2649-2}

\bibitem[{Higuera \& Cary(2017)}]{HigueraCary2017}
Higuera, A.~V., \& Cary, J.~R. 2017, Physics of Plasmas, 24, 052104,
  \dodoi{10.1063/1.4979989}

\bibitem[{Huang {et~al.}(2010)Huang, Schmitt, Lu, Fougairolles, Gagne, \&
  Liu}]{HuangEA2010}
Huang, Y.~X., Schmitt, F.~G., Lu, Z.~M., {et~al.} 2010, \pre, 82, 026319,
  \dodoi{10.1103/PhysRevE.82.026319}

\bibitem[{Hunter(2007)}]{Hunter2007}
Hunter, J.~D. 2007, Computing in Science \& Engineering, 9, 90,
  \dodoi{10.1109/MCSE.2007.55}

\bibitem[{Isenberg(2005)}]{Isenberg2005}
Isenberg, P.~A. 2005, \apj, 623, 502, \dodoi{10.1086/428609}

\bibitem[{Ivascenko {et~al.}(2016)Ivascenko, Lange, Spanier, \&
  Vainio}]{IvascenkoEA2016}
Ivascenko, A., Lange, S., Spanier, F., \& Vainio, R. 2016, \apj, 833, 223,
  \dodoi{10.3847/1538-4357/833/2/223}

\bibitem[{Jokipii(1966)}]{Jokipii1966}
Jokipii, J.~R. 1966, \apj, 146, 480, \dodoi{10.1086/148912}

\bibitem[{Kaiser {et~al.}(1978)Kaiser, Birmingham, \& Jones}]{KaiserEA1978}
Kaiser, T.~B., Birmingham, T.~J., \& Jones, F.~C. 1978, Physics of Fluids, 21,
  361, \dodoi{10.1063/1.862234}

\bibitem[{{le Roux} \& Webb(2007)}]{leRouxWebb2007}
{le Roux}, J.~A., \& Webb, G.~M. 2007, \apj, 667, 930, \dodoi{10.1086/520954}

\bibitem[{Matsumoto \& Nishimura(1998)}]{MatsumotoNishimura1998}
Matsumoto, M., \& Nishimura, T. 1998, ACM Transactions on Modeling and Computer
  Simulation (TOMACS), 8, 3

\bibitem[{Matthaeus {et~al.}(1990)Matthaeus, Goldstein, \&
  Roberts}]{MatthaeusEA1990}
Matthaeus, W.~H., Goldstein, M.~L., \& Roberts, D.~A. 1990, \jgr, 95, 20673,
  \dodoi{10.1029/JA095iA12p20673}

\bibitem[{Matthaeus {et~al.}(2003)Matthaeus, Qin, Bieber, \&
  Zank}]{MatthaeusEA2003}
Matthaeus, W.~H., Qin, G., Bieber, J.~W., \& Zank, G.~P. 2003, \apjl, 590, L53,
  \dodoi{10.1086/376613}

\bibitem[{Minnie(2006)}]{Minnie2006}
Minnie, J. 2006, PhD thesis, North-West University, South Africa

\bibitem[{Minnie {et~al.}(2007)Minnie, Bieber, Matthaeus, \&
  Burger}]{MinnieEA2007}
Minnie, J., Bieber, J.~W., Matthaeus, W.~H., \& Burger, R.~A. 2007, \apj, 663,
  1049, \dodoi{10.1086/518765}

\bibitem[{Oughton {et~al.}(2011)Oughton, Matthaeus, Smith, Breech, \&
  Isenberg}]{OughtonEA2011}
Oughton, S., Matthaeus, W.~H., Smith, C.~W., Breech, B., \& Isenberg, P.~A.
  2011, Journal of Geophysical Research (Space Physics), 116, A08105,
  \dodoi{10.1029/2010JA016365}

\bibitem[{Owens(1978)}]{Owens1978}
Owens, A.~J. 1978, \jgr, 83, 1673, \dodoi{10.1029/JA083iA04p01673}

\bibitem[{P{\'e}tri(2017)}]{Petri2017}
P{\'e}tri, J. 2017, Journal of Plasma Physics, 83, 705830206,
  \dodoi{10.1017/S0022377817000307}

\bibitem[{Pine {et~al.}(2020)Pine, Smith, Hollick, Argall, Vasquez, Isenberg,
  Schwadron, Joyce, Sok{\'o}{\l}, Bzowski, Kubiak, Hamilton, {McLaurin}, \&
  Leamon}]{PineEA2020}
Pine, Z.~B., Smith, C.~W., Hollick, S.~J., {et~al.} 2020, \apj, 900, 92,
  \dodoi{10.3847/1538-4357/abab0f}

\bibitem[{Pleumpreedaporn \& Snodin(2019)}]{PleumpreedapornSnodin2019}
Pleumpreedaporn, C., \& Snodin, A.~P. 2019, in Journal of Physics Conference
  Series, Vol. 1380, Journal of Physics Conference Series, 012141,
  \dodoi{10.1088/1742-6596/1380/1/012141}

\bibitem[{Qin(2002)}]{Qin2002}
Qin, G. 2002, PhD thesis, University of Delaware, USA

\bibitem[{Qin \& Shalchi(2009)}]{QinShalchi2009}
Qin, G., \& Shalchi, A. 2009, \apj, 707, 61, \dodoi{10.1088/0004-637X/707/1/61}

\bibitem[{Qin \& Shalchi(2014)}]{QinShalchi2014}
---. 2014, Applied Physics Research, 6, 1

\bibitem[{Riordan \& {Pe'er}(2019)}]{RiordanPeer2019}
Riordan, J.~D., \& {Pe'er}, A. 2019, \apj, 873, 13,
  \dodoi{10.3847/1538-4357/aaffd2}

\bibitem[{Ripperda {et~al.}(2018)Ripperda, Bacchini, Teunissen, Xia, Porth,
  Sironi, Lapenta, \& Keppens}]{RipperdaEA2018}
Ripperda, B., Bacchini, F., Teunissen, J., {et~al.} 2018, \apjs, 235, 21,
  \dodoi{10.3847/1538-4365/aab114}

\bibitem[{Ruffolo {et~al.}(2006)Ruffolo, Chuychai, \&
  Matthaeus}]{RuffoloEA2006}
Ruffolo, D., Chuychai, P., \& Matthaeus, W.~H. 2006, \apj, 644, 971,
  \dodoi{10.1086/503625}

\bibitem[{Ruffolo {et~al.}(2012)Ruffolo, Pianpanit, Matthaeus, \&
  Chuychai}]{RuffoloEA2012}
Ruffolo, D., Pianpanit, T., Matthaeus, W.~H., \& Chuychai, P. 2012, \apjl, 747,
  L34, \dodoi{10.1088/2041-8205/747/2/L34}

\bibitem[{Sakai \& Kato(1984)}]{SakaiKato1984}
Sakai, T., \& Kato, M. 1984, Journal of Geomagnetism and Geoelectricity, 36,
  33, \dodoi{10.5636/jgg.36.33}

\bibitem[{Schlickeiser(1989)}]{Schlickeiser1989}
Schlickeiser, R. 1989, \apj, 336, 243, \dodoi{10.1086/167009}

\bibitem[{Shalchi(2005)}]{Shalchi2005}
Shalchi, A. 2005, Physics of Plasmas, 12, 052905, \dodoi{10.1063/1.1895805}

\bibitem[{Shalchi(2009)}]{Shalchi2009}
---. 2009, {Nonlinear cosmic ray diffusion theories}, Vol. 362
  (Springer-Verlag), \dodoi{10.1007/978-3-642-00309-7}

\bibitem[{Shalchi(2020)}]{Shalchi2020}
---. 2020, \ssr, 216, 23, \dodoi{10.1007/s11214-020-0644-4}

\bibitem[{Shalchi {et~al.}(2004{\natexlab{a}})Shalchi, Bieber, \&
  Matthaeus}]{ShalchiEA2004a}
Shalchi, A., Bieber, J.~W., \& Matthaeus, W.~H. 2004{\natexlab{a}}, \apj, 604,
  675, \dodoi{10.1086/382128}

\bibitem[{Shalchi {et~al.}(2004{\natexlab{b}})Shalchi, Bieber, Matthaeus, \&
  Qin}]{ShalchiEA2004b}
Shalchi, A., Bieber, J.~W., Matthaeus, W.~H., \& Qin, G. 2004{\natexlab{b}},
  \apj, 616, 617, \dodoi{10.1086/424839}

\bibitem[{Shalchi {et~al.}(2012)Shalchi, Webb, \& {le Roux}}]{ShalchiEA2012}
Shalchi, A., Webb, G.~M., \& {le Roux}, J.~A. 2012, \physscr, 85, 065901,
  \dodoi{10.1088/0031-8949/85/06/065901}

\bibitem[{Smith {et~al.}(2006)Smith, Isenberg, Matthaeus, \&
  Richardson}]{SmithEA2006}
Smith, C.~W., Isenberg, P.~A., Matthaeus, W.~H., \& Richardson, J.~D. 2006,
  \apj, 638, 508, \dodoi{10.1086/498671}

\bibitem[{Sun {et~al.}(2016)Sun, Jokipii, \& Giacalone}]{SunEA2016}
Sun, P., Jokipii, J.~R., \& Giacalone, J. 2016, \apj, 827, 16,
  \dodoi{10.3847/0004-637X/827/1/16}

\bibitem[{Tautz(2013)}]{Tautz2013}
Tautz, R.~C. 2013, \aap, 558, A148, \dodoi{10.1051/0004-6361/201322143}

\bibitem[{Tautz {et~al.}(2013)Tautz, Dosch, Effenberger, Fichtner, \&
  Kopp}]{TautzEA2013}
Tautz, R.~C., Dosch, A., Effenberger, F., Fichtner, H., \& Kopp, A. 2013, \aap,
  558, A147, \dodoi{10.1051/0004-6361/201322142}

\bibitem[{Tautz {et~al.}(2014)Tautz, Shalchi, \& Dosch}]{TautzEA2014}
Tautz, R.~C., Shalchi, A., \& Dosch, A. 2014, \apj, 794, 138,
  \dodoi{10.1088/0004-637X/794/2/138}

\bibitem[{Teufel \& Schlickeiser(2002)}]{TeufelSchlickeiser2002}
Teufel, A., \& Schlickeiser, R. 2002, \aap, 393, 703,
  \dodoi{10.1051/0004-6361:20021046}

\bibitem[{Usmanov {et~al.}(2016)Usmanov, Goldstein, \&
  Matthaeus}]{UsmanovEA2016}
Usmanov, A.~V., Goldstein, M.~L., \& Matthaeus, W.~H. 2016, \apj, 820, 17,
  \dodoi{10.3847/0004-637X/820/1/17}

\bibitem[{{van den Berg}(2023)}]{vandenBerg2023}
{van den Berg}, J.~P. 2023, PhD thesis, North-West University, South Africa

\bibitem[{Vay(2008)}]{Vay2008}
Vay, J.~L. 2008, Physics of Plasmas, 15, 056701, \dodoi{10.1063/1.2837054}

\bibitem[{Vay(2020)}]{Vay2020}
---. 2020, arXiv e-prints, arXiv:2008.07300, \dodoi{10.48550/arXiv.2008.07300}

\bibitem[{Virtanen {et~al.}(2020)Virtanen, Gommers, Oliphant, Haberland, Reddy,
  Cournapeau, Burovski, Peterson, Weckesser, Bright, {van der Walt}, Brett,
  Wilson, Millman, Mayorov, Nelson, Jones, Kern, Larson, Carey, Polat, Feng,
  Moore, {VanderPlas}, Laxalde, Perktold, Cimrman, Henriksen, Quintero, Harris,
  Archibald, Ribeiro, Pedregosa, {van Mulbregt}, \& {SciPy 1. 0
  Contributors}}]{VirtanenEA2020}
Virtanen, P., Gommers, R., Oliphant, T.~E., {et~al.} 2020, Nature Methods, 17,
  261, \dodoi{10.1038/s41592-019-0686-2}

\bibitem[{Weidl {et~al.}(2015)Weidl, Jenko, Teaca, \&
  Schlickeiser}]{WeidlEA2015}
Weidl, M.~S., Jenko, F., Teaca, B., \& Schlickeiser, R. 2015, \apj, 811, 8,
  \dodoi{10.1088/0004-637X/811/1/8}

\bibitem[{Wiengarten {et~al.}(2016)Wiengarten, Oughton, Engelbrecht, Fichtner,
  Kleimann, \& Scherer}]{WiengartenEA2016}
Wiengarten, T., Oughton, S., Engelbrecht, N.~E., {et~al.} 2016, \apj, 833, 17,
  \dodoi{10.3847/0004-637X/833/1/17}

\bibitem[{Williams \& Zank(1994)}]{WilliamsZank1994}
Williams, L.~L., \& Zank, G.~P. 1994, \jgr, 99, 19229,
  \dodoi{10.1029/94JA01657}

\bibitem[{Zimbardo \& Perri(2020)}]{ZimbardoPerri2020}
Zimbardo, G., \& Perri, S. 2020, \apj, 903, 105,
  \dodoi{10.3847/1538-4357/abb951}

\bibitem[{Zimbardo {et~al.}(2015)Zimbardo, Amato, Bovet, Effenberger, Fasoli,
  Fichtner, Furno, Gustafson, Ricci, \& Perri}]{ZimbardoEA2015}
Zimbardo, G., Amato, E., Bovet, A., {et~al.} 2015, Journal of Plasma Physics,
  81, 495810601, \dodoi{10.1017/S0022377815001117}

\end{thebibliography}
\bibliographystyle{aasjournal}

\end{document}